\documentclass[11pt,letterpaper]{article}
\usepackage[utf8]{inputenc}
\usepackage{amsmath, amsthm, amssymb}
\usepackage[protrusion=false]{microtype}
\AddToHook{para/begin}{\looseness=-1\relax}
\makeatletter
\g@addto@macro\define@newfont{\fontdimen2\font@name=0.95\fontdimen2\font@name\relax}
\makeatother
\usepackage{geometry}
\usepackage{xcolor}
\usepackage{needspace}
\usepackage{tikz}
\usetikzlibrary{positioning}
\usepackage{xspace}
\usepackage{algorithm}
\usepackage{algpseudocode}
\newtheorem{lemma}{Lemma}
\newtheorem{theorem}{Theorem}

\definecolor{darkgreen}{rgb}{0,0.5,0}
\definecolor{darkblue}{rgb}{0,0,0.6}
\usepackage{hyperref}
\hypersetup{
	unicode=false,
	colorlinks=true,
	linkcolor=darkblue,
	citecolor=darkgreen,
	filecolor=magenta,
	urlcolor=cyan
}

\usepackage{graphicx}
\newtheorem{definition}{Definition}
\usepackage{cleveref}
\usepackage{enumerate}

\newlength{\CMparagraphskip}
\makeatletter
\renewcommand\paragraph{\@startsection{paragraph}{4}{\z@}{\CMparagraphskip}{-1em}{\normalfont\normalsize\bfseries}}
\makeatother

\makeatletter
\renewcommand{\theHALG@line}{\arabic{algorithm}.\arabic{ALG@line}}
\makeatother

\tikzset{
  scfail/.style={circle,draw=red!75!black,fill=red!20,
    minimum size=3.4mm,inner sep=0pt,font=\tiny},
  sccrashed/.style={fill=red!70!black,text=white},
  scsurv/.style={circle,draw=green!40!black,fill=green!20,
    minimum size=2.2mm,inner sep=0pt},
  scpath/.style={green!40!black,line width=.8pt},
  scspoke/.style={black!45,line width=.5pt},
  scdropped/.style={red!60!black,dash pattern=on 1pt off 1pt,line width=.5pt},
  scclosing/.style={purple!75!black,dashed,line width=.8pt},
  scwitness/.style={circle,draw=black,line width=1.1pt,
    minimum size=3.6mm,inner sep=0pt},
  scstep/.style={->,>=stealth,black!70,line width=.7pt},
  scgraph/.style={font=\footnotesize\bfseries},
  sccaption/.style={font=\tiny,text=black!70,align=center},
}

\newlength{\scsp}
\newcommand{\scAtomicPanel}[6]{\begin{scope}[shift={(#1,#2)}]
  \foreach \i in {0,...,7}
    \node[scsurv] (q\i) at ({\i*\scsp},0) {};
  \foreach \i/\j in {0/1,1/2,2/3,3/4,4/5,5/6,6/7} \draw[scpath] (q\i)--(q\j);
  \node[scfail] (x1) at ({1.5*\scsp},1.0) {$x_1$};
  \node[scfail] (x2) at ({5.5*\scsp},1.0) {$x_2$};
  \draw[scspoke] (x1)--(x2);
  \foreach \i in {0,1,2,3} \draw[scspoke] (x1)--(q\i);
  \foreach \i in {4,5,6,7} \draw[scspoke] (x2)--(q\i);
  \ifnum#3=1
    \draw[scclosing] (q0) .. controls ({0*\scsp},-.6) and ({7*\scsp},-.6) .. (q7);
  \fi
  \if\relax\detokenize{#4}\relax\else
    \node[scfail,sccrashed] at (x1) {$x_1$};
    \node[sccaption,above=1pt,text=red!75!black] at (x1.north) {round #4};
  \fi
  \if\relax\detokenize{#5}\relax\else
    \node[scfail,sccrashed] at (x2) {$x_2$};
    \node[sccaption,above=1pt,text=red!75!black] at (x2.north) {round #5};
  \fi
  \ifnum#6>-1
    \node[scwitness] at ({#6*\scsp},0) {};
  \fi
\end{scope}}

\newcommand{\scStandardPanel}[6]{\begin{scope}[shift={(#1,#2)}]
  \foreach \i in {0,...,7}
    \node[scsurv] (p\i) at ({\i*\scsp},0) {};
  \foreach \i/\j in {0/1,1/2,2/3,3/4,4/5,5/6,6/7} \draw[scpath] (p\i)--(p\j);
  \node[scfail] (z) at ({3.5*\scsp},1.0) {$z$};
  \foreach \i in {0,...,7} \draw[scspoke] (z)--(p\i);
  \if\relax\detokenize{#5}\relax\else
    \foreach \i in {#5} {
      \draw[white,line width=1.6pt] (z)--(p\i);
      \draw[scdropped] (z)--(p\i);
    }
  \fi
  \ifnum#3=1
    \draw[scclosing] (p0) .. controls ({0*\scsp},-.6) and ({7*\scsp},-.6) .. (p7);
  \fi
  \if\relax\detokenize{#4}\relax\else
    \node[scfail,sccrashed] at (z) {$z$};
    \node[sccaption,above=1pt,text=red!75!black] at (z.north) {round #4};
  \fi
  \ifnum#6>-1
    \node[scwitness] at ({#6*\scsp},0) {};
  \fi
\end{scope}}

\newcommand{\scArrowR}[3]{\draw[scstep] ({#1*1cm+2.52cm+1mm},#3) -- ({#2*1cm-1mm},#3);}
\newcommand{\scDots}[2]{\node[font=\large,text=black!60] at (#1,#2) {$\cdots$};}

\newcommand{\CN}{\ensuremath{\Pi_{\mathsf{CN}}}}
\newcommand{\calE}{\mathcal{E}}

\newcommand{\local}{\ensuremath{\mathsf{LOCAL}}\xspace}

\DeclareMathOperator{\dist}{dist}
\newcommand{\diam}{\operatorname{diam}}
\newcommand{\rightchild}{\mathsf{right}}
\newcommand{\leftchild}{\mathsf{left}}

\newtheorem{observation}{Observation}
\crefname{observation}{observation}{observations}
\Crefname{observation}{Observation}{Observations}
\newtheorem{corollary}{Corollary}
\newtheorem{remark}{Remark}
\title{Local Distributed Computation under Crash Failures}
\author{\begin{tabular}{@{}c@{\hspace{2em}}c@{}}
    Diana Ghinea\textsuperscript{1} & Fabian Kuhn\textsuperscript{2} \\
    {\small\texttt{diana.ghinea@hslu.ch}} &
    {\small\texttt{kuhn@cs.uni-freiburg.de}} \\[0.8em]
    Chen-Da Liu-Zhang\textsuperscript{1} & Zahra Parsaeian\textsuperscript{2} \\
    {\small\texttt{chen-da.liuzhang@hslu.ch}} &
    {\small\texttt{zahra.parsaeian@cs.uni-freiburg.de}}
  \end{tabular}\\[1em]
  {\small\textsuperscript{1}School of Computer Science and Information Technology}\\
  {\small Lucerne University of Applied Sciences and Arts, Switzerland}\\[0.4em]
  {\small\textsuperscript{2}Department of Computer Science}\\
  {\small University of Freiburg, Germany}}
\date{}
\begin{document}
\pagenumbering{gobble}
\hypersetup{pageanchor=false}
\maketitle
\thispagestyle{empty}

% Keep the abstract at the main text size rather than the smaller article default.
\begingroup
\let\small\normalsize
\begin{abstract}
We study how crash failures affect the round complexity of deterministic algorithms for locally checkable graph problems in the synchronous \local model. A node that crashes during a round may send messages to only some neighbors, leaving surviving nodes with incompatible local views and invalidating the usual local simulation of fault-free algorithms.

Our main tool is the consistent neighborhoods problem $\CN(R)$: surviving nodes construct partial radius-$R$ views that satisfy local consistency across surviving edges and are complete without crashes. With $R=T+\ell+1$, where $T$ is the fault-free running time and $\ell$ the checking radius, these views support local simulation of algorithms that remain correct on subgraphs. The outputs satisfy local constraints wherever the checking neighborhood contains no crashed node.

On graphs with at most $n$ nodes and maximum degree at most $\Delta\geq3$, we solve $\CN(R)$ in $O(\min\{n,(\Delta-1)^R,R\Delta f\})$ rounds when a crash bound $f\geq1$ is globally known; the first two bounds require no knowledge of $f$. For $\Delta=2$ or girth greater than $2R$, including forests, $O(R)$ rounds suffice. Conversely, one crash can force $\Omega(n)$ rounds even for $R=3$ on diameter-two graphs. For each fixed $\Delta\geq3$ and sufficiently large $R$, we prove a worst-case lower bound of $(\Delta-1)^{R-O(\sqrt R)}$, even when crashes occur only at the beginning of rounds. We further prove an $\Omega(\Delta R f)$ lower bound when $R\geq4\lceil\log_{\Delta-1}f\rceil+4$ and $n\geq\Delta R f$. Thus all our upper bounds for consistent neighborhoods are essentially tight in suitable parameter regimes.

For locally checkable labeling (LCL) problems, we preserve $O(\log^* n)$ complexity under any number of crashes whenever this bound holds without faults on all graphs satisfying a fixed degree bound. In contrast, we construct an LCL on graphs of maximum degree at most five with fault-free complexity $\Theta(\log n)$ and crash-tolerant complexity $\Theta(n)$. Finally, we transform any fault-free sinkless-orientation algorithm into one tolerating any number of crashes with only one additional round, guaranteeing correctness even at surviving nodes adjacent to crashes.
\end{abstract}

\endgroup

\clearpage
\begingroup
\pagestyle{empty}
\setcounter{tocdepth}{2}
\small
\tableofcontents
\clearpage
\endgroup
\pagenumbering{arabic}
\hypersetup{pageanchor=true}

\section{Introduction}\label{sec:introduction}

What can a distributed system of $n$ processes achieve despite process failures? This is a central question in distributed computing~\cite{PSL80,FLP1985,Lynch1996,AttiyaWelch2004}. For a given distributed task, the answer depends on the number and the nature of failures one tolerates and the communication assumptions. Even when a task remains solvable, tolerating failures can substantially increase the cost of doing so. One therefore seeks to characterize the limits and the cost of fault tolerance.

Another major line of research in distributed computing studies the round complexity of distributed graph algorithms~\cite{Linial1992,Peleg2000}. In the standard setting, the network is represented by a graph $G=(V,E)$ with at most $n$ nodes and maximum degree at most $\Delta$. Nodes initially have only local information and exchange messages along the edges of $G$. Each node must compute its part of a global solution, such as its color in a node coloring, its incident edges in a minimum spanning tree, or its distance from a designated source.

Our goal is to understand how crash failures affect the round complexity of the widely studied class of locally checkable graph problems~\cite{Linial1992,NaorStockmeyer1995,ChangKopelowitzPettie2019,STOC:GhaKuhMau17,FOCS:ChaPet17,STOC:BHKLOS18,PODC:Brandt19,PODC:BBCORS19,Chang2020Landscape,BalliuBrandtOlivettiSuomela2020,STOC:RozGha20}. A graph problem is called \emph{locally checkable} if it is specified by a \emph{checking radius} $\ell$ and a rule by which each node accepts or rejects a proposed solution based only on its induced radius-$\ell$ neighborhood, including the input and proposed output labels. Both the radius and the rule may depend on $n$. A proposed solution is valid if and only if every node accepts. This notion of local verification has roots in work on self-stabilization, where violations of global correctness are detected through local checks~\cite{AfekKuttenYung1991,FOCS:AwePatVar91}. An extensively studied subclass is that of \emph{locally checkable labeling} (LCL) problems, defined on graphs of constant maximum degree $\Delta$, with constant checking radius $\ell$, with input and output labels drawn from fixed finite alphabets, and with a fixed validity rule independent of $n$ and identifiers~\cite{NaorStockmeyer1995}. For example, proper node coloring and maximal independent set are both locally checkable with checking radius one. For coloring, each node checks that its color belongs to the prescribed palette and differs from all neighboring colors. For maximal independent set, each selected node checks that no neighbor is selected, and each unselected node checks that some neighbor is selected. Both are LCL problems when $\Delta=O(1)$, provided that the palette also has size $O(1)$ for coloring.

We work in the \emph{\local model}, the standard model for studying locality in distributed graph algorithms~\cite{Linial1992,Peleg2000}. Nodes have unique identifiers. We assume that each node initially knows its own identifier, its input label (if any), the identifiers of its neighbors, and the bounds $n$ and $\Delta$. Computation proceeds in synchronous communication rounds. In each round, a node can send a separate message to each neighbor based on its current state. It then receives the messages sent to it in this round and updates its state. Message sizes and internal computation at the nodes are unrestricted. Time complexity is the number of rounds until all nodes output.

We study the \local model with \emph{crash failures}.
We use the standard crash model for synchronous systems~\cite{LamportFischer1982,Lynch1996}: a node follows its algorithm until it crashes, after which it performs no further computation and sends no further messages. In its crash round, an arbitrary subset of its prescribed outgoing messages may be delivered. In the fault-free setting, every deterministic $T$-round \local algorithm can be put into a canonical form: each node first spends $T$ rounds collecting the initial states of all nodes within distance $T$, and then simulates the algorithm locally to determine its own output, without further communication. The collected states include the nodes' identifiers, input labels, and neighbor lists. Partial delivery complicates this approach: adjacent surviving nodes may obtain views that disagree on the same part of the graph. Independently simulating a fault-free algorithm on these views can therefore produce outputs that violate the local constraints.

A node is called \emph{surviving} if it does not crash during the execution. We extend the notion of local checkability to the setting with crashes as follows. For a locally checkable problem with checking radius $\ell$, a surviving node $v$ is \emph{locally satisfied} if and only if either some node in its radius-$\ell$ neighborhood in the original graph $G$ crashes or the local constraint at $v$ specified by the fault-free definition of the problem is satisfied. A crash-tolerant algorithm must ensure that every surviving node is locally satisfied.

We study general methods for transforming fault-free \local algorithms into crash-tolerant ones. We focus on deterministic algorithms. The formal model appears in \Cref{sec:preliminaries}.

\subsection{Contributions}\label{sec:contributions}

We introduce the \emph{consistent neighborhoods} problem $\CN(R)$ as a replacement for the neighborhood-collection phase of fault-free algorithms.\footnote{Related neighborhood-view consistency specifications for wireless sensor networks were studied in~\cite{SRDS:JhMo09,TOSN:JhMo16}; see \Cref{sec:related-work}.} For an integer $R\geq1$, each surviving node $v$ must output a view $H_v$: a labeled subgraph of its radius-$R$ neighborhood in $G$, containing $v$ and retaining the original identifiers and input labels. The views of adjacent surviving nodes must contain their common edge and agree on the subgraphs induced by the nodes within distance $R-1$ of either endpoint, with distances measured within each view. These requirements ensure that each view contains every node reachable from its center by a path of at most $R$ edges through surviving nodes. Information involving crashed nodes may be omitted, subject to consistency. Without crashes, every view must be the complete radius-$R$ neighborhood. The formal definition appears in \Cref{def:cn}.

We show that consistent neighborhoods allow surviving nodes to simulate fault-free local algorithms for locally checkable problems without further communication. Each surviving node simulates the algorithm on its own view and adopts the output assigned to it. For sufficiently large $R$, the outputs ensure local satisfaction. This requires the fault-free algorithm to remain correct on every subgraph of $G$, since the views used for simulation may omit nodes or edges. The simulations, checking radius, and validity rule all use the original global bounds $n$ and $\Delta$.
The next theorem gives this guarantee and the optimal fault-free round complexity of $\CN(R)$.

\begin{samepage}
\begin{theorem}[Consistent neighborhoods and fault-free algorithms]
\label{thm:cn-to-lcl}
The following statements hold in the deterministic \local model.

\begin{enumerate}[(i)]
    \item For every integer $R\geq1$, $\CN(R)$ can be solved in
    $R$ rounds in the fault-free setting. This bound is optimal in the worst
    case even for $\Delta=2$.

    \item Let $\Pi$ be a locally checkable problem with checking radius $\ell\geq1$, and let $\mathcal A$ be a deterministic algorithm that solves $\Pi$ within $T$ rounds on every subgraph of $G$ in the fault-free setting, with inherited labels and the original bounds $n$, $\Delta$, where $T\geq0$. Given a solution to $\CN(R)$ on $G$ with
    $
        R\geq T+\ell+1,
    $
    each surviving node can compute its output without communication so
    that every surviving node is locally satisfied for $\Pi$.
\end{enumerate}
\end{theorem}
\end{samepage}

This transformation makes the round complexity of $\CN(R)$ the central question. Can consistent neighborhoods still be constructed in $O(R)$ rounds under crashes, or in a number of rounds depending only on $R$ and the crash bound $f$? More generally, we study their complexity as a function of $R$, $f$, $\Delta$, and $n$, obtaining upper and lower bounds that match in several parameter regimes. The bound $f$ is assumed known to the nodes only when explicitly stated.

Our upper bounds hold in the standard crash model described above. Several lower bounds hold even in the more restrictive \emph{atomic crash-fault model}, in which a node crashes at the beginning of a round, before sending any messages.

The following theorem rules out bounds depending only on $R$ and $f$ on general graphs: even when $R=3$ and $f=1$, every deterministic algorithm for $\CN(R)$ requires $\Omega(n)$ rounds in the worst case. This construction has maximum degree $\Theta(n)$, so we also establish lower bounds for smaller degrees, with explicit dependence on $\Delta$, $R$, and $f$.

\begin{theorem}[Lower bounds for $\CN(R)$]\label{thm:lower-bound}
Let $R\geq1$ and $f\geq1$ be integers. The following worst-case lower bounds
hold for deterministic algorithms solving $\CN(R)$ in the synchronous
\local model with standard crash failures, even when the bounds $n$, $\Delta$, and $f$ are globally known.
\begin{enumerate}[(i)]
    \item $\Omega(n)$ rounds on general graphs with at most $n$ nodes
    for $R\geq3$, even when $f=1$.
    \item On graphs with at most $n$ nodes and maximum degree at most $\Delta$, $\Omega\bigl((\Delta-1)^{\lfloor(R-1)/2\rfloor}\bigr)$ rounds are necessary, provided $\Delta\geq3$, $R\geq4$, $n\geq4(\Delta-1)^{\lfloor(R-1)/2\rfloor}$, and $f\geq(\Delta-1)^{\lfloor(R-1)/2\rfloor-1}$. Further, for every $\Delta\geq3$, there exists an integer $R_0(\Delta)\geq4$ such that, for every $R\geq R_0(\Delta)$, the stronger lower bound $(\Delta-1)^{R-3\lceil\sqrt R\rceil}$ holds, provided $n\geq(\Delta-1)^R$ and $f\geq\lceil(\Delta-1)^R/\Delta\rceil$.
    \item On graphs with at most $n$ nodes and maximum degree at most
    $\Delta$, $\Omega\bigl((R-2\log_{\Delta-1}f)\cdot f\cdot\Delta\bigr)$
    rounds are necessary, provided $\Delta\geq3$,
    $R\geq2\lceil\log_{\Delta-1}f\rceil+4$, and $n\geq R\cdot f\cdot\Delta$.
\end{enumerate}
In the atomic crash-fault model, (ii) and (iii) hold with the additional assumption $f\geq2$, and the $\Omega(n)$
lower bound in (i) holds for $R\geq4$, even when $f=2$.
\end{theorem}

We complement these lower bounds with an $O(n)$-round algorithm and two bounds independent of $n$: $O((\Delta-1)^R)$ for $\Delta\geq3$, and $O(\Delta \cdot R\cdot f)$ when a bound $f\geq1$ on the number of crashes is known. Moreover, $O(R)$ rounds suffice under any number of crashes when $\Delta=2$ or the graph has girth greater than $2R$, including forests. On these graph classes, \Cref{thm:cn-to-lcl} transforms $T$-round \local algorithms for locally checkable problems with checking radius $\ell$ into crash-tolerant algorithms running in $O(T+\ell)$ rounds, provided they remain correct on all subgraphs of the input graph.

We state the following upper bounds for $\Delta\geq2$. The case $\Delta\leq1$ is immediate: every connected component has at most two nodes, and $\CN(R)$ can be solved in at most one round.

\begin{theorem}[Upper bounds for $\CN(R)$]\label{thm:upper-bound}
Let $R\geq1$ be an integer. In the synchronous \local model with crash
failures and the globally known bounds $n$ and $\Delta\geq2$, $\CN(R)$ can be solved on any graph $G$ with at most $n$ nodes and maximum degree at most $\Delta$
deterministically in
\begin{enumerate}[(i)]
    \item $O(n)$ rounds, for any number of crashes;
    \item $O((\Delta-1)^R)$ rounds if $\Delta\geq3$, for any number of crashes;
    \item $O(\Delta\cdot R\cdot f)$ rounds, if a bound $f\geq1$ on the number of
    crashes is globally known;
    \item $O(R)$ rounds if $\Delta=2$ or the graph has girth greater than
    $2R$, for any number of crashes. In particular, this applies to forests.
\end{enumerate}
\end{theorem}

These bounds also let us relate crash tolerance to the complexity theory of
LCL problems. On general graphs of constant maximum degree, the deterministic
fault-free \local complexity of an LCL problem is either $O(\log^* n)$ or
$\Omega(\log n)$~\cite{ChangKopelowitzPettie2019}. We show that every LCL problem
in the former regime retains an $O(\log^* n)$-round algorithm in the presence
of crashes. At the logarithmic threshold, however, tolerating crashes can
already increase the complexity to $\Theta(n)$.

\begin{theorem}[Crash tolerance and the LCL complexity gap]
\label{thm:lcl-crash-complexity}
The following statements hold in the synchronous \local model with $O(\log n)$-bit identifiers.
\begin{enumerate}[(i)]
    \item For every fixed $\Delta\geq2$ and every LCL problem $\Pi$, if $\Pi$ admits a deterministic $O(\log^* n)$-round algorithm on all graphs of maximum degree at most $\Delta$ in the fault-free setting, then $\Pi$ admits a deterministic $O(\log^* n)$-round algorithm tolerating any number of crash failures.

    \item There is an explicit input-labeled LCL problem $\Pi$ with checking radius two, defined on all graphs of maximum degree at most five, whose deterministic fault-free complexity is $\Theta(\log n)$ and whose deterministic crash-tolerant complexity is $\Theta(n)$. The lower bound holds already with $\Theta(n)$ atomic crashes, and the upper bound holds in the standard crash-fault model.
\end{enumerate}
\end{theorem}

To prove \Cref{thm:lcl-crash-complexity}(i), we use the normal form of deterministic $O(\log^* n)$-round LCL algorithms: a coloring phase followed by constant-round computation~\cite{ChangKopelowitzPettie2019,BalliuEtAl2021}. The coloring can be made crash tolerant, and our consistent neighborhoods algorithms handle the remaining computation in $O(1)$ rounds. For part~(ii), we encode agreement on a root's input bit as a fixed LCL and apply a similarity-chain argument to obtain the linear lower bound. Subsection~\ref{sec:overview-lcl} gives the main ideas. The separation in \Cref{thm:lcl-crash-complexity}(ii) also holds without problem-specific input labels, with possibly larger constant degree and checking radius; see Subsection~\ref{sec:lcl-no-input-labels}.

\Cref{thm:lcl-crash-complexity}(ii) shows that no general transformation can preserve the asymptotic round complexity of every deterministic fault-free LCL algorithm. Problem-specific transformations can nevertheless preserve this complexity. Sinkless orientation requires every node of degree at least three to have an outgoing edge. Under crashes, adjacent surviving nodes must agree on their common edge's orientation, and every surviving node of input degree at least three must have an outgoing edge, possibly towards a crashed neighbor. Our transformation adds only one round and tolerates any number of crashes.
These guarantees are stronger than local satisfaction: they also apply to surviving nodes with crashed neighbors, i.e., crashed nodes within the checking radius.

\begin{theorem}[Sinkless orientation under crash faults]
\label{thm:so-transformation}
On graphs of arbitrary maximum degree, any deterministic $T$-round \local algorithm for sinkless orientation in the fault-free setting can be transformed into a deterministic $(T+1)$-round \local algorithm satisfying the crash-tolerant requirements above for any number of crash failures.
\end{theorem}

For fixed $\Delta\geq3$, sinkless orientation has deterministic fault-free complexity $\Theta(\log n)$~\cite{STOC:BFHKLR16,ChangKopelowitzPettie2019,SODA:GhaSu17}. Thus, this LCL in the $\Theta(\log n)$ class retains its complexity under crashes.

\subsection{Related Work}\label{sec:related-work}

Our work lies at the intersection of distributed graph algorithms \cite{Linial1992,NaorStockmeyer1995} and fault-tolerant distributed computing \cite{Lynch1996}. Although fault-free complexity is well studied, less is known about how failures affect round complexity. Several lines of work address related questions in graph algorithms and networks with arbitrary topologies. We review the most relevant ones below.

\paragraph{Agreement in general networks.}
The classical literature on agreement problems is especially relevant to our work. In consensus, each of $n$ parties holds an input and up to $f$ parties may fail. All non-faulty parties must terminate and agree on an output satisfying a validity condition. For crash failures, the usual validity condition requires the output to be an input of some party. Deterministic consensus is impossible in asynchronous networks even with one possible crash~\cite{FLP1985}. Randomized protocols circumvent this impossibility under suitable resilience assumptions~\cite{PODC:BenOr83,FOCS:Rabin83}. In complete synchronous networks, deterministic crash consensus has tight worst-case round complexity $\min\{f+1,n-1\}$ for every $f<n$~\cite{LamportFischer1982,Lynch1996,AttiyaWelch2004}.\footnote{Under validity requiring that a unanimous input of the non-faulty parties be preserved, Byzantine consensus in complete synchronous networks requires $n>3f$ without digital signatures or $n>2f$ with signatures, and $f+1$ rounds suffice in either case~\cite{PSL80,DolStr83,FCT:Fischer83}.}

Global consensus in general networks also requires connectivity~\cite{Dolev1982,DistComp:Hadzilacos87,AFT:WanWat20,Arxiv:TseVai15}. In undirected networks, no permitted crash set may separate surviving nodes~\cite{DistComp:Hadzilacos87}. For a known graph $G$ and a known crash bound $f$ below its node connectivity, \cite{CastanedaEtAl2023} study global synchronous crash consensus. Their round bound is a radius accounting for crash patterns.

Our consistent neighborhoods problem does not require global agreement and therefore remains solvable even when crashes disconnect the network. Related relaxations of global agreement have also been studied. For example, \cite{SSS:CKOO23} study \emph{disconnected agreement} in synchronous networks with non-faulty nodes and omission-prone links. They require agreement within components formed by links that have not omitted messages. Fraigniaud, Nguyen, and Paz~\cite{STACS:FrNgPa25} study \emph{local consensus} under crash failures: each surviving component agrees on some node's initial proposal, possibly from outside that component. Their $n-1$-round flooding algorithm provides the componentwise agreement argument underlying our $O(n)$-round $\CN$ algorithm in \Cref{thm:upper-bound}(i).

All our crash-tolerant lower bounds use constructions requiring consistency among a global set of nodes in fault-free executions. This lets us adapt classical indistinguishability-chain arguments~\cite{FischerLynch1982,LamportFischer1982,Lynch1996}. Related arguments for general networks appear in~\cite{STACS:FrNgPa25}.

\paragraph{Topology and neighborhood discovery.}
Nesterenko and Tixeuil~\cite{SIROCCO:NesTix06} study topology discovery in asynchronous networks with Byzantine faults. Dolev, Liba, and Schiller~\cite{NETYS:DolLibSch13} give self-stabilizing solutions and consider constant-radius neighborhood discovery. These results require enough connectivity for reliable communication despite faults.

Jhumka and Mottola~\cite{SRDS:JhMo09,TOSN:JhMo16} study neighborhood-view consistency in wireless sensor networks as a service for higher-level protocols. Their specifications concern maintaining consistent one-hop neighbor information under crashes, communication omissions, and transient state faults. We study terminating algorithms for partial radius-$R$ graph views with consistency across surviving edges under permanent crashes, and use these views to simulate fault-free local algorithms.

\paragraph{Fault-tolerant graph algorithms in fully connected networks.}
Byzantine graph computation has also been studied in the \emph{congested clique}, where every pair of processors can communicate directly, regardless of the input graph. Augustine, Molla, Pandurangan, and Vasudev~\cite{DISC:AMPV22} study connectivity testing, while Cifuentes-Núñez, Montealegre, and Rapaport~\cite{OPODIS:CiMoRa25} study acyclicity and bipartiteness. Faulty processors can hide or misreport parts of the graph, so either answer is permitted on some inputs.

Deterministic transformations make fault-free congested-clique algorithms tolerate a constant fraction of node crashes~\cite{OPODIS:CenSot25,DISC:CFGS25,PODC:FisGel26}. They either protect inputs during initial failure-free rounds or assume inputs are already encoded and stored redundantly across the network. Complete communication lets surviving processors take over the computations of crashed ones.

\paragraph{Wait-free locally checkable problems.}
Closely related to our work are wait-free algorithms for locally checkable problems, where processes are asynchronous and any number may crash. Wait-freedom requires every surviving process to terminate regardless of others' progress or failures. Wait-free $3$-coloring of rings was first obtained in the DECOUPLED model, whose communication remains synchronous and reliable: routers keep forwarding messages after their processes crash~\cite{TCS:CDFRR19}. The abstract ABST-DECOUPLED model additionally provides snapshots of the local topology. This enables generic transformations from deterministic \local algorithms for input-free LCL problems to wait-free algorithms~\cite{TCS:DFFR25}.

In the ASYNC \local model, processes communicate through shared registers: each process writes to its own register and reads its neighbors' registers. A crashed process's last written state remains readable. Wait-free algorithms color cycles~\cite{DISC:FLR22} and general graphs~\cite{DISC:BFLOR24}. Yet deterministic wait-free MIS is impossible on cycles of at least seven nodes even when correctness is required only in fault-free executions~\cite{DISC:BFLOR24}, ruling out generic wait-free transformations. Our synchronous \local model admits a generic transformation ensuring local satisfaction under \Cref{thm:cn-to-lcl}(ii), although substantial asymptotic slowdown can be unavoidable (\Cref{thm:lcl-crash-complexity}(ii)).

\paragraph{Repairing configurations.}
Other work \emph{repairs} or \emph{maintains} valid configurations after state corruptions or network changes. A classical approach is self-stabilization~\cite{ACM:Dijkstra74,FOCS:AwePatVar91,WDAG:APVD94,SSS:LeSuWa09,PODC:BaElGo18,SIROCCO:Turau19}, which handles arbitrary \emph{transient} state corruptions. After the last fault, the system must reach a valid configuration in finite time and remain valid thereafter; this recovery time is the stabilization time. Strict stabilization~\cite{SRDS:NesAro02,FUN:DuTiZh12,SSS:CoPiSe21} also tolerates permanent Byzantine faults, eventually guaranteeing correctness at nodes sufficiently far from faulty ones.

Generic transformations convert deterministic $T$-round \local algorithms with $T\geq1$ into synchronous self-stabilizing algorithms with stabilization time $O(T)$~\cite{FOCS:AweVar91,SSS:LeSuWa09}. Recent work gives asynchronous self-stabilizing compilers for terminating synchronous algorithms~\cite{PODC:JDMI24} and, under additional assumptions on the input algorithm, randomized synchronous LCL compilers whose recovery time depends on the number of affected nodes~\cite{Arxiv:BEIK24}. These guarantees concern recovery after transient faults cease, whereas our algorithms terminate despite crashes during execution.

Fault-local recovery seeks repair times governed by the number of affected nodes rather than network size. Kutten and Peleg~\cite{FOCS:KutPel95} give such transformations for MIS after transient state corruptions. Dynamic distributed algorithms~\cite{OPODIS:KonWat13,PODC:CeHaKa16,SODA:PaPeSo16} update an existing solution after node or edge insertions and deletions. Highly dynamic networks~\cite{IEEE:BaKuMa19,OPODIS:CDKPS20} allow changes throughout the computation. Mayer, Naor, and Stockmeyer~\cite{ISTCS:MayNaoSto95} already studied robust local computation under continually failing and reappearing links.

Mendability~\cite{SIROCCO:BHMORS22,OPODIS:MeSuVi23} concerns extending locally correct partial LCL labelings. The task is to label another node while preserving local correctness, possibly changing existing labels. The cost measures how far changes extend from that node or how many nodes are relabeled.

\paragraph{Preprocessing for future failures.}
Another line of work constructs fault-tolerant subgraphs in a failure-free preprocessing phase. Fault-tolerant spanners preserve approximate distances after failures~\cite{STOC:CLPR09,PODC:DinKra11,STOC:Parter22}; fault-tolerant tree structures and tree augmentation preserve shortest-path, spanning-tree, or connectivity guarantees~\cite{SPAA:GhaPar16,DISC:Parter20,OPODIS:CenDor17}. For local graph problems, \cite{CasteigtsEtAl2020Robustness} study which graphs admit an MIS that remains valid under edge deletions preserving connectivity; \cite{DuboisEtAl2025Robustness} extend this to bounded deletions, domination, and matching.

\subsection{Organization of the Paper}

We give a technical overview in \Cref{sec:technical-overview} and introduce the model and notation in \Cref{sec:preliminaries}. \Cref{sec:cn-problem} defines consistent neighborhoods and applies them to local simulation. \Cref{sec:cn-algorithms,sec:lower-bound} give upper and lower bounds. \Cref{sec:lcl-complexity} studies how crashes affect LCL complexity, and \Cref{sec:edge-grabbing-lemmas} gives transformations for edge grabbing and sinkless orientation. We conclude in \Cref{sec:conclusion}.

\section{Technical Overview}\label{sec:technical-overview}

This section outlines the main technical ideas behind our results. We revisit the consistent neighborhoods problem in Subsection~\ref{sec:overview-cn-problem} and discuss its upper and lower bounds in Subsections~\ref{sec:overview-upper-bounds} and~\ref{sec:overview-lower-bounds}. We then turn to crash tolerance for LCL problems in Subsection~\ref{sec:overview-lcl} and the sinkless-orientation transformation in Subsection~\ref{sec:overview-sinkless-orientation}.

\subsection{The Consistent Neighborhoods Problem}\label{sec:overview-cn-problem}

Every deterministic $T$-round algorithm in the fault-free \local model can be put into the following canonical form. First, each node collects the initial states of all nodes within distance $T$ in $T$ rounds. It then simulates the algorithm locally to determine its output, without further communication. This directly solves the consistent neighborhoods problem in the fault-free setting: the collected views describe $T$-hop neighborhoods of the same labeled graph $G$ and therefore agree on their overlaps. The precise relation between the view radius $R$ and the collection time $T$ depends on the initial knowledge and consistency requirements.
For our model and definition of $\CN(R)$, \Cref{thm:cn-to-lcl}(i) establishes that the optimal worst-case fault-free complexity is exactly $R$ rounds.

In the presence of crashes, however, information may spread unevenly, so neighboring nodes may collect views of overlapping parts of different labeled graphs. When locally simulating an algorithm for a locally checkable problem, such views may not yield consistent, valid outputs. To solve $\CN(R)$, we must therefore ensure that neighboring views agree on their overlaps.

Consistency is required only between neighboring views, but it propagates along surviving paths, with the agreement radius decreasing by one at each hop. For a locally checkable problem with checking radius $\ell$, taking $R\geq T+\ell+1$ ensures that every node in a crash-free checking neighborhood computes from its own view the output assigned to it by the $T$-round simulation on the center's view. The local check thus sees the same neighborhood and outputs as in this single fault-free simulation. If the algorithm is correct on every subgraph of $G$, these outputs satisfy the local constraints, as stated in \Cref{thm:cn-to-lcl}(ii).

\subsection{The Consistent Neighborhoods Upper Bounds}\label{sec:overview-upper-bounds}

\paragraph{A warm-up: obtaining an $\boldsymbol{O(n)}$-round algorithm.}
Consider a graph with at most $n$ nodes and assume for now that the surviving nodes remain connected after all crashes. Each node initially knows its labeled star: its own identifier and input label (if any), together with its neighbors' identifiers. Flooding these stars for $n-1$ rounds gives every surviving node the stars of all surviving nodes, since any two are joined by a surviving path of length at most $n-1$. Each node can therefore reconstruct a graph containing the surviving graph, possibly with additional information from crashed nodes. This suggests a simple strategy: gather information and then use the reconstructed graphs as proposals in a standard consensus algorithm (see~\cite{CastanedaEtAl2023} for earlier work on arbitrary graphs). Each node forwards its smallest known proposal, under a fixed total order, for another $n-1$ rounds and then outputs its current minimum; cf., e.g.,~\cite[full version, Lemma~24]{STACS:FrNgPa25}.

Crashes may, however, disconnect the surviving graph. In this case, agreement is needed only within each surviving component. This is the goal of the \emph{local consensus} problem studied by Fraigniaud, Nguyen, and Paz~\cite{STACS:FrNgPa25}. Their elementary algorithm floods identifier--proposal pairs for $n-1$ rounds, after which the nodes in each surviving component have the same collection and can select the same proposal. Applying this flooding principle to labeled stars combines gathering and agreement in a single phase of $n-1$ rounds, proving \Cref{thm:upper-bound}(i).

\paragraph{Why shorter flooding does not ensure agreement.}
The linear-time algorithm solves $\CN(R)$ for every $R$, but for smaller $R$ we would like to have faster algorithms with bounds ideally depending only on $R$, $\Delta$, and, when known, $f$. Simply stopping flooding after $R$ rounds does not suffice. Consider a cycle of length $R+2$, for $R\geq2$, with consecutive nodes $w,u,v$. Suppose that only $w$ crashes, in round~1, sending its star only to its other neighbor. The star travels around the cycle and reaches $v$ in round~$R$, but has not yet reached $u$. If nodes output the graphs induced by their collected star sources, $v$ includes $w$ as a neighbor of $u$, whereas $u$ omits it, violating consistency across $\{u,v\}$. Simply discarding this late-arriving star at $v$ also fails if its other neighbor retains it: that neighbor received the star one round earlier and includes $w$ within distance $R-1$, so the disagreement moves to their common edge. Thus, whether to include or omit information propagated in only one direction is itself an agreement problem.\footnote{As we show later, on cycles, this issue can be resolved in $O(R)$ rounds (\Cref{thm:upper-bound}(iv)). Such a bound however does not extend to general graphs.}

\paragraph{Flooding the gathered information.}
After gathering, surviving nodes must coordinate which nodes and edges to include so that adjacent outputs agree on their required overlaps. A natural next step is to freeze and flood each node's collection for a prescribed number of rounds. Crashes may destroy short routes and force long detours, so a short flooding phase may miss nodes that must agree on whether to include a source. Flooding longer does not resolve this: unrestricted flooding may bring an additional collection to one endpoint of a surviving edge only in the final round. If its source belongs to the required overlap, the other endpoint must make a compatible inclusion decision without having received the collection. We therefore restrict each source's information to a fixed propagation graph that stays within distance $R$ of its source while retaining every edge across which agreement about that source is required.

\paragraph{Propagation graphs.}
Each node floods its initial labeled star for $R$ rounds. For each node $v$ that completes gathering, define $v$'s \emph{propagation graph} $\Gamma_v$ as the subgraph of $G$ induced by $v$ and all nodes that received $v$'s star during this phase, including nodes that later crash. Thus, $\Gamma_v$ records where information about $v$ has spread. Flooding for $R$ rounds keeps $\Gamma_v$ within distance $R$ of $v$.

After gathering, each node $v$ initiates a fresh token containing only its identifier and broadcasts it within $\Gamma_v$. Only $v$ initially holds this token, and only members of $\Gamma_v$ retain and forward it. The identifier suffices because these nodes already know $v$'s labeled star. All sources broadcast in parallel, without enlarging the propagation graphs. We run this phase long enough that, in every connected component of the surviving part of $\Gamma_v$, either all nodes store $v$'s token or none does. This also holds if $v$ crashes during the second phase. \Cref{sec:parameter-cn} formalizes this for every $R\geq1$.

\paragraph{Why propagation graphs help.}
Each surviving node outputs the labeled graph induced by itself and the sources whose tokens it receives in the second phase. The collected stars supply the edges and labels. Surviving neighbors store each other's tokens, so their common edge is included. Crucially, every node that is a source of the second phase completed the first phase. Thus every path in an output was available throughout the first phase, even if some of its nodes crashed later.

Consider a surviving edge $\{u,v\}$ and a path in $u$'s output of length at most $R-1$ from $x$ to either endpoint of $\{u,v\}$. During gathering, $x$'s star reaches that endpoint along this path by round $R-1$ and the other endpoint across $\{u,v\}$ by round $R$. Thus, $\{u,v\}$ is an edge of $\Gamma_x$. Since $x$ belongs to $u$'s output, $u$ stores $x$'s token. Agreement within the surviving components of $\Gamma_x$ ensures that $v$ also stores it and includes $x$ in its output.

Applying this argument to every node on the path transfers the entire path to $v$'s output. Applying it in both directions makes the outputs agree on the required overlap, including edges and labels. Without crashes, $\Gamma_x$ contains exactly the nodes within distance $R$ of $x$, so the outputs are complete radius-$R$ neighborhoods.

\paragraph{Bounding the number of rounds.}
Tracing successive first receptions of a token backwards gives a simple path, with one edge per round. For $\Delta\geq3$, each propagation graph has $O((\Delta-1)^R)$ nodes, so no first reception occurs later than this many rounds. Each surviving component therefore receives the token everywhere or nowhere by then, proving \Cref{thm:upper-bound}(ii). For $\Delta=2$, the same argument gives $O(R)$ rounds, as in \Cref{thm:upper-bound}(iv).

Alternatively, every node of a propagation graph has a path of length at most $R$ to its source, so its diameter before crashes is at most $2R$. Crashes can destroy these shortcuts, but there are at most $\Delta f$ surviving neighbors of crashed nodes. If a crash occurs in the propagation graph, assign every surviving node to one such neighbor by following a shortest path toward a crashed node and stopping just before the first crash. The surviving prefix has length $O(R)$, so nodes assigned to the same neighbor are connected by surviving routes of length $O(R)$. Once a node receives the token, the token reaches every other node in its group within $O(R)$ further rounds. Consecutive nodes on a first-reception path receive it one round apart, so the path contains at most $O(R)$ nodes from each group. Adding at most $f$ crashed nodes on this simple path and accounting for the crash-free case gives $O(R\cdot\Delta\cdot f)$ rounds when $f\geq1$ is known, proving \Cref{thm:upper-bound}(iii). Both bounds include gathering and are independent of $n$.

\paragraph{Graphs of large girth.}
Finally, on graphs of girth greater than $2R$, including forests, flooding stars for $R$ rounds suffices. Each node outputs the graph induced by the sources of its received stars. Between any two nodes, there is at most one simple path of length at most $R$. Consider a source $x$ within distance $R-1$ of either endpoint of a surviving edge $\{u,v\}$ in $u$'s view. By uniqueness, its first-reception path to $u$ either has length at most $R-1$, leaving time to forward the star to $v$, or passes through $v$. Thus both endpoints receive the star by round~$R$. Applying this to every node in the required overlap proves consistency, yielding the $R$-round bound in \Cref{thm:upper-bound}(iv).

\subsection{The Consistent Neighborhoods Lower Bounds}\label{sec:overview-lower-bounds}

The lower bounds for $\CN(R)$ in \Cref{thm:lower-bound} show how crashes can force coordination over distances much larger than $R$. Their proofs combine consistency with indistinguishability: even when crashes destroy short communication routes, a witness that cannot detect the change forces the relevant outputs to remain unchanged. We first explain this pattern, then describe the three constructions.

\begin{figure}[t]
\centering
\begin{tikzpicture}[scale=.85,x=1cm,y=1cm,
  corevertex/.style={circle,draw=blue!65!black,fill=blue!25,
    minimum size=4mm,inner sep=0pt},
  failvertex/.style={circle,draw=red!75!black,fill=red!25,
    minimum size=3.6mm,inner sep=0pt},
  pathvertex/.style={circle,draw=green!40!black,fill=green!25,
    minimum size=3.6mm,inner sep=0pt},
  coreedge/.style={draw=blue!50!black,line width=.65pt},
  attachment/.style={draw=black!45,line width=.65pt},
  pathedge/.style={draw=green!40!black,line width=1pt}]
  \draw[blue!55!black,fill=blue!4,line width=.75pt]
    (6,2.7) ellipse [x radius=2.65,y radius=.7];
  \draw[red!60!black,fill=red!4,line width=.75pt]
    (6,1.35) ellipse [x radius=5.25,y radius=.45];
  \draw[green!40!black,fill=green!4,line width=.75pt,
    rounded corners=7pt] (-.15,-.55) rectangle (12.15,.55);
  \node[text=blue!65!black,font=\large] at (6,3.1) {$C$};
  \node[text=red!75!black,font=\large] at (.3,1.35) {$F$};
  \node[text=green!40!black,font=\small] at (6,-.28)
    {$S$: the surviving path};

  \foreach \i/\x/\y in {1/4.2/2.5,2/5.1/2.6,3/6/2.3,
                            4/6.9/2.6,5/7.8/2.5}
    \node[corevertex] (c\i) at (\x,\y) {};
  \draw[coreedge] (c1)--(c2)--(c3)--(c4)--(c5) (c2)--(c4);
  \foreach \i/\x in {1/1.8,2/3.9,3/6,4/8.1,5/10.2} {
    \node[failvertex] (f\i) at (\x,1.35) {};
    \draw[attachment] (c\i)--(f\i);
  }
  \draw[red!60!black,line width=.75pt] (f1)--(f2) (f4)--(f5);

  \foreach \i in {0,...,14}
    \node[pathvertex] (s\i) at ({.4+.8*\i},.15) {};
  \foreach \i in {0,...,13} {
    \pgfmathtruncatemacro{\j}{\i+1}
    \draw[pathedge] (s\i)--(s\j);
  }
  \foreach \i/\first/\last in {1/0/2,2/3/5,3/6/8,4/9/11,5/12/14}
    \foreach \j in {\first,...,\last}
      \draw[attachment] (f\i)--(s\j);
  \node[below right=1pt,font=\small,inner sep=1pt] at (s0.south east) {$a$};
  \node[below left=1pt,font=\small,inner sep=1pt] at (s14.south west) {$b$};
  \draw[purple!75!black,dashed,line width=1pt,rounded corners=9pt]
    (s0.south)--(.4,-1.05)--(11.6,-1.05)--(s14.south);
  \node[fill=white,inner sep=3pt,text=purple!75!black,font=\small]
    at (6,-1.05) {closing edge $\{a,b\}$, present only in $G^+$};
\end{tikzpicture}
\caption{The common lower-bound framework. Only the red nodes $F$ may crash. The green nodes $S$ form a path in $G^-$ and a cycle in $G^+$, differing only in the dashed closing edge. Shortcuts through $C\cup F$ put the entire graph within distance $R-1$ of every node of $S$; crashing $F$ removes them.}
\label{fig:lb-framework}
\end{figure}
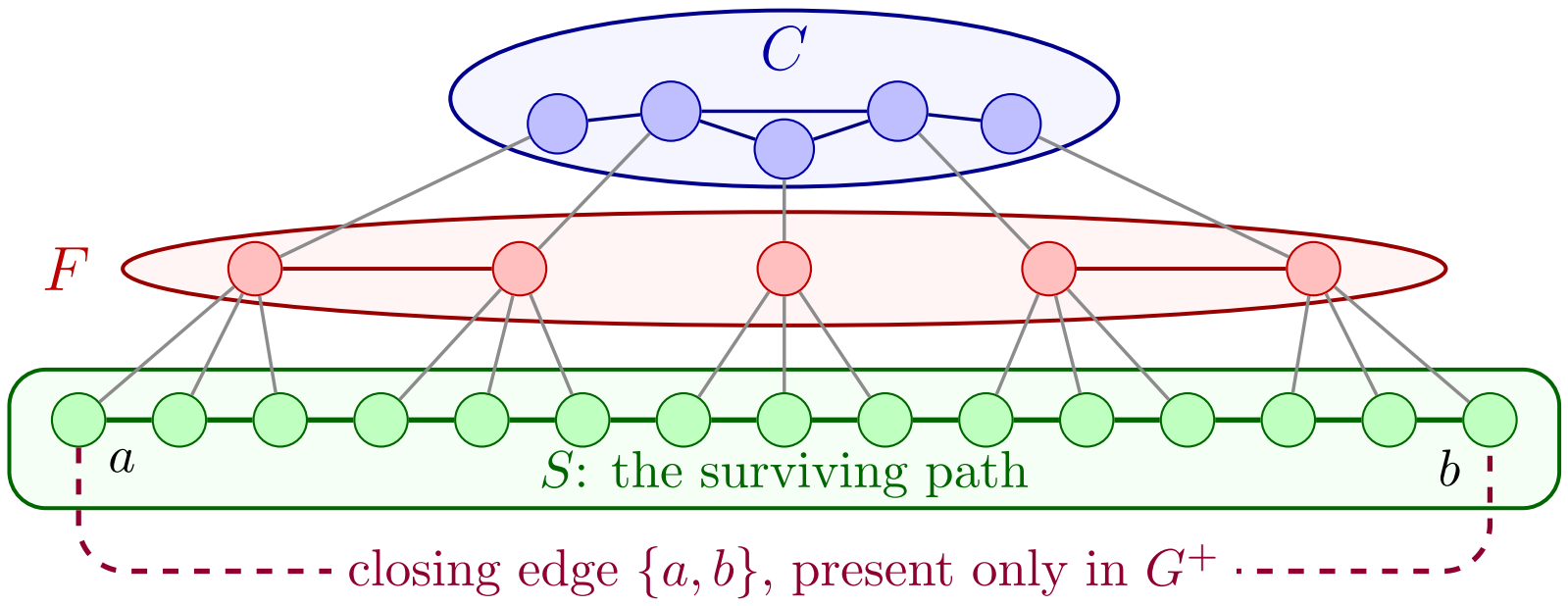

\paragraph{The common framework.}
We compare two graphs $G^-$ and $G^+$ with the same node set, identifiers, labels, and known global bounds. Their nodes are partitioned into $C$, $F$, and $S$ as in \Cref{fig:lb-framework}. The nodes of $S$, which always survive, form a path from $a$ to $b$ in $G^-$; adding the closing edge $\{a,b\}$ gives $G^+$. Only nodes in $F$ may crash. Shortcuts through $C\cup F$ place every node of $G^+$ within distance $R-1$ of each node of $S$. Without crashes, completeness requires each node to output its full radius-$R$ neighborhood (\Cref{def:cn}), which is all of $G^+$ for nodes in $S$. Every path from $S$ to $C$ passes through $F$. Once all nodes of $F$ have crashed, communication between nodes of $S$ must therefore follow the long path or cycle.

A large distance after crashes does not by itself prove a lower bound, since $\CN(R)$ permits incomplete views. Consistency makes the complete output $G^+$ an \emph{all-or-nothing} choice on $S$ (\Cref{lem:lb-propagation}). If one node $p\in S$ outputs $G^+$, every node of its view is within distance $R-1$ of $p$. Consistency forces each neighbor in $S$ to include this graph and hence output $G^+$. Applying this along $S$ forces every node of $S$ to output $G^+$, even if some nodes have crashed. Once $F$ stops sending messages, a difference between two executions can spread along $S$ by at most one hop per round (\Cref{lem:lb-spreading}), so reaching nodes $\Theta(|S|)$ hops away requires $\Omega(|S|)$ rounds. The similarity chain below shows why choosing incomplete views cannot avoid this coordination.

\paragraph{Similarity chains.}
We use the \emph{similarity chain} technique from consensus lower bounds~\cite{LamportFischer1982,Lynch1996}. For a deterministic $T$-round algorithm, a similarity chain is a sequence of executions $\calE_1,\dots,\calE_K$ in which each consecutive pair is \emph{indistinguishable} through round $T$ at a \emph{witness} node $v$ surviving both: its initial state and received messages are identical. The witness may change between pairs. For binary consensus, validity requires output $0$ in a fault-free all-zero execution and output $1$ in a fault-free all-one execution. Along a chain between them, indistinguishability fixes each witness's output across the pair,
and agreement forces the same output at all surviving nodes in each execution.
Thus, the common decision cannot change along the chain, contradicting validity.

For a hypothetical $T$-round algorithm for $\CN(R)$, the all-or-nothing property plays the role of agreement. We start with the fault-free execution on $G^+$, where completeness forces output $G^+$ throughout $S$, and move the crashes of $F$ before round~1. At each step, a witness in $S$ retains output $G^+$, so consistency forces all of $S$ to retain it. We then switch to $G^-$. Only $a$ and $b$ initially know whether the closing edge exists. With $F$ silent, this difference travels along $S$ at most one hop per round (\Cref{lem:lb-spreading}). A node farther than $T$ from both endpoints along the cycle therefore still outputs $G^+$, including the nonexistent closing edge. It remains to give a witness for each step.

\paragraph{Linear bounds via similarity chains (\Cref{thm:lower-bound}(i)).}
With no degree restriction, let $F=\{z\}$, $|S|=n-1$, and connect $z$ to all nodes of $S$ (\Cref{fig:lb-linear}). Both have diameter $2$, so $R\geq3$ suffices.

Starting from the fault-free execution, we first let $z$ crash in round $T$ after sending all its messages. No surviving node can notice this change through round $T$. We then keep the crash round fixed and, in each successive execution, withhold one additional round-$T$ message of $z$, leaving all earlier rounds unchanged. Once none is delivered, this looks to the surviving nodes exactly like a crash in round $T-1$ after sending all its messages. We repeat for rounds $T-1$, $T-2$, and so on, until $z$ sends nothing in round~1. Removing a round-$r$ message affects only its recipient initially. Since $z$ sends nothing afterwards, any effect spreads along $S$ for at most $T-r$ further rounds. For $T<|S|/2$, a node on the opposite side of the cycle cannot distinguish the two consecutive executions and thus serves as a witness (\Cref{lem:lb-standard-chain}). The final switch to $G^-$ has a witness for $T<|S|/2-1$, giving $\Omega(n)$ rounds with one standard crash.

In the atomic model, a crash suppresses all messages of a node at once. A witness must therefore be far along the cycle from \emph{all} neighbors in $S$ of the node whose crash moved. A single node adjacent to all of $S$ does not work. We instead use two adjacent nodes in $F$, each attached to one half of the $n-2$ nodes of $S$. The diameter becomes three, requiring $R\geq4$.

We move both crashes to rounds $T,T-1,\ldots,1$, one at a time. In each step, neither node sends further messages in the remaining rounds, so the effect again spreads only along $S$. For $T<|S|/4$, a node in the middle of the opposite half of $S$ cannot distinguish the two consecutive executions and is a witness (\Cref{lem:lb-atomic-chain}). This gives the $\Omega(n)$ bound with two atomic crashes.

The remaining bounds use the same chains, moving all crashes one round earlier, one node at a time in the atomic model. Keeping each crashing node attached within a short consecutive block of $S$ leaves a distant witness at each step. This gives an $\Omega(|S|)$ lower bound, so we seek the longest $S$ while keeping every node within distance $R-1$ of each node in $S$ and respecting $n$, $\Delta$, and $f$.

\paragraph{Exponential bounds in the radius (\Cref{thm:lower-bound}(ii)).}
For $\Delta\geq3$, only $O((\Delta-1)^R)$ nodes fit within distance $R-1$, limiting the size of $S$. To obtain an exponentially long cycle for $R\geq5$, we use a complete $(\Delta-1)$-ary tree of height $h=\lfloor(R-1)/2\rfloor$. Its leaves, connected in left-to-right order, form the path $S$; their parents form $F$ (\Cref{fig:lb-degree-tree}). Tree paths have length at most $2h\leq R-1$, while each crashing node attaches to consecutive leaves. Hence, $|S|=(\Delta-1)^h$ gives the first exponential lower bound. The factor two in the exponent comes from paths going up and down the tree.

To improve the exponent, we use a $\Delta$-regular core $H$ of diameter at most $D$ with $N=\Omega_\Delta((\Delta-1)^D/D)$ nodes, obtained from random regular graphs~\cite{BollobasFernandez1982}. Its nodes have no spare degree, so we subdivide selected edges, placing a node of $F$ on each and connecting it to a new node of $S$ (\Cref{fig:lb-sparse-core}). Subdividing every edge would double distances. Instead, we greedily select a maximal set of edges at pairwise distance at least $\ell$. A shortest core path then crosses only $O(1+D/\ell)$ selected edges, while maximality guarantees $\Omega(N/(\Delta-1)^\ell)$ selected edges in total. Thus, greater separation limits the increase in the core's diameter but allows fewer nodes in $S$. Choosing $\ell=\lceil\sqrt R\rceil$ balances these losses, allowing $D=R-O(\sqrt R)$ and giving $|S|=(\Delta-1)^{R-O(\sqrt R)}$ for every fixed $\Delta$ and all sufficiently large $R$. Each crashing node has one neighbor in $S$, preserving distant witnesses for the similarity chain.

\paragraph{Dependence on the crash bound (\Cref{thm:lower-bound}(iii)).}

With at most $f$ crashes, at most $f\Delta$ nodes of $S$ can be adjacent to $F$. To gain a factor depending on $R$, we extend $S$ by spacing these attachment nodes apart along the path (\Cref{fig:lb-spaced-cycle}). For $f\geq2$, the $f$ crashing nodes form the leaves of a tree with at most $\Delta-1$ children per node and height at most $h=\lceil\log_{\Delta-1}f\rceil$. The set $S$ consists of a set $S_F$ of $f(\Delta-1)$ nodes and paths of length $L$ between any two consecutive nodes in $S_F$ (so $|S|=\Theta(f\Delta L)$). We attach each leaf of the tree (i.e., each node in $F$) to a disjoint block of $\Delta-1$ consecutive nodes of $S_F$. Every node of $S$ is within $L/2$ hops of an attachment node, so reaching such a node at each end and passing through the tree takes at most $L+2h+2$ hops. Choosing $L=R-2h-3$ keeps this within $R-1$ and yields $|S|=\Theta((R-2\log_{\Delta-1}f)f\Delta)$ under the theorem's assumptions. These consecutive groups provide distant witnesses for atomic crashes. For $f=1$, a single hub with $\Delta$ spaced-out attachment nodes gives $\Omega(R\Delta)$ under standard crashes.

\subsection{Crash Tolerance for Locally Checkable Labelings}\label{sec:overview-lcl}

The gap theorem of Chang, Kopelowitz, and Pettie~\cite{ChangKopelowitzPettie2019} states that, on graphs of fixed maximum degree, the deterministic fault-free complexity of an LCL problem is either $O(\log^* n)$ or $\Omega(\log n)$. \Cref{thm:lcl-crash-complexity} relates these regimes to crash tolerance: every LCL with a deterministic $O(\log^* n)$-round algorithm on all such graphs admits such an algorithm tolerating any number of crashes. In contrast, a problem of fault-free complexity $\Theta(\log n)$ can require $\Theta(n)$ rounds under crashes.

\paragraph{Isolating the symmetry-breaking step (\Cref{thm:lcl-crash-complexity}(i)).}
We use the normal form of deterministic $O(\log^* n)$-round LCL algorithms on bounded-degree graphs~\cite{ChangKopelowitzPettie2019,BalliuEtAl2021}. For a constant $t$ depending only on the problem and the degree bound, such an algorithm first computes a coloring with a constant number of colors in which any two distinct nodes at distance at most $t$ have different colors. It then uses this coloring to solve the LCL in at most $t$ rounds. Thus, only the coloring has a round complexity dependent on $n$. We make this step crash tolerant directly and use consistent neighborhoods only for the constant-round second step.

We use Linial's color reduction~\cite{Linial1992}, starting with identifiers as colors. Each old color determines a candidate set of new colors. After exchanging colors for $t$ rounds, a node chooses from its own set while avoiding those of received colors other than its own. Missing messages only leave more candidates available. Let $A$ be the set of nodes that complete the coloring phase. Nodes within distance $t$ in $G[A]$ receive each other's colors at every step, so mutual exclusion preserves distinct colors. This gives the required coloring on $G[A]$ in $O(\log^* n)$ rounds (\Cref{lem:linial-crashes}).

\paragraph{Finishing on consistent neighborhoods.}
We keep the colors fixed and run \Cref{alg:parameter-cn} with fresh labeled stars that include them. Every view is a subgraph of $G[A]$, so the coloring is inherited by all nodes in a view even if some of its nodes crash during the construction. Two distinct nodes within distance $t$ in a view are also within distance $t$ in $G[A]$, so their inherited colors differ. The constant-round algorithm can thus be simulated on every view. Taking $R=t+\ell+1$, where $\ell$ is the checking radius, ensures local satisfaction for the original LCL by the simulation argument of \Cref{thm:cn-to-lcl}(ii). For constant degree bound and $R$, \Cref{thm:upper-bound} constructs these views in \mbox{$O(1)$} rounds. Together with coloring, this takes $O(\log^* n)$ rounds.

\paragraph{A linear lower bound for an LCL problem (\Cref{thm:lcl-crash-complexity}(ii)).}
We first define a locally checkable task on a graph family where the similarity-chain argument from Subsection~\mbox{\ref{sec:overview-lower-bounds}} yields a linear crash lower bound, although the task is solvable in logarithmic time without crashes. We then extend the task to arbitrary input-labeled graphs of maximum degree five. Following earlier LCL constructions~\cite{BalliuBrandtOlivettiSuomela2018,BalliuBrandtOlivettiSuomela2020,BalliuEtAl2021,ICALP:BGKMOR25,HadadNaor2025}, we encode the intended structure by finite input labels and verify it locally. Nodes that fail a structural check form the \emph{defect set}. On malformed components, nodes can use locally verifiable \emph{error outputs} that certify defects or point toward them. To preserve the separation, these outputs must be impossible on correctly labeled components without crashes, yet computable in $O(\log n)$ rounds on malformed components. This preserves the fault-free upper bound and the required outputs at both ends of the chain.

We use complete binary trees whose levels are joined into horizontal paths. Each node must output the root's input bit. This is enforced locally by requiring neighbors to agree and the root to match its input. The fault-free complexity is $\Theta(\log n)$: the diameter is logarithmic, as is the distance from the root to the leaves. After a level near the leaves crashes, however, the surviving bottom levels form a long strip without the tree's short communication paths.

Structural input labels specify parents, children, and horizontal neighbors, with radius-two checks characterizing the intended family. The extended LCL $\Pi_{\mathrm{bit}}$ requires adjacent nodes either to output the same bit or both to use error outputs. An error output certifies a structural defect at its node or points along a mutually consistent parent--child edge to another error output; two adjacent nodes may not point to each other. On a correctly labeled component without crashes, error outputs are impossible: there are no defects, and the parent--child tree admits no pointer cycle without two nodes pointing to each other.

In a malformed component, every node is within $O(\log n)$ hops of the defect set along mutually consistent parent--child edges (\Cref{lem:lcl-tree-defects}). The structural checks force successive internal levels to double in size; if no nearby defect exists, they instead force a complete, correctly labeled component. Nodes at defects certify failed checks, and other nodes point along shortest parent--child paths to the defect set, giving the $O(\log n)$ fault-free upper bound.

For the crash lower bound, only nodes in the level $F$ three edges above the leaves may crash; let $S$ be the bottom three levels (\Cref{fig:lcl-crash-layer}). The set $F$ has $\Theta(n)$ nodes. Each leaf's radius-two checking neighborhood survives, so its local constraints must still hold. Adjacent leaves must either output the same bit or both use error outputs. The leaf path thus forces a common bit or error outputs at every leaf.

For a contradiction, consider a deterministic algorithm that always terminates within $T:=|F|$ rounds. Starting from the fault-free execution with root bit~$0$, move the atomic crashes of $F$ to rounds $r=T,T-1,\ldots,1$, one node at a time. Once all of $F$ crashes before round~1, $S$ is cut off from the root, so we switch the root bit to~$1$ and traverse the crash schedules in reverse order (\Cref{fig:lcl-chain}). For a step moving a crash into round $r$, no node of $F$ sends in rounds $r+1,\ldots,T$; differences can then propagate from the changed node's children only through $G[S]$, for at most $T-r<|F|$ hops. For each node of $F$, one of the two extreme leaves is at distance at least $|F|$ from that node's children in $G[S]$ (\Cref{obs:lcl-distant-leaf} and \Cref{lem:lcl-chain}), providing a witness for the corresponding crash-time changes through round $T$. Initially all leaves output $0$. Each witness retains $0$, forcing all leaves to retain $0$ throughout the chain. The fault-free endpoint with root bit~$1$ yields a contradiction. Since $|F|=\Theta(n)$, this proves an $\Omega(n)$ lower bound already for atomic crashes. The general simulation using \Cref{thm:upper-bound}(i) and \Cref{thm:cn-to-lcl}(ii) gives the matching $O(n)$ upper bound.

\subsection{Problem-Specific Transformation for Sinkless Orientation}\label{sec:overview-sinkless-orientation}
By \Cref{thm:lcl-crash-complexity}(ii), no general transformation can preserve the asymptotic round complexity of every deterministic fault-free LCL algorithm. For particular problems, however, the asymptotic round complexity can sometimes be maintained by exploiting the problem's structure directly. We illustrate this for sinkless orientation, which requires nodes to orient their edges so that every node of input degree $\geq 3$ (i.e., in the input graph $G$) has at least one outgoing edge. We use edge grabbing as an intermediate problem: every node of input degree $\geq 3$ grabs exactly one incident edge so that no edge is grabbed by both endpoints.

Any $T$-round fault-free sinkless orientation algorithm yields a $T$-round edge grabbing algorithm: each such node grabs an outgoing edge. We run it while tracking failures. On first detecting a missing message or receiving a failure notification, a node records that edge, grabbing it if its input degree is at least three, and notifies its neighbors in subsequent rounds. Nodes that detect no failure within $T$ rounds keep their fault-free views and outputs. Surviving endpoints cannot grab the same edge: either both follow the fault-free algorithm, or the first to detect a failure does so at a different neighbor, since the other endpoint survives and has not yet sent a notification.

In one additional round, nodes announce their grabbed edges. Each such edge is oriented away from the node that grabbed it, and other edges are pointed towards the smaller identifier. Surviving neighbors agree on the orientation of their common edge, and every surviving node of input degree $\geq3$ has an outgoing edge, possibly towards a crashed neighbor, proving \Cref{thm:so-transformation}. For both edge grabbing and sinkless orientation, the correctness guarantees also hold at surviving nodes whose checking neighborhoods contain a crash.

\section{Preliminaries}\label{sec:preliminaries}
In this section, we introduce the graph notation and the communication and crash-fault models used throughout the paper. Moreover, we discuss the initial knowledge available to each node and we formally define the notion of local satisfaction under crashes.

\subsection{Graphs and Related Notations}
Throughout the paper, we consider finite, simple, undirected graphs $G=(V,E)$ with at most $n$ nodes and maximum degree at most $\Delta$. We assume $\Delta\geq2$ without loss of generality: since $\Delta$ is only an upper bound, graphs of maximum degree at most one are covered by taking $\Delta=2$.
Each node has a unique identifier and, if required by the problem, an input label. Subgraphs retain these identifiers and input labels.
We say that two labeled graphs are \emph{equal} if they have the same node set and edge set, and the identifiers and input labels agree at every node.

For a node $v\in V$, its \emph{open and closed neighborhoods} are $N_G(v):=\{u\in V:\{u,v\}\in E\}$ and $N_G[v]:=N_G(v)\cup\{v\}$ respectively, and its \emph{degree} is $\deg_G(v):=|N_G(v)|$.

These notations extend naturally to a set of nodes $X\subseteq V$: its open and closed neighborhoods are
\[
    N_G(X):=\Bigl(\bigcup_{v\in X}N_G(v)\Bigr)\setminus X,\qquad N_G[X]:=N_G(X)\cup X,
\]
respectively. The \emph{induced subgraph} of $G$ on $X$ is denoted by $G[X]$. We omit graph subscripts when they are clear from the context.

We write $\dist_G$ for the \emph{shortest-path distance} in $G$, with distance infinity between nodes in different components. For $v \in V$ and $X\subseteq V$, the \emph{distance from $v$ to $X$} is $\dist_G(v,X):=\min_{x\in X}\dist_G(v,x)$, with $\min\emptyset=\infty$.
For an integer $r\geq0$, the \emph{radius-$r$ balls} around $v\in V$ and $X\subseteq V$ are defined as follows:
\[
    B_G(v,r):=\{u:\dist_G(u,v)\leq r\}
    \qquad\text{and}\qquad
    B_G(X,r):=\bigcup_{v\in X}B_G(v,r).
\]

The \emph{eccentricity} of $v$ is $\max_{u\in V}\dist_G(v,u)$, and the \emph{diameter} $\diam(G)$ of a nonempty connected graph is the maximum eccentricity of its nodes. The \emph{girth} of $G$ is the length of a shortest cycle, and acyclic graphs have infinite girth.

\subsection{Model} \label{sec:model}

\paragraph{Communication and parameters.}
We work in the synchronous \local model~\cite{Linial1992}. Computation proceeds in rounds: in each round, every node that has not yet crashed may send a separate message to each neighbor. Nodes that complete the round receive the messages delivered to them and update their local states. Message size and local computation are unrestricted, and the running time is measured by the number of communication rounds. All algorithms considered in this paper are deterministic.

All nodes know upper bounds $n$ and $\Delta$ on the number of nodes and the maximum degree of $G$, respectively. These bounds do not need to be tight and remain unchanged when an algorithm is run on a subgraph. When solving $\CN(R)$, all nodes also know the radius $R$. We state explicitly whenever we additionally assume that nodes know an upper bound $f$ on the number of crashes. Identifiers are distinct integers in $\{1,\ldots,\max\{2,n\}^{c_{\mathrm{ID}}}\}$, where $c_{\mathrm{ID}}\geq1$ is a fixed known integer.

\paragraph{Crash faults.}
An adversary may crash up to $f<n$ nodes over the entire execution. Crashes may occur in any round or before the initial local computation.
Once a node crashes, it performs no further computation and sends no further messages. We call an execution with at most $f$ crashes \emph{admissible}.
If  $f=0$, the execution is fault-free, and \Cref{thm:cn-to-lcl}(i) gives the complexity of $\CN(R)$. Unless stated otherwise, we assume $f\geq1$; correctness is still required in executions without crashes.

In the \emph{standard crash-fault model}, if a node crashes in round $t$, an arbitrary subset of its prescribed round-$t$ messages may be delivered. In the \emph{atomic crash-fault model}, a node crashes at the beginning of a round, before sending any messages in that round. Atomic crashes are therefore a special case of standard crashes. Apart from crashes, communication is reliable: a message sent in a round is delivered in that round whenever both its sender and its recipient complete the round. Unless stated otherwise, we use the standard model throughout the paper.

A node is called \emph{surviving} if it does not crash during the execution. Let $\mathcal F$ be the set of nodes that crash and $\mathcal S:=V\setminus\mathcal F$ the set of surviving nodes. The \emph{surviving graph} is the induced subgraph $G[\mathcal S]$~\cite{Lynch1996}.

\paragraph{Correctness.}
An algorithm solves a problem in $T$ rounds if, in every admissible execution, all surviving nodes output by the end of round $T$ and their outputs satisfy the problem's correctness conditions. Nodes that crash have no output requirement. When we refer to the \emph{fault-free setting}, these requirements apply only to executions without crashes.

\paragraph{Locally checkable problems.}
A locally checkable problem is specified by a checking radius $\ell$ and local validity conditions: the local constraint at each node depends only on its induced radius-$\ell$ neighborhood, including the input and output labels.
The radius, label domains, and validity conditions may depend on the known bounds $n$ and $\Delta$. The conditions are independent of identifiers unless these are explicitly part of the problem's inputs; algorithms may use identifiers freely.
We retain the original bounds throughout the execution and in every subgraph simulation, so the checking radius and rule remain unchanged.
An LCL is specified by a fixed degree bound, fixed finite input and output alphabets, a constant checking radius, and a fixed set of allowed rooted input/output-labeled neighborhoods (including local ports when used), independent of $n$ and identifiers~\cite{NaorStockmeyer1995}.

\begin{definition}[Local satisfaction under crashes]
\label{def:lcl-away-from-crashes}
Let $\Pi$ be a locally checkable problem with checking radius $\ell$. In an execution on the original graph $G$ with crash set $\mathcal F$, a surviving node $v$ is \emph{locally satisfied} if $B_G(v,\ell)\cap\mathcal F\neq\emptyset$; otherwise, it is locally satisfied if and only if the input and output labels on $G[B_G(v,\ell)]$ satisfy $\Pi$'s local constraint at $v$. The surviving outputs are valid if and only if every surviving node is locally satisfied.
\end{definition}

Distinct locally specified problems may have the same valid global fault-free solutions but different requirements under this definition.

\subsection{Initial Knowledge}
\label{sec:initial-knowledge}
We now discuss the information available to each node at the start of an execution.

In addition to the global parameters of \Cref{sec:model}, each node $v$ initially knows its identifier, its input label (if any), and the identifiers of its neighbors in $G$.
We denote its initial topological information by the \emph{star} $\sigma_v=(v,N_G[v])$ with \emph{root} $v$, consisting of $v$, its neighbors, and all edges incident to $v$. This does not include edges between neighbors of $v$ or their input labels. The root's input label, if present, is included when the star is communicated but omitted from the notation.

If required by the problem definition, we additionally assume that each node knows distinct local port numbers for its incident edges. Its star records these numbers at the root. They are included when comparing labeled graphs and are preserved without renumbering in subgraphs.

For $\CN(R)$ and locally checkable problems under \Cref{def:lcl-away-from-crashes}, the assumption that stars are initially known can be relaxed at the cost of one additional round. We may instead assume that each node initially knows only its own identifier, its input label (if any), the globally known parameters, and any port numbers required by the problem. A node can send messages to each neighbor and associate received messages with their incident edges even before knowing the neighbors' identifiers. In a fault-free execution, one round of exchanging identifiers reveals every star. The following transformation shows that one additional round also suffices with crashes: it simulates an execution with initially known stars on a spanning subgraph whose omitted edges all have crashed endpoints.

\paragraph{Transformation.}
Let $\mathcal A$ be an algorithm that assumes initially known stars. In one preliminary round, each node sends its identifier to every neighbor. Each node $v$ that completes this round constructs the star $\widehat\sigma_v=(v,L_v\cup\{v\})$, where $L_v$ is the set of neighbors whose identifiers it received, retaining any associated local port numbers. It then starts $\mathcal A$ with this star, its original input label, and the unchanged global parameters, and adopts its output. The round counter is reset: round $r$ of $\mathcal A$ is executed in physical round $r+1$.
Throughout this simulation, $\mathcal A$ treats $L_v$ as the neighbor set of $v$: it uses degree $|L_v|$ and sends messages only to neighbors in $L_v$.

Let $C$ be the set of nodes that crash before completing the preliminary round. These crashes count toward the overall bound $f$. A node of $C$ may fail to send its identifier to some neighbors, so the constructed stars can omit edges incident to $C$. The following lemma shows that, after the preliminary round, the transformed algorithm simulates $\mathcal A$ on a spanning subgraph of $G$ that retains every edge with both endpoints outside $C$.

\begin{lemma}[One-round star initialization]
\label{lem:star-initialization}
For every deterministic algorithm $\mathcal A$ with initially known stars and every execution on $G$, the transformation simulates an execution of $\mathcal A$ on a labeled spanning subgraph $G'\subseteq G$ with $G'[V\setminus C]=G[V\setminus C]$, in the same crash model and with the same crashed nodes. Thus, if $\mathcal A$ tolerates up to $f$ crashes and terminates within $T$ rounds on every labeled spanning subgraph of $G$, the transformation inherits its output guarantees on $G'$ within $T+1$ rounds, with at most $f$ crashes and unchanged global parameters.
\end{lemma}

\begin{proof}
Let $A:=V\setminus C$, and let $G'$ be the labeled graph on the full node set $V$, with the original identifiers and input labels, whose edges are those reported in the constructed stars:
\[
    G'=(V,E'),\qquad
    E':=\bigcup_{v\in A}\{\{u,v\}:u\in L_v\}.
\]
Every reported edge belongs to $G$, and both endpoints of an edge in $G[A]$ receive each other's identifiers. Hence $G'\subseteq G$ and $G'[A]=G[A]$.

Consider the execution of $\mathcal A$ on $G'$ in which all nodes of $C$ crash before initialization and, for each $r \geq 1$, the crashes and crash-round delivery choices of physical round $r+1$ are copied to round $r$. This execution has the same crashed nodes as the transformation and respects the same crash model. We show by induction on $r\geq0$ that, after $r$ simulated rounds, every node that has not yet crashed has the same $\mathcal A$-state in both executions, including any output already produced.

For the base case $r=0$, the nodes that have not crashed are exactly those of $A$. Fix $v\in A$. Every $u\in L_v$ contributes the edge $\{u,v\}$ to $E'$, so $L_v\subseteq N_{G'}(v)$. Conversely, an edge $\{u,v\}\in E'$ is reported either by $v$, in which case $u\in L_v$, or by $u\in A$. In the latter case, both endpoints complete the preliminary round and exchange their identifiers, so again $u\in L_v$. Hence, $N_{G'}(v)=L_v$, and the constructed star of $v$ is exactly its initial star in $G'$. The identifiers, input labels, and global parameters also agree, so $v$ starts $\mathcal A$ in the same state in both executions.

For the induction step, let $r\geq1$ and assume that the claim holds after $r-1$ simulated rounds. Every node $v$ that has not crashed before round $r$, including one that crashes during the round, starts the round in the same $\mathcal A$-state in both executions. Since $\mathcal A$ is deterministic and $v$ uses the same neighbor set $L_v$ in both executions, it prescribes the same messages for round $r$. Nodes that crashed earlier send nothing. The copied crash-round delivery choices give the same partial deliveries in the standard model, while nodes crashing atomically in that round send nothing in either execution. Thus, every node completing the round receives the same messages and makes the same state update.

Consequently, surviving nodes produce identical outputs. If $\mathcal A$ terminates within $T$ rounds, the transformation produces these outputs within $T+1$ physical rounds.
\end{proof}

\paragraph{Correctness on the original graph.}\leavevmode
By \Cref{lem:star-initialization}, the transformation produces the same surviving outputs as $\mathcal A$ on $G'$. Assuming that $\mathcal A$ is correct on $G'$, it remains to show that these outputs also satisfy the requirements on $G$. We therefore verify that restoring the omitted edges incident to $C$ preserves correctness for $\CN(R)$ and for locally checkable problems under \Cref{def:lcl-away-from-crashes}.

For $\CN(R)$, a view in $G'$ is still a labeled subgraph of $G$, and its radius remains valid because $\dist_G\leq\dist_{G'}$. Every edge between surviving nodes is retained, and consistency depends only on the unchanged views. If no node crashes, then $C=\emptyset$ and $G'=G$, so fault-free completeness is preserved.

For a locally checkable problem with checking radius $\ell$, fix a surviving node $v$ with $B_G(v,\ell)\cap\mathcal F=\emptyset$. Every node in this ball belongs to $A$, and $G'[A]=G[A]$. Thus, every edge within the ball and every shortest path from $v$ to a node in the ball is retained in $G'$. Since deleting edges cannot shorten distances, we obtain that
$$
    G'[B_{G'}(v,\ell)]=G[B_G(v,\ell)]
$$
as labeled graphs. This common checking neighborhood contains no crashed node, so correctness on $G'$ implies that its input and output labels satisfy the original local constraint at $v$. The same constraint therefore holds in $G$. All other surviving nodes are locally satisfied by definition.

In the following sections, exact round complexities are computed under the assumption that the stars are initially known to nodes. For $\CN(R)$ and locally checkable problems under \Cref{def:lcl-away-from-crashes}, this convention changes the complexity by at most one round, provided that the algorithm's guarantees hold on every labeled spanning subgraph of the input graph.

\section{The Consistent Neighborhoods Problem}
\label{sec:cn-problem}
\label{sec:cn-to-lcl}

This section formally defines the consistent neighborhoods problem $\CN(R)$ and proves \Cref{thm:cn-to-lcl}.
\Cref{sec:fault-free-cn} shows that, without crashes, $\CN(R)$ can be solved in $R$ rounds and that this is optimal, proving part~(i). Afterwards, \Cref{sec:cn-local-simulation} focuses on part~(ii) and shows that $\CN(R)$ with $R\geq T+\ell+1$ enables surviving nodes to simulate a deterministic fault-free $T$-round algorithm for a locally checkable problem with checking radius $\ell$ without further communication. Finally, in \Cref{sec:cn-limits}, we show that views of radius $T+\ell$ do not suffice in general, even for $T=\ell=1$, and we note that the simulation only guarantees the original local constraints of the problem at surviving nodes whose checking neighborhoods contain no crash.

Without crashes, every node can learn its complete radius-$R$ neighborhood within $R$ rounds by forwarding labeled stars. Crashes disrupt this approach: (a) the views may omit nodes or edges of the complete radius-$R$ neighborhoods, and (b) two adjacent surviving nodes may end up with conflicting views. Before presenting the formal definition of $\CN(R)$, we discuss the properties it requires in order to address these issues.

To address issue (a), $\CN(R)$ permits incomplete views when crashes occur but specifies what each view may and must contain.
On the one hand, views stay local: $\CN(R)$ provides surviving nodes with \emph{radius-$R$ views}, where a \emph{radius-$R$ view centered at $v$} is a labeled subgraph $H$ of $G$ such that $v \in V(H)$ and $V(H) \subseteq B_G(v,R)$.
Here, distances are measured in the communication graph $G$, not in the view $H$. This keeps views local with respect to $G$, while allowing a view to include a node $w$ even if crashes cut the shortest paths from $v$ to $w$.
On the other hand, $\CN(R)$ limits how incomplete views can be: adjacent surviving nodes must include their common edge in their views (\emph{Edge Validity}), and, when no node crashes, every view must be the complete radius-$R$ neighborhood (\emph{Fault-Free Completeness}).

To address issue (b), $\CN(R)$ requires the views of adjacent surviving nodes to induce identical labeled subgraphs on their respective comparison sets (\emph{Consistency}). The comparison sets are defined for an edge $\{u,v\}$ at the fixed view radius $R$.

\begin{definition}[Comparison sets for $\{u, v\}$]\label{def:comparison-sets}
Let $H_u$ and $H_v$ be labeled graphs that both contain the edge $\{u,v\}$. For the fixed integer radius $R\geq1$, define
\begin{align*}
    S_u(v) &:= B_{H_u}(u,R-1) \cup B_{H_u}(v,R-1),\\
    S_v(u) &:= B_{H_v}(u,R-1) \cup B_{H_v}(v,R-1).
\end{align*}
Thus, $S_u(v)$ and $S_v(u)$ contain the nodes within distance $R-1$ of either endpoint, measured in $H_u$ and $H_v$, respectively. The dependence on the fixed radius $R$ is suppressed in the notation.
\end{definition}

Putting these requirements together, we formally define $\CN(R)$ as follows.
\begin{definition}[Consistent Neighborhoods Problem]\label{def:cn}
For a communication graph $G=(V,E)$, an integer $R\geq1$, and a crash bound $f$, an algorithm
solves $\CN(R)$ on $G$ if every execution on $G$ with at most $f$ crashes satisfies
the following properties.
\begin{itemize}
    \item \textbf{Termination and radius.} Every surviving node $v$ outputs a
    radius-$R$ view $H_v$ centered at $v$, with the center labeled by $v$'s
    own identifier.

    \item \textbf{Edge validity.} If $u$ and $v$ are adjacent surviving
    nodes, then both $H_u$ and $H_v$ contain the edge $\{u,v\}$.

    \item \textbf{Consistency.} If $u$ and $v$ are adjacent surviving nodes,
    then, using the comparison sets from \Cref{def:comparison-sets}, $H_u[S_u(v)]=H_v[S_v(u)]$
    as labeled graphs.

    \item \textbf{Fault-free completeness.} If no node crashes, then every
    node $v$ outputs its complete radius-$R$ neighborhood
    $H_v=G[B_G(v,R)]$.
\end{itemize}
\end{definition}

An algorithm solves $\CN(R)$ on a graph class if it satisfies these requirements on every graph in the class.

We highlight that the consistency property of \Cref{def:cn} is deliberately local to a surviving edge. We do not require all surviving nodes to output the same graph, and we do not require nodes in distinct connected components of the final surviving graph to agree. As we show in \Cref{sec:cn-local-simulation}, this local agreement already suffices for simulating algorithms for locally checkable problems.

We will show in \Cref{lem:cn-exact-checking} that the first three properties imply that, for every surviving node $v$ and integer $0\leq\ell<R$, the labeled induced radius-$\ell$ neighborhood of $v$ in its view coincides with that in $G$ whenever no node in $B_G(v,\ell)$ crashes during the execution, i.e., $B_G(v,\ell)\cap \mathcal F = \emptyset$.

We also note that, in fault-free executions, the first three properties of \Cref{def:cn} already come close to fault-free completeness: they force each view $H_v$ to contain all nodes of $B_G(v,R)$ and every edge with an endpoint at distance less than $R$ from $v$ (this follows from \Cref{lem:cn-view-propagation}). However, they do not force the presence of edges between two nodes at distance exactly $R$ from $v$.
For example, on a triangle with $R=1$, every node can output just its own star.
These views satisfy radius, edge validity, and consistency, but each omits the edge between its two neighbors and is therefore not complete.
Fault-free completeness rules out such omissions when no node crashes, ensuring that the primitive recovers ordinary collection of complete induced radius-$R$ neighborhoods.

\subsection{Consistent Neighborhoods without Crashes}
\label{sec:fault-free-cn}
We verify that \Cref{def:cn} matches the fault-free approach described at the beginning of this section: without crashes, flooding labeled stars for $R$ rounds solves $\CN(R)$. This is also optimal in general. Fault-free completeness requires each view to contain the edges between nodes at distance exactly $R$. Since information travels only one hop per round, learning these edges requires $R$ rounds in the worst case. The following lemma establishes both bounds and proves part~(i) of \Cref{thm:cn-to-lcl}.

\begin{lemma}[Fault-free consistent neighborhoods]
\label{lem:fault-free-cn}
For every integer $R\geq1$, $\CN(R)$ can be solved deterministically in $R$ rounds without crashes. In the worst case, $R$ rounds are necessary even on graphs with $2R+1$ nodes and maximum degree at most two.
\end{lemma}
\begin{proof}
For the upper bound, every node starts with its star, together with its input label (if any), and, in each of $R$ rounds, sends to its neighbors all stars it knows, each with the input label of its source.
Afterwards, the roots of the stars collected by a node $v$, including its own initial star, form exactly $B_G(v,R)$. These stars reveal every edge between their roots, including edges whose two endpoints are at distance exactly $R$ from $v$. Node $v$ retains only these roots and the reported edges between them, discarding any nodes at distance $R+1$ mentioned in stars rooted at distance $R$. Node $v$ then outputs $H_v = G[B_G(v,R)]$.

These views satisfy all properties of \Cref{def:cn}. Termination and radius, edge validity, and fault-free completeness hold by construction. For consistency, consider two adjacent nodes $u$ and $v$. Every path of length at most $R-1$ from $u$ or $v$ stays within distance $R$ of both endpoints, and hence lies in both $H_u$ and $H_v$. Therefore, both comparison sets equal $S := B_G(u,R-1) \cup B_G(v,R-1)$, and since both views are induced subgraphs of $G$, we obtain $H_u[S] = G[S] = H_v[S]$.

For the lower bound, let $P$ be a path on $2R+1$ nodes with middle node $v$ and endpoints $a$ and $b$, and let $C$ be the cycle obtained from $P$ by adding the edge $\{a,b\}$. We assign the same identifiers and input labels to the nodes of both graphs and, if used, the same local port numbers to their common edges. Fix a deterministic algorithm for $\CN(R)$ that terminates within $T < R$ rounds without crashes, and let $\mathcal E_P$ and $\mathcal E_C$ denote its fault-free executions on $P$ and on $C$, respectively, both with global bounds $n = 2R+1$ and $\Delta = 2$. In these executions, only $a$ and $b$ have different initial states, and both are at distance $R$ from $v$ in both graphs. Since information about an initially different state can traverse at most one edge per round, induction on the number of rounds shows that the state of $v$ is identical in $\mathcal E_P$ and $\mathcal E_C$ through round $R-1$, and hence $v$ obtains the same output in both.

However, by fault-free completeness, $v$ must output $H_v = P$ in $\mathcal E_P$ and $H_v = C$ in $\mathcal E_C$, and these outputs differ in the edge $\{a,b\}$. Hence, no deterministic algorithm can always solve $\CN(R)$ in fewer than $R$ fault-free rounds.
\end{proof}

\subsection{Local Simulation from Consistent Neighborhoods}
\label{sec:cn-local-simulation}

Consider a locally checkable problem $\Pi$, and let $\mathcal{A}$ be a deterministic $T$-round algorithm for $\Pi$ in the fault-free setting. We now describe how a solution to $\CN(R)$ enables us to transform $\mathcal{A}$ into a crash-tolerant algorithm for $\Pi$. Here, a solution to $\CN(R)$ on $G$ consists of views $H_v$ of the surviving nodes of an execution on $G$ that satisfy the properties of \Cref{def:cn}. The transformation is as follows: every surviving node $v$ simulates the algorithm on its view $H_v$ (treating its view as the entire graph) and adopts its simulated output. We present the pseudocode in \Cref{alg:cn-simulation}. Afterwards, \Cref{lem:cn-local-simulation} shows that, for sufficiently large $R$, the resulting outputs make every surviving node locally satisfied.

\begin{algorithm}[H]
\caption{Crash-Tolerant Transformation of $\mathcal{A}$ at Node $v$}
\label{alg:cn-simulation}
\begin{algorithmic}[1]
\Statex \textbf{Parameters:} problem $\Pi$ with checking radius $\ell$, deterministic algorithm $\mathcal A$, round bound $T$, global bounds $n,\Delta$, and view radius $R$.
\Statex \textbf{Preconditions:} $\mathcal A$ solves $\Pi$ within $T$ rounds on every subgraph of $G$; $R\geq T+\ell+1$.
\Statex \textbf{Input:} labeled view $H_v$ of $v$ in a solution to $\CN(R)$ on $G$.
\Statex \textbf{Output:} output label of $v$ for $\Pi$.
\State Initialize each node of $H_v$ with its identifier, its input label (if any), and its neighbor identifiers in $H_v$. If ports are used, preserve their inherited names without renumbering.
\State Simulate $T$ rounds of $\mathcal{A}$ on $H_v$ with these initial states and the global bounds $n$ and $\Delta$.
\State Return the output assigned to $v$ in the simulated execution of $\mathcal{A}$.
\end{algorithmic}
\end{algorithm}

\begin{lemma}[Local simulation from consistent neighborhoods]
\label{lem:cn-local-simulation}
Let $\Pi$ be a locally checkable problem with checking radius $\ell\geq1$, and let $\mathcal A$ be a deterministic algorithm that solves $\Pi$ within $T$ rounds on every subgraph of $G$ in the fault-free setting, where $T\geq0$. Given a solution to $\CN(R)$ on $G$ with $R\geq T+\ell+1$, each surviving node can compute its output with no further communication so that every surviving node is locally satisfied for $\Pi$.
\end{lemma}

We require $\mathcal A$ to remain correct within $T$ rounds on every subgraph of $G$ because the views may omit nodes and edges of $G$. This assumption holds, for example, if $\mathcal A$ solves $\Pi$ within $T$ rounds on every graph with at most $n$ nodes and maximum degree at most $\Delta$. Every subgraph of $G$ satisfies these bounds. In each simulation, we retain the inherited node labels and use the original bounds $n$ and $\Delta$, as specified in \Cref{sec:preliminaries}. The checking radius $\ell$ and the local checking rule are also kept unchanged. In contrast, algorithms that rely on properties lost under taking subgraphs, such as connectivity, need not satisfy this assumption.

The proof of \Cref{lem:cn-local-simulation} builds on the following fact: since $\mathcal A$ is correct on every subgraph of $G$, simulating it on a view $H_v$ produces a valid solution on $H_v$. A surviving node whose checking neighborhood contains a crashed node is locally satisfied by definition (\Cref{def:lcl-away-from-crashes}). It therefore remains to verify the original constraints at surviving nodes whose checking neighborhoods do not contain any crashed nodes. To establish local satisfaction at such a node, it suffices to show that its actual checking neighborhood, including the output labels, coincides with the corresponding neighborhood in the simulation on $H_v$.

We establish this correspondence in two steps. First, we show that the checking neighborhood of $v$ in $H_v$ coincides with its checking neighborhood in $G$. Second, we show that each node $u$ in this neighborhood computes from its own view $H_u$ the same output that the simulation on $H_v$ assigns to $u$. Both steps use the propagation of consistency along surviving paths (\Cref{lem:cn-view-propagation}); the second also uses the locality of $T$-round algorithms.

We first formalize how consistency propagates along surviving paths. Consistency only constrains neighboring views, but together with edge validity it extends along surviving paths: if $v$ and $w$ are joined by a surviving path of length $d \le R$, then $w \in V(H_v)$, and $H_v$ agrees with $H_w$ within radius $R-d$ of $w$. We prove this in \Cref{lem:cn-view-propagation}, using the following observation about induced subgraphs.

\begin{observation}[Matching neighborhoods in equal induced subgraphs]
\label{lem:ball-transfer}
Let $H$ and $H'$ be labeled graphs with $H[X] = H'[X']$ for some node sets $X$ and $X'$. Let $r \ge 0$, and consider a node $v \in X$. If $B_H(v,r) \subseteq X$ and $B_{H'}(v,r) \subseteq X'$, then $H[B_H(v,r)] = H'[B_{H'}(v,r)]$.
\end{observation}

\begin{lemma}[Propagation of consistent views]
\label{lem:cn-view-propagation}
Let $p_0, p_1, \ldots, p_d$ be a path of  surviving nodes of length $d \leq R$. Suppose that the views $H_v$ of the surviving nodes $v$ satisfy the termination and radius, edge validity, and consistency requirements of $\CN(R)$. Then, $p_d \in V(H_{p_0})$ and $
    H_{p_0}\bigl[B_{H_{p_0}}(p_d, R-d)\bigr]
    = H_{p_d}\bigl[B_{H_{p_d}}(p_d, R-d)\bigr]$
as labeled graphs, with distances measured in the respective views.
\end{lemma}
\begin{proof}
We argue by induction on $d$. For $d=0$, the claim follows because $H_{p_0}$ contains its center $p_0$ and the two sides of the claimed equality coincide. Let $1\leq d\leq R$ and assume that the claim holds for paths of length $d-1$.

Set $X:=B_{H_{p_0}}(p_{d-1},R-d+1)$ and $X':=B_{H_{p_{d-1}}}(p_{d-1},R-d+1)$.
By the induction hypothesis, we know that $p_{d-1}\in V(H_{p_0})$ and $H_{p_0}[X]=H_{p_{d-1}}[X']$.

Edge validity places the edge $\{p_{d-1},p_d\}$ in $H_{p_{d-1}}$. Since $R-d+1\geq1$, both endpoints belong to $X'$, so the edge belongs to $H_{p_{d-1}}[X']$ and, by the induction hypothesis, also to $H_{p_0}[X]$. In particular, $p_d\in V(H_{p_0})$.

Since both views contain this edge, $p_{d-1}$ is at distance at most $1$ from $p_d$ in both of these views. The triangle inequality implies that \mbox{$ B_{H_{p_0}}(p_d,R-d)\subseteq X$} and $B_{H_{p_{d-1}}}(p_d,R-d)\subseteq X'$. Applying \Cref{lem:ball-transfer}, we obtain the following:
\[
H_{p_0}[B_{H_{p_0}}(p_d,R-d)]
=
H_{p_{d-1}}[B_{H_{p_{d-1}}}(p_d,R-d)].
\]

Next, consider the comparison sets $Y:=S_{p_{d-1}}(p_d)$ and $Y':=S_{p_d}(p_{d-1})$. Edge validity ensures that both endpoint views contain $\{p_{d-1},p_d\}$, so these sets are defined, and consistency gives $H_{p_{d-1}}[Y]=H_{p_d}[Y']$. Since $R-d\leq R-1$, the radius-$(R-d)$ balls around $p_d$ in these two views lie in $Y$ and $Y'$, respectively. Applying \Cref{lem:ball-transfer} again yields:
\[
H_{p_{d-1}}[B_{H_{p_{d-1}}}(p_d,R-d)]
=
H_{p_d}[B_{H_{p_d}}(p_d,R-d)].
\]
Combining the two neighborhood equalities completes the induction.
\end{proof}

We now show that, for every surviving node $v$ and integer $0\leq\ell<R$, the labeled induced radius-$\ell$ neighborhoods of $v$ in $H_v$ and $G$ coincide if no node in $B_G(v,\ell)$ crashes.

\begin{lemma}[Exact checking neighborhoods]
\label{lem:cn-exact-checking}
Let $v$ be a surviving node, and suppose that the views $H_u$ of all surviving nodes $u$ satisfy the termination and radius, edge validity, and consistency requirements of $\CN(R)$. Let $0\leq\ell<R$ be an integer. If $B_G(v,\ell)\cap\mathcal F=\emptyset$, then $H_v[B_{H_v}(v,\ell)]=G[B_G(v,\ell)]$ as labeled graphs.
\end{lemma}
\begin{proof}
We show that each of the two graphs is a labeled subgraph of the other. We first show that $H_v[B_{H_v}(v,\ell)]$ is a labeled subgraph of $G[B_G(v,\ell)]$. By termination and radius, $H_v$ is a labeled subgraph of $G$, which implies that every edge of $H_v$ is an edge of $G$, and distances in $H_v$ are at least those in $G$. Hence $B_{H_v}(v,\ell) \subseteq B_G(v,\ell)$ and  $H_v[B_{H_v}(v,\ell)]$ is a labeled subgraph of $G[B_G(v,\ell)]$.

For the reverse inclusion, fix $a\in B_G(v,\ell)$ and choose a shortest $v$--$a$ path of length $d=\dist_G(v,a)\leq\ell$. This path lies entirely in $B_G(v,\ell)$, so all its nodes survive. By \Cref{lem:cn-view-propagation}, $a\in V(H_v)$, and the labeled induced radius-$(R-d)$ neighborhoods of $a$ in $H_v$ and $H_a$ coincide. For every edge $\{a,b\}$ of $G[B_G(v,\ell)]$, both endpoints survive, so edge validity places it in $H_a$. Since $R-d\geq R-\ell\geq1$, this edge also belongs to $H_v$. Thus $H_v$ contains every node and every edge of $G[B_G(v,\ell)]$.

It remains to verify that these nodes lie within distance $\ell$ of $v$ in $H_v$. The shortest paths considered above lie entirely in $G[B_G(v,\ell)]$ and are therefore retained in $H_v$. Hence every $a\in B_G(v,\ell)$ satisfies $\dist_{H_v}(v,a)\leq\dist_G(v,a)\leq\ell$. Together with the edge inclusion and the inherited labels, this proves that $G[B_G(v,\ell)]$ is a labeled subgraph of $H_v[B_{H_v}(v,\ell)]$.
\end{proof}

Having established equality of the checking neighborhoods, it remains to compare their output labels. For every surviving node $v$ whose checking neighborhood contains no crashed node, we show that each $u\in B_G(v,\ell)$ computes from its own view $H_u$ the same output that the simulation of $\mathcal A$ on $H_v$ assigns to $u$.

We use the following locality property: if two labeled graphs have identical induced radius-$(T+1)$ neighborhoods around a node $u$, then the fault-free $T$-round executions of $\mathcal A$ on these graphs, with the same global parameters, assign the same output to $u$. Indeed, these neighborhoods determine the initial states of all nodes within distance $T$ of $u$, and information propagates by at most one edge per round. We will apply this fact to $H_v$ and $H_u$, centered at $u$, after using propagation of consistency to establish the required neighborhood equality.

This is where the condition $R \ge T+\ell+1$ comes from: a node $u$ in the checking neighborhood of $v$ is at most $\ell$ hops away, so $H_v$ and $H_u$ agree within distance $R-\ell \ge T+1$ of $u$, which suffices to determine its output. \Cref{sec:cn-limits} shows that this bound cannot be improved in general.
\begin{lemma}[Matching simulated outputs]
\label{lem:cn-matching-outputs}
Let $\mathcal A$ be a deterministic $T$-round algorithm, and suppose that all simulations use the same global bounds $n$ and $\Delta$. Suppose that the views $H_v$ of the surviving nodes $v$ satisfy the termination and radius, edge validity, and consistency requirements of $\CN(R)$. Consider two surviving nodes $u$ and $v$ that are joined by a path of surviving nodes of length $d \leq R - (T + 1)$. Then, the output of $u$ in the simulation of $\mathcal A$ on $H_v$ equals the output of $u$ in the simulation of $\mathcal A$ on $H_u$.
\end{lemma}
\begin{proof}
By \Cref{lem:cn-view-propagation}, applied to the path from $v$ to $u$, we have $u \in V(H_v)$ and $H_v[B_{H_v}(u,R-d)] = H_u[B_{H_u}(u,R-d)]$. Since $T+1 \le R-d$, the balls of radius $T+1$ around $u$ are contained in the balls of radius $R-d$ around $u$, in both views. Applying \Cref{lem:ball-transfer} with these balls of radius $R-d$ as $X$ and $X'$ gives $H_v[B_{H_v}(u,T+1)] = H_u[B_{H_u}(u,T+1)]$. Finally, by the locality property above, both simulations assign $u$ the same output.
\end{proof}

We now combine these lemmas to prove \Cref{lem:cn-local-simulation}, and afterwards we prove \Cref{thm:cn-to-lcl}.
\begin{proof}[Proof of \Cref{lem:cn-local-simulation}]
We verify that \Cref{alg:cn-simulation}, applied to the given solution to $\CN(R)$, has the claimed properties. By termination and radius, every surviving node $v$ obtains a view $H_v$ that is a labeled subgraph of $G$. Since $\mathcal A$ solves $\Pi$ within $T$ rounds on every subgraph of $G$, its fault-free simulation on $H_v$ produces an output for $v$. This simulation is performed locally and requires no communication.

It remains to argue that every surviving node is locally satisfied. This holds by definition for surviving nodes with a crashed node in their checking neighborhood, so fix a surviving node $v$ with $B_G(v,\ell) \cap \mathcal F = \emptyset$. Since $\mathcal A$ is correct on $H_v$, the simulated outputs satisfy the local constraints of $\Pi$ at $v$ in $H_v$, with the same checking rule and radius $\ell$ as in $G$. It therefore suffices to show that the checking neighborhood $H_v[B_{H_v}(v,\ell)]$, together with the simulated outputs, coincides with the checking neighborhood $G[B_G(v,\ell)]$, together with the actual outputs.

Since $\ell < T+\ell+1 \le R$, \Cref{lem:cn-exact-checking} gives $H_v[B_{H_v}(v,\ell)] = G[B_G(v,\ell)]$ as labeled graphs. Moreover, every $u \in B_G(v,\ell)$ is joined to $v$ by a shortest path of length $d \le \ell \le R-(T+1)$ whose nodes all survive, so by \Cref{lem:cn-matching-outputs}, the output that $u$ computes from $H_u$ equals its output in the simulation on $H_v$. Consequently, $v$ is locally satisfied.
\end{proof}

\begin{proof}[Proof of \Cref{thm:cn-to-lcl}]
Part~(i) follows from \Cref{lem:fault-free-cn}, and part~(ii) from \Cref{lem:cn-local-simulation}. Both results use the initial-knowledge and global-bound conventions of \Cref{sec:preliminaries}.
\end{proof}

\subsection{Limits of Local Simulation}\label{sec:cn-limits}
We conclude the section by discussing two limits of simulating algorithms on consistent neighborhoods as in \Cref{lem:cn-local-simulation}. First, the following lemma shows that the additional hop in the radius requirement $R\geq T+\ell+1$ is necessary in general: views of radius $T+\ell$ need not suffice, even when $T=\ell=1$.

\begin{lemma}[Radius-two views do not suffice]\label{lem:cn-radius-two-insufficiency}
There exists a locally checkable problem with checking radius $\ell=1$ that is solvable in one fault-free round on every graph. Consider a deterministic rule that computes each surviving node's output using only its view and initial local information. For this problem, no such rule guarantees local satisfaction at every surviving node for every family of views satisfying $\CN(2)$.
\end{lemma}
\begin{proof}
We define a locally checkable problem, construct views satisfying $\CN(2)$ in an execution with crashes, and show that every rule that is correct without crashes violates the check at a node with no nearby crash.

For this auxiliary problem, we explicitly treat node identifiers as problem inputs, retaining the initial-knowledge convention of \Cref{sec:initial-knowledge}. Consider the locally checkable problem in which each node $u$ outputs a table indexed by $N_G[u]$, consisting of $u$ and its direct neighbors, where the entry for a node $w$ is its labeled star $\sigma_w$. The radius-one check at $u$ requires that its table be indexed by exactly $N_G[u]$, that its entry for itself is its actual star, and that all tables output by nodes in $N_G[u]$ agree on every entry they share. In particular, this check compares the tables of two distinct neighbors of $u$, even when their common entry describes a node outside $N_G[u]$.

Without crashes, one round solves this problem on any graph: each node sends its star to its neighbors and builds its table from its own star and the received ones. Conversely, in every valid fault-free solution, an entry describing a neighbor $w$ must agree with $w$'s own entry, which is checked against $w$'s actual star. Hence, every entry must be the actual star of the node it names.

We now construct the counterexample. Let $G^+$ be the four-cycle $v$--$a$--$x$--$b$--$v$ with an additional node $y$ joined only to $x$, and let $G^-$ be obtained from $G^+$ by deleting the edge $\{x,y\}$ while retaining $y$ as an isolated node. Suppose that $x$ and $y$ crash in $G^+$, leaving $v,a,b$ as the surviving nodes, and consider the views $H_a=G^+$ and $H_v=H_b=G^+[\{v,a,b,x\}].$

These views satisfy $\CN(2)$. Every included node is within two hops of its view's center, and both surviving edges, $\{v,a\}$ and $\{v,b\}$, appear in their endpoint views. Consistency uses radius one around the endpoints of each surviving edge. For either edge, both comparison graphs are the four-cycle on $\{v,a,b,x\}$, and the node $y$ lies outside the comparison sets. Thus, $a$'s view may include $y$ and $\{x,y\}$ while $b$'s view omits them.

However, no deterministic rule can produce valid outputs from these views. The view $H_a = G^+$ is exactly $a$'s complete radius-two neighborhood in a fault-free execution on $G^+$, and the four-cycle $H_b$ is exactly $b$'s complete radius-two neighborhood in a fault-free execution on $G^-$. The initial neighbor list is $\{v,x\}$ at both nodes, all retained nodes have the same identifiers and input labels, and both graphs can use the same global bounds $n=5$ and $\Delta=3$. Hence, a rule that is correct without crashes must give $a$ and $b$ their respective fault-free outputs: $a$'s entry for $x$ lists the neighbors $\{a,b,y\}$, whereas $b$'s lists $\{a,b\}$. Since $a$ and $b$ are both neighbors of $v$, the check at $v$ compares these entries and rejects, although $B_{G^+}(v,1)=\{v,a,b\}$ contains no crashed node. Thus, $v$ is not locally satisfied.
\end{proof}

Second, \Cref{lem:cn-local-simulation} only guarantees the local constraints at surviving nodes whose radius-$\ell$ checking neighborhood contains no crash. For particular problems, useful correctness conditions may also be achievable at surviving nodes whose checking neighborhoods contain crashes. Such stronger guarantees rely on arguments that use the constraints of the
particular problem, and we give examples of such algorithms in \Cref{sec:sinkless-orientation}.

\section{Algorithms for Consistent Neighborhoods}
\label{sec:cn-algorithms}

We now turn to solving $\CN(R)$ in the presence of crashes, proving all four bounds in \Cref{thm:upper-bound}. Without crashes, $\CN(R)$ can be solved in $R$ rounds by forwarding labeled stars, as shown in \Cref{lem:fault-free-cn}. With crashes, however, the main difficulty is issue~(b) from \Cref{sec:cn-problem}: a node that crashes may deliver messages to some neighbors but not to others, so adjacent surviving nodes may end up with conflicting views.

We address this difficulty by using two standard information-propagation primitives: \textit{full-information gathering}, which collects the local topology (\Cref{sec:full-information-gathering}), and \textit{flooding on fixed subgraphs}, which ensures that, within each surviving component, either all nodes receive a message or none does (\Cref{sec:flooding-fixed-subgraphs}). We then use these primitives to solve $\CN(R)$ in three settings. In \Cref{sec:linear-cn-from-propagation}, every node gathers information for $n-1$ rounds, so that all nodes in a surviving component reconstruct the same graph; this proves part~(i) of \Cref{thm:upper-bound}. In \Cref{sec:parameter-cn}, we restrict the propagation of information to local regions, which yields bounds independent of $n$ and proves parts~(ii) and~(iii), together with the case $\Delta=2$ of part~(iv). In \Cref{sec:linear-radius-cn}, we show that $R$ rounds of gathering suffice on graphs of girth greater than $2R$, which proves the remaining case of part~(iv). Finally, \Cref{sec:cn-algorithms-end} combines these results and presents the proof of \Cref{thm:upper-bound}.

We assume the standard crash-fault model described in \Cref{sec:preliminaries}. Our algorithms execute the primitives in consecutive phases, with round numbers measured from the beginning of each phase. Nodes that crashed in earlier phases remain inactive and count toward the overall crash bound $f$.

\subsection{Full-Information Gathering}
\label{sec:full-information-gathering}

Fix an integer $t\geq0$. In the full-information gathering primitive, each node starts with its labeled star $\sigma_v=(v,N_G[v])$, including its own input label (if any), and sends all stars it knows to all neighbors for $t$ rounds. The pseudocode is presented in \Cref{alg:full-information-gathering}.

\begin{algorithm}[ht]
\caption{Full-Information Gathering at Node $v$}
\label{alg:full-information-gathering}
\begin{algorithmic}[1]
    \Statex \textbf{Input:} labeled star $\sigma_v$ and number of rounds $t$.
    \Statex \textbf{Output:} collection $\mathcal K_v$ of labeled stars.
    \State $\mathcal K_v\gets\{\sigma_v\}$
    \For{$r=1,\ldots,t$}
        \State Send $\mathcal K_v$ to every neighbor in $G$.
        \State Add all stars received in this round to $\mathcal K_v$.
    \EndFor
    \State \Return $\mathcal K_v$.
\end{algorithmic}
\end{algorithm}

To analyze the information collected in \Cref{alg:full-information-gathering}, we denote by $A$ the set of nodes that complete the primitive without crashing: these nodes may crash in later phases. In addition, for every node $v \in A$, we define the set $Q_v:=\{x\in V:\sigma_x\in\mathcal K_v\}$, which records the sources of the collected stars.

The next lemma describes both the extent of $Q_v$ and the paths along which its stars reach $v$.

\begin{lemma}[Full-information gathering]\label{lem:full-information-gathering}
Let $t\geq0$ be an integer, and consider an execution of \Cref{alg:full-information-gathering} for $t$ rounds in the standard crash-fault model, with any number of nodes crashing before or during its execution. For every node $v\in A$, the following properties hold.
\begin{enumerate}
    \item \textbf{Coverage and locality.} $B_{G[A]}(v,t)\subseteq Q_v\subseteq B_G(v,t).$
    \item \textbf{Propagation paths.} If $v$ first learns $\sigma_x$ in round $r$, where $0\leq r\leq t$, then there is a simple path $x=p_0,p_1,\ldots,p_r=v$ such that $p_i$ first receives $\sigma_x$ from $p_{i-1}$ in round $i$ for every $1\leq i\leq r$. Moreover, $v$ collects the star $\sigma_{p_i}$ of every node on this path, so the entire path lies in $G[Q_v]$. In particular, $\dist_{G[Q_v]}(v,x)\leq t$ for every $x\in Q_v$.
\end{enumerate}
The first-reception path in item~2 also exists for every node that learns $\sigma_x$ before crashing, even if it does not complete the primitive. Moreover, every $v\in A$ can reconstruct $G[Q_v]$ from its collected stars. In a fault-free execution, $Q_v=B_G(v,t)$.
\end{lemma}

\begin{proof}
We first prove the coverage and locality bounds. Let $v\in A$. Every node on a path in $G[A]$ does not crash during the $t$ rounds. Thus, for every $x\in B_{G[A]}(v,t)$, the star $\sigma_x$ reaches $v$ along such a path in at most $t$ rounds. Therefore, $B_{G[A]}(v,t)\subseteq Q_v$. In addition, every node that first learns $\sigma_x$ in round $r$ is at distance at most $r\leq t$ from $x$ in $G$, so $Q_v\subseteq B_G(v,t)$.

We now prove the propagation-path property. Let $v$ be any node that first receives $\sigma_x$ in round $r\geq1$, whether or not it completes the primitive. If a node first receives this star in round $j>1$, its sender must have first received it in round $j-1$. Otherwise, the sender would have sent the star earlier, before either node had crashed. In round $1$, the star can only be sent by $x$. Following these senders backwards gives a path $x=p_0,p_1,\ldots,p_r=v$, where $p_i$ first receives $\sigma_x$ in round $i$ for every $1\leq i\leq r$. No node is repeated, since each node first learns the star only once.

For $v\in A$, we next show that every node on this path belongs to $Q_v$. For each $i<r$, node $p_i$ sends its own star together with $\sigma_x$. The star $\sigma_{p_i}$ then follows the remaining path and reaches $v$ by round $r$. Since $v$ also knows its own star, all nodes on the path belong to $Q_v$. Therefore, $\dist_{G[Q_v]}(v,x)\leq r\leq t$. The nodes on this path may crash after forwarding their stars. For $x=v$, the claim holds with $r=0$ and the path consisting only of $v$.

Finally, we show that $v$ can reconstruct $G[Q_v]$. This follows because each collected star gives its source's label and complete neighbor list. In a fault-free execution, $A=V$, and coverage and locality imply that $Q_v=B_G(v,t)$.
\end{proof}

\subsection{Flooding on Fixed Subgraphs} \label{sec:flooding-fixed-subgraphs}

We now describe the second information-propagation primitive, namely flooding on a fixed subgraph. While full-information gathering lets each node learn its surroundings, flooding on a fixed subgraph controls who receives a message: if a message spreads within a fixed subgraph $H$ for sufficiently many rounds, then, in each surviving component of $H$, either all nodes receive it or none does.

Fix a subgraph $H\subseteq G$ and a set $X\subseteq V(H)$ of nodes that initially hold a message $m$. Each node $v\in V(H)$ stores a value $m_v$, initially equal to $m$ if $v\in X$ and to $\bot$ otherwise, where $\bot$ denotes the absence of a message. We say that a node is \emph{informed} once $m_v=m$. In each round, every node that has not yet crashed and holds $m$ at the start of the round sends it to all of its neighbors in $H$. A node that first receives $m$ in round~$r$ stores it and starts forwarding it in round~$r+1$ if it has not crashed. Nodes outside $X$ learn $m$ only through these transmissions. A finite flooding phase of $T$ rounds consists of the first $T$ rounds of this process.
The pseudocode is given in \Cref{alg:flooding-fixed-subgraphs}.

\begin{algorithm}[ht]
\caption{Flooding a Message at Node $v\in V(H)$}
\label{alg:flooding-fixed-subgraphs}
\begin{algorithmic}[1]
    \Statex \textbf{Input:} neighbor set $N_H(v)$ and initial value $m_v\in\{m,\bot\}$.
    \For{round $r=1,2,\ldots$}
        \If{$m_v\neq\bot$}
            \State Send $m_v$ to every node in $N_H(v)$.
        \EndIf
        \If{at least one copy of $m$ is received in round $r$}
            \State $m_v\gets m$.
        \EndIf
    \EndFor
\end{algorithmic}
\end{algorithm}

For the next two lemmas, we consider an execution in which flooding continues indefinitely. Let $\mathcal S$ be the set of nodes that survive the entire execution. We call each connected component of $H[V(H)\cap\mathcal S]$ a \emph{surviving component}. Since no node of such a component crashes, if one of them learns $m$, all of them eventually learn $m$. The next lemma bounds the number of rounds after which each surviving component is either entirely informed or entirely uninformed, using the lengths of simple paths in $H$ starting at a node of $X$.

\begin{lemma}[Flooding and simple paths]\label{lem:flooding-simple-paths}
Let $L\geq0$ be an integer such that every simple path in $H$ starting at a node of $X$ has length at most $L$. Then no node becomes informed about $m$ after round~$L$. Moreover, each surviving component is either entirely informed about $m$ by the end of round~$L$ or remains entirely uninformed throughout the execution. In particular, if $H$ is nonempty, flooding for $|V(H)|-1$ rounds guarantees that all nodes of each surviving component agree on whether they hold $m$.
\end{lemma}

\begin{proof}
We first prove that no node becomes informed about $m$ after round~$L$. Every node in $X$ is informed in round~$0$. Suppose $v\notin X$ first learns $m$ in round~$r>0$. By the same argument used in the proof of \Cref{lem:full-information-gathering}, its sender first learned $m$ in round~$r-1$. Otherwise, the sender would already have sent $m$ to $v$ in round~$r-1$. Tracing first receptions backwards therefore gives a path of length $r$ in $H$ starting in $X$. The rounds in which its nodes first become informed strictly increase along the path, so it is simple and $r\leq L$.

We next prove the claim for surviving components. Let $C$ be such a component. If no node of $C$ ever holds $m$, then $C$ remains entirely uninformed. Otherwise, flooding along paths in $C$, whose nodes never crash, eventually informs every node of $C$. The preceding bound shows that all of them are informed by the end of round~$L$.

Every simple path in a nonempty graph $H$ has length at most $|V(H)|-1$. Taking $L=|V(H)|-1$ gives the claimed agreement within each surviving component.
\end{proof}

\Cref{lem:flooding-simple-paths} guarantees agreement within every surviving component after $|V(H)|-1$ rounds when $H$ is nonempty. This bound can be large, but when a bound on the number of crashes is known, we obtain another bound. We next prove a bound for the same flooding process in terms of the maximum degree of $H$, the diameters of its connected components, and the number of crashes. These diameters are measured in $H$, before removing any crashed nodes.

A small diameter $D$ before crashes does not by itself bound the flooding time by $D$, since crashes may remove the paths witnessing that diameter. The proof below uses the at most $\Delta \cdot f$ surviving neighbors of crashed nodes to bound the resulting delay.

\begin{lemma}[Flooding with few crashes]\label{lem:flooding-few-crashes}
Let $f\geq1$. Suppose $H$ has maximum degree at most $\Delta$, every connected component of $H$ has diameter at most $D\geq1$, and at most $f$ nodes crash before or during the execution of \Cref{alg:flooding-fixed-subgraphs}. Every surviving node $v\notin X$ that ever learns $m$ does so within $O(D\cdot\Delta\cdot f)$ rounds. Within the same bound, each surviving component is either entirely informed about $m$ or remains entirely uninformed throughout the execution.
\end{lemma}

\begin{proof}
Fix a surviving node $v\notin X$ that first learns $m$ in round~$r>0$. By the proof of \Cref{lem:flooding-simple-paths}, there is a simple path $P=(p_0,\ldots,p_r=v)$ in $H$, where $p_0\in X$ and $p_i$ first holds $m$ in round~$i$ for $0\leq i\leq r$. Let $C$ be the connected component of $H$ containing $P$. If no node of $C$ crashes, flooding along a shortest path from $p_0$ to $v$ gives $r\leq D$.

We now consider the case where $C$ contains a crashed node. Let $B$ be the set of surviving nodes of $C$ adjacent to a crashed node. Since at most $f$ nodes crash and each has degree at most $\Delta$, we have $|B|\leq\Delta f$. For each surviving node $x$ of $C$, choose a shortest path in $C$ from $x$ to a crashed node; its length is at most $D$. Let $b$ be the node immediately before the first crashed node on this path. Then $b\in B$, and the path from $x$ to $b$ contains only surviving nodes and has length at most $D$. Assign $x$ to one such $b$.

We next bound the number of nodes of $P$ assigned to each $b\in B$. If $p_i$ and $p_j$, with $i<j$, are assigned to the same $b$, their assigned paths give a surviving walk of length at most $2D$ between them. Once $p_i$ holds $m$ in round~$i$, flooding along this walk informs $p_j$ by round~$i+2D$. Thus $j-i\leq2D$. For any fixed $b$, the indices assigned to $b$ therefore lie in an interval of length at most $2D$, which contains at most $2D+1$ integers. Since $P$ is simple, it also contains at most $f$ crashed nodes. Consequently, \[r+1\leq(2D+1)|B|+f\leq(2D+1)\Delta f+f.\] Together with the case where no node of $C$ crashes, and using $f\geq1$, this gives $r=O(D\cdot\Delta\cdot f)$.

Finally, we prove the claim for surviving components. If no node of a surviving component ever holds $m$, then that component remains entirely uninformed. Otherwise, flooding along paths in the component eventually informs all of its nodes, since none of them crashes. Nodes in $X$ are informed initially, and every other node is informed within the bound just proved. Hence the entire component is informed within $O(D\cdot\Delta\cdot f)$ rounds.
\end{proof}

\subsection{Consistent Neighborhoods in \texorpdfstring{$O(n)$}{O(n)} Rounds}
\label{sec:linear-cn-from-propagation}

We now prove part~(i) of \Cref{thm:upper-bound}: we describe an $O(n)$-round algorithm for $\CN(R)$ for every integer $R\geq1$. Each node runs full-information gathering (\Cref{alg:full-information-gathering}) for $n-1$ rounds, reconstructs the labeled graph induced by the sources of its collected stars, and outputs its radius-$R$ neighborhood in that graph. The algorithm uses only the initial labeled star and the known parameters $n$ and $R$, and does not require a bound on the number of crashes. The pseudocode is given in \Cref{alg:linear-cn-from-propagation}.

\begin{algorithm}[ht]
\caption{Consistent Neighborhoods in $n-1$ Rounds at Node $v$}
\label{alg:linear-cn-from-propagation}
\begin{algorithmic}[1]
    \Statex \textbf{Input:} labeled star $\sigma_v$, upper bound $n\geq |V|$, and radius $R\geq1$.
    \Statex \textbf{Output:} radius-$R$ view $H_v$ centered at $v$.
    \State Run \Cref{alg:full-information-gathering} for $n-1$ rounds to obtain $\mathcal K_v$.
    \State $Q_v\gets\{x:\sigma_x\in\mathcal K_v\}$.
    \State Construct the labeled graph $J_v\gets G[Q_v]$ from $\mathcal K_v$.
    \State $H_v\gets J_v[B_{J_v}(v,R)]$.
    \State \Return $H_v$.
\end{algorithmic}
\end{algorithm}

Although step~3 of \Cref{alg:linear-cn-from-propagation} is phrased in terms of $G$, node $v$ can carry it out locally: for each $x \in Q_v$, the collected star $\sigma_x$ gives the label of $x$ and its complete neighbor list in $G$. Hence $v$ knows all edges of $G$ between nodes of $Q_v$, and thus $J_v = G[Q_v]$. The computation of $H_v$ is then local as well.

The main property of \Cref{alg:linear-cn-from-propagation} is that all nodes in the same connected component of the surviving graph reconstruct the same graph, which contains their entire component. Intuitively, gathering (\Cref{alg:full-information-gathering}) floods every star on $G$ for $n-1$ rounds, which is long enough for each star to reach either all or none of the nodes of a surviving component. Consequently, adjacent surviving nodes compute their views in the same graph, which makes consistency straightforward. We make this argument formal in the lemma below, which states the guarantees of \Cref{alg:linear-cn-from-propagation}.

\begin{lemma}[Linear-round consistent neighborhoods]
\label{lem:linear-cn-from-propagation}
Let $G=(V,E)$ be a graph with at most $n$ nodes, and let $R\geq1$ be an integer. \Cref{alg:linear-cn-from-propagation} deterministically solves $\CN(R)$ in $n-1$ rounds in the synchronous \local model under standard crash faults, with any number of nodes crashing before or during its execution. Each node uses only its initial labeled star and the globally known parameters $n$ and $R$; no bound on the number of crashes is required.
\end{lemma}

\begin{proof}
Let $A$ be the set of nodes that complete \Cref{alg:full-information-gathering} with $t = n - 1$, and fix a connected component $C$ of $G[A]$. For the analysis, extend gathering indefinitely without further crashes; then $A$ is exactly the set of surviving nodes in this extension. For each initial star $\sigma_x$, apply \Cref{lem:flooding-simple-paths} with $H=G$, $X=\{x\}$, and $L=n-1$. Either every node of $C$ knows that star by the end of round $n-1$, or none does. Thus all nodes of $C$ finish with the same collection $\mathcal K$, source set $Q$, and reconstructed graph $J=G[Q]$. By \Cref{lem:full-information-gathering}, $C\subseteq Q$, so $G[C]\subseteq J\subseteq G$.

For $v\in C$, the output $H_v=J[B_J(v,R)]$ contains its root and all incident surviving edges. As $J\subseteq G$, it is a radius-$R$ view in $G$. For adjacent nodes $u,v\in C$, the edge $\{u,v\}$ belongs to $J$, so every path in $J$ of length at most $R-1$ from either endpoint lies in both radius-$R$ balls. Consequently, their comparison sets satisfy:
\[
    S_u(v)=S_v(u)=B_J(u,R-1)\cup B_J(v,R-1).
\]
Both outputs induce $J$ on this set, proving consistency.

If no node crashes, \Cref{lem:full-information-gathering} gives $Q_v=B_G(v,n-1)$. Since $G$ has at most $n$ nodes, this ball contains exactly the nodes of the connected component of $G$ containing $v$. Hence $J_v$ is that component's complete labeled graph, and each output is $G[B_G(v,R)]$. Finally, gathering runs for $n-1$ rounds and is followed only by local computation, establishing termination and the claimed round complexity.
\end{proof}

\begin{remark}[Relation to local consensus]
Our $(n-1)$-round algorithm uses the same flooding principle as the elementary local-consensus algorithm of Fraigniaud, Nguyen, and Paz~\cite[full version, Lemma~24]{STACS:FrNgPa25}. Local consensus requires the nodes in each surviving component to agree on an initial proposal. Their algorithm floods identifier--proposal pairs for $n-1$ rounds, so nodes in the same surviving component obtain the same collection and select the same proposal. Our algorithm floods labeled stars and uses the common collection to reconstruct a graph, from which each node extracts its radius-$R$ view.
\end{remark}

\subsection{Consistent Neighborhoods with Parameter-Dependent Bounds}
\label{sec:parameter-cn}

We now prove parts~(ii) and~(iii) of \Cref{thm:upper-bound}, together with the $\Delta=2$ case of part~(iv).
The linear-round algorithm behind part~(i) makes all nodes in each surviving component reconstruct the same graph. To obtain bounds independent of $n$, we instead restrict the propagation of each source's information to a local region, while preserving the agreement required between adjacent output views.

For every $R\geq1$, our algorithm has two phases. In Phase~I (full-information gathering), each node runs \Cref{alg:full-information-gathering} for $R$ rounds. As in \Cref{sec:full-information-gathering}, let $A$ denote the nodes that complete this phase. Each $v\in A$ obtains a collection of stars $\mathcal K_v$ with source set $Q_v$. These collections and source sets remain fixed throughout Phase~II.

In Phase~II (restricted flooding), each node $v\in A$ maintains a set of tokens $W_v$, initially $W_v=\{v\}$. In each of $T$ rounds, it sends $W_v$ to every neighbor and retains a received token $x$ only if $x\in Q_v$, meaning that it collected $\sigma_x$ in Phase~I. Stored tokens are kept and forwarded in subsequent rounds. We choose $T$ later using \Cref{lem:flooding-simple-paths,lem:flooding-few-crashes}.

Even if $v$ collected $\sigma_x$ in Phase~I, it does not start Phase~II with token $x$ unless $x=v$. It must receive that token during Phase~II before adding it to $W_v$. Since only nodes that complete Phase~I start with their own tokens, every identifier in $W_v$ belongs to $A$.

At the end of Phase~II, node $v$ outputs $H_v:=G[W_v]$. Since $W_v\subseteq Q_v$, the stars collected in Phase~I give the label and complete neighbor list of every node in $W_v$, so $v$ can reconstruct this labeled induced subgraph locally. Phase~II therefore needs to transmit only identifiers. The pseudocode is given in \Cref{alg:parameter-cn}.

\begin{algorithm}[H]
\caption{Consistent Neighborhoods with Restricted Propagation at Node $v$}
\label{alg:parameter-cn}
\begin{algorithmic}[1]
    \Statex \textbf{Input:} labeled star $\sigma_v$, radius $R\geq1$, and number of propagation rounds $T\geq1$.
    \Statex \textbf{Output:} radius-$R$ view $H_v$ centered at $v$.
    \State \textbf{Phase I: full-information gathering}
    \State Run \Cref{alg:full-information-gathering} for $R$ rounds to obtain $\mathcal K_v$.
    \State $Q_v\gets\{x:\sigma_x\in\mathcal K_v\}$; $W_v\gets\{v\}$.
    \State \textbf{Phase II: restricted propagation}
    \For{$r=1,\ldots,T$}
        \State Send $W_v$ to every neighbor in $G$.
        \State Add to $W_v$ all identifiers received in this round that belong to $Q_v$.
    \EndFor
    \State Construct the labeled graph $H_v\gets G[W_v]$ from the stars in $\mathcal K_v$.
    \State \Return $H_v$.
\end{algorithmic}
\end{algorithm}

\paragraph{Propagation graphs.}
We now turn to the analysis of Phase~II. Recall that $A$ is the set of nodes that complete Phase~I. Only these nodes start Phase~II with their own tokens, so $W_v\subseteq A$ for every node that completes Phase~II. For each $x\in A$, let $V_x$ consist of $x$ and all nodes that receive $\sigma_x$ during Phase~I. Note that $V_x$ may include nodes that crash before Phase~I ends or during Phase~II.

By the coverage and locality properties proven in \Cref{lem:full-information-gathering}, we obtain
\begin{equation}\label{eq:propagation-coverage}
    B_{G[A]}(x,R)\subseteq V_x\subseteq B_G(x,R).
\end{equation}
The right inclusion implies that every node that stores token $x$ is within distance $R$ of $x$, since it stores the token only if it collected $\sigma_x$ in Phase~I. The left inclusion will be used to establish the consistency requirement of \Cref{def:cn}. More precisely, if an edge of $G[A]$ has an endpoint within distance $R-1$ of $x$ in $G[A]$, then both endpoints belong to $V_x$.

We now define the fixed subgraph along which token $x$ spreads.\footnote{The use of propagation graphs in these upper bounds was suggested by LLMs; see \Cref{sec:ai-methodology}.} For each $x\in A$, the \emph{propagation graph} $\Gamma_x=(V_x,E_x)$ has node set $V_x$ as defined above and edge set $E_x:=\{\{a,b\}\in E:a,b\in V_x\}$. Equivalently, $\Gamma_x=G[V_x]$. This graph remains unchanged throughout Phase~II; nodes are retained even if they crash before or during this phase.

For every $v\in A$, we have $v\in V_x$ if and only if $x\in Q_v$, so a node can determine locally whether to retain token $x$. Thus every copy of token $x$ received along an edge of $\Gamma_x$ is stored, while copies received outside $V_x$ are discarded.  Token $x$ therefore spreads by flooding on $\Gamma_x$, with $x$ as its only initial holder. We will show that, if Phase~II runs for sufficiently many rounds, the surviving endpoints of every edge of $\Gamma_x$ either both store token $x$ or neither does.

The next lemma shows that token $x$ spreads in Phase~II exactly as it would under flooding on $\Gamma_x$. It also bounds the size and diameter of each propagation graph.

\begin{lemma}[Flooding on propagation graphs]\label{lem:cn-propagation-graphs}
Fix an execution of \Cref{alg:parameter-cn} with $R\geq1$, and let $x\in A$. The following properties hold.
\begin{enumerate}[(i)]
    \item Consider \Cref{alg:flooding-fixed-subgraphs} on $\Gamma_x$ for $T$ rounds, with token $x$ initially held only by $x$. Nodes that crashed before Phase~II remain inactive, and every other node has the same crash round, if any, as in Phase~II. In a crash round, a flooding message is delivered to precisely those neighbors in $\Gamma_x$ that receive the sender's Phase~II transmission. At the end of each round, every node in $V_x$ that has not crashed stores token $x$ in Phase~II if and only if it is informed in this flooding execution. Only nodes in $V_x$ can store token $x$ during Phase~II.
    \item The graph $\Gamma_x$, including nodes that crash before or during Phase~II, is a connected subgraph of $G$, has maximum degree at most $\Delta$, and satisfies $\diam(\Gamma_x)\leq2R$. Moreover, $V_x\subseteq B_G(x,R)$ and $|V_x|\leq1+\Delta\sum_{i=0}^{R-1}(\Delta-1)^i$ for $\Delta\geq3$, while for $\Delta=2$, we have $|V_x|\leq2R+1$.
\end{enumerate}
\end{lemma}

\begin{proof}

Initially only $x$ holds token $x$, and $x\in V_x$. During Phase~II, every node $v$ that has not yet crashed stores a received copy of token $x$ exactly when $x\in Q_v$, or equivalently, when $v\in V_x$. Every subsequent holder is therefore in $V_x$. Each node holding the token at the start of a round sends it to every neighbor in $\Gamma_x$, and every copy received along these edges is stored. Copies received at nodes $u \notin V_x$ are discarded since $x \notin Q_u$. The initial holders, forwarding rule, crash rounds, and delivery choices along $\Gamma_x$ thus agree with the flooding execution in~(i). Induction on the rounds of Phase~II gives the same holders among nodes that have not crashed by the end of each round. Nodes that crashed before Phase~II remain inactive and count toward any crash bound. This concludes the proof of~(i).

We next prove~(ii). Since $\Gamma_x\subseteq G$, its maximum degree is at most $\Delta$. By \eqref{eq:propagation-coverage}, $V_x\subseteq B_G(x,R)$. Counting nodes by distance from $x$ gives $|V_x|\leq1+\Delta\sum_{i=0}^{R-1}(\Delta-1)^i$ for $\Delta\geq3$, while for $\Delta=2$ the same count gives $|V_x|\leq2R+1$.

For every $w\in V_x$, \Cref{lem:full-information-gathering} gives a first-reception path of length at most $R$ from $x$ to $w$, even if $w$ crashes before Phase~I ends. Every node on this path held $\sigma_x$ during Phase~I and hence belongs to $V_x$. Since $\Gamma_x$ is induced, the entire path lies in $\Gamma_x$. Thus $\dist_{\Gamma_x}(x,w)\leq R$ for every $w\in V_x$, so $\Gamma_x$ is connected and $\diam(\Gamma_x)\leq2R$.
\end{proof}

For $R\geq1$, we set the number $T$ of propagation rounds for parts~(ii),~(iii), and the $\Delta=2$ case of part~(iv) of \Cref{thm:upper-bound} to the following values, respectively:
\begin{equation}\label{eq:parameter-cn-rounds}
T=\begin{cases}
\Delta\displaystyle\sum_{i=0}^{R-1}(\Delta-1)^i & \Delta\geq3,\ \text{any number of crashes},\\
c\cdot\Delta\cdot R\cdot f & \text{a globally known upper bound }f\geq1\text{ on crashes},\\
2R & \Delta=2,\ \text{any number of crashes}.
\end{cases}
\end{equation}
 Here $c$ is a sufficiently large positive integer constant. When a crash bound $f$ is assumed known, it bounds the total number of crashes, including those before Phase~II. The following lemma shows that, for every $x\in A$, these choices suffice for all nodes in each surviving component of $\Gamma_x$ to agree on whether they store token $x$ at the end of Phase~II.

\begin{lemma}[Agreement on stored tokens]\label{lem:parameter-cn-agreement}
Let $R\geq1$, choose $T$ as in \eqref{eq:parameter-cn-rounds}, and let $\mathcal S\subseteq A$ be the set of nodes that complete \Cref{alg:parameter-cn}. For every $x\in A$ and every connected component of $\Gamma_x[\mathcal S\cap V_x]$, either all its nodes store token $x$ at the end of Phase~II, or none of them does. In particular, the endpoints of every surviving edge of $\Gamma_x$ agree on whether they store token $x$.
\end{lemma}

\begin{proof}
Fix $x\in A$. By \Cref{lem:cn-propagation-graphs}(i), the nodes storing token $x$ in Phase~II are exactly those informed by flooding on $\Gamma_x$, with $x$ as the only initial holder and the same crashes and message deliveries. For the analysis, continue this flooding execution beyond round $T$ without further crashes. This leaves the first $T$ rounds unchanged, and its surviving nodes are precisely $\mathcal S\cap V_x$.

For the two choices of $T$ that do not require a crash bound, \Cref{lem:cn-propagation-graphs}(ii) gives $|V_x|-1\leq T$. By \Cref{lem:flooding-simple-paths}, for each connected component of $\Gamma_x[\mathcal S\cap V_x]$, either all its nodes store token $x$ at the end of round~$T$, or none of them does.

For the choice of $T$ using the known crash bound $f$, \Cref{lem:cn-propagation-graphs}(ii) shows that $\Gamma_x$ is connected, has maximum degree at most $\Delta$, and has diameter at most $2R$. The upper bound $f$ also counts nodes that crashed before Phase~II. Applying \Cref{lem:flooding-few-crashes} with $D=2R$ shows that, for each connected component of $\Gamma_x[\mathcal S\cap V_x]$, either all its nodes store token $x$ within $O(\Delta\cdot R\cdot f)$ rounds, or none of them ever holds it. Choosing $c$ sufficiently large makes this number of rounds at most $T$. In every case, the endpoints of a surviving edge of $\Gamma_x$ lie in the same connected component of $\Gamma_x[\mathcal S\cap V_x]$ and therefore agree on whether they store token $x$.
\end{proof}

We next verify the radius and edge-validity requirements of \Cref{def:cn} for the graphs reconstructed from the stored tokens and collected stars.

\begin{lemma}[Radius and edge validity]\label{lem:parameter-cn-validity}
Let $R\geq1$, and consider an execution of \Cref{alg:parameter-cn}. If a node $v$ outputs $H_v$, then $H_v$ is a labeled induced subgraph of $G$ and a radius-$R$ view centered at $v$. If two adjacent nodes $u$ and $v$ both output, then both $H_u$ and $H_v$ contain the edge $\{u,v\}$.
\end{lemma}

\begin{proof}
The output is $H_v=G[W_v]$, and $v\in W_v$ because $v$ stores its own token. Since $W_v\subseteq Q_v$, the labels, including the root's identifier, are inherited from the stars collected in Phase~I. For every $x\in W_v$, \Cref{lem:cn-propagation-graphs} gives $v\in V_x\subseteq B_G(x,R)$, so $x\in B_G(v,R)$. Thus $v\in V(H_v)\subseteq B_G(v,R)$, as required for a radius-$R$ view centered at $v$.

Let $u$ and $v$ be adjacent nodes that both output. In the first round of Phase~I, they exchange their own stars, so $u\in Q_v$ and $v\in Q_u$. Each endpoint therefore stores the other endpoint's token in the first round of Phase~II. Both output views contain $u$ and $v$. Since they are induced subgraphs of $G$, they also contain the edge $\{u,v\}$.
\end{proof}

We now use \Cref{lem:parameter-cn-agreement} to prove that the output views agree on the comparison subgraphs around every surviving edge.

\begin{lemma}[Consistency]\label{lem:parameter-cn-consistency}
Let $R\geq1$, choose $T$ as in \eqref{eq:parameter-cn-rounds}, and consider an execution of \Cref{alg:parameter-cn}. If two adjacent nodes $u$ and $v$ both output, then $H_u[S_u(v)]=H_v[S_v(u)]$ as labeled graphs, where $S_u(v)$ and $S_v(u)$ are the comparison sets from \Cref{def:comparison-sets}, computed in $H_u$ and $H_v$, respectively.
\end{lemma}

\begin{proof}
Fix adjacent nodes $u$ and $v$ that both output. By \Cref{lem:parameter-cn-validity}, this edge belongs to both $H_u$ and $H_v$. Let $P$ be a path in $H_u$ of length at most $R-1$ starting at an endpoint $a\in\{u,v\}$. Since $W_u\subseteq A$, we have $H_u\subseteq G[A]$. Thus every node $x$ of $P$ satisfies $\dist_{G[A]}(x,a)\leq R-1$. The other endpoint of $\{u,v\}$ also belongs to $A$ and is at distance at most $R$ from $x$ in $G[A]$. By \eqref{eq:propagation-coverage}, both $u$ and $v$ belong to $V_x$, so their edge belongs to the induced graph $\Gamma_x$. Since $u$ stores token $x$, \Cref{lem:parameter-cn-agreement} implies that $v$ also stores token $x$. Hence every node of $P$ belongs to $W_v$. Since $H_v=G[W_v]$, the entire path $P$ occurs in $H_v$.

Every path defining membership in $S_u(v)$ therefore also occurs in $H_v$, which proves $S_u(v)\subseteq S_v(u)$. Interchanging $u$ and $v$ gives equality. Both comparison graphs are the labeled subgraph of $G$ induced by this common set, so $H_u[S_u(v)]=H_v[S_v(u)]$.
\end{proof}

When no node crashes, the exact gathering sets from \Cref{lem:full-information-gathering} yield fault-free completeness.

\begin{lemma}[Fault-free completeness]\label{lem:parameter-cn-completeness}
For $R\geq1$ and $T$ chosen as in \eqref{eq:parameter-cn-rounds}, if no node crashes in \Cref{alg:parameter-cn}, then every node $v$ outputs $H_v=G[B_G(v,R)]$.
\end{lemma}

\begin{proof}
If no node crashes, \Cref{lem:full-information-gathering} gives $Q_v=B_G(v,R)$ for every $v\in V$. Hence $V_x=B_G(x,R)$ for every $x\in V$. By \Cref{lem:cn-propagation-graphs}(ii), $\Gamma_x$ is connected. Since $x$ stores its own token, \Cref{lem:parameter-cn-agreement} implies that every node of $\Gamma_x$ stores token $x$ by the end of Phase~II. By \Cref{lem:cn-propagation-graphs}(i), no node outside $V_x$ can store token $x$. Thus $v$ stores token $x$ exactly when $v\in B_G(x,R)$, equivalently when $x\in B_G(v,R)$. Therefore $W_v=B_G(v,R)$ and $H_v=G[B_G(v,R)]$.
\end{proof}

We finish checking the requirements of \Cref{def:cn} by verifying termination.

\begin{lemma}[Termination]\label{lem:parameter-cn-termination}
For every $R\geq1$ and $T \ge 1$, every surviving node in \Cref{alg:parameter-cn} produces its output at the end of round~$R+T$.
\end{lemma}

\begin{proof}
Phase~I executes \Cref{alg:full-information-gathering} for $R$ rounds, and Phase~II consists of exactly $T$ rounds. Constructing $H_v$ from the stored tokens and collected stars requires only local computation. Every surviving node therefore outputs $H_v$ at the end of round~$R+T$.
\end{proof}

Substituting the three choices of $T$ from \eqref{eq:parameter-cn-rounds} into \Cref{lem:parameter-cn-termination} gives the following round complexities.

\begin{lemma}[Round complexity]\label{lem:parameter-cn-round-complexity}
For $R\geq1$ and $T$ chosen as in \eqref{eq:parameter-cn-rounds}, the round complexity of \Cref{alg:parameter-cn} is $O((\Delta-1)^R)$, $O(\Delta\cdot R\cdot f)$, or $3R=O(R)$ for the three choices of $T$, respectively.
\end{lemma}

\begin{proof}
By \Cref{lem:parameter-cn-termination}, the algorithm uses $R+T$ rounds. For the choice with $\Delta\geq3$ and no assumed crash bound, we have $T=\Delta((\Delta-1)^R-1)/(\Delta-2)\leq3(\Delta-1)^R$. Since $R\leq2^{R-1}\leq(\Delta-1)^R$, the algorithm uses $O((\Delta-1)^R)$ rounds in total.

For the choice using the known crash bound $f$, $R+T=R+c\cdot\Delta\cdot R\cdot f=O(\Delta\cdot R\cdot f)$. For the choice with $\Delta=2$ and no assumed crash bound, $R+T=3R=O(R)$.
\end{proof}

\subsection{Consistent Neighborhoods on Graphs of Large Girth}
\label{sec:linear-radius-cn}

We now prove the large-girth case of \Cref{thm:upper-bound}(iv). Here a simple gathering algorithm suffices.

On graphs of girth greater than $2R$, run \Cref{alg:full-information-gathering} for $R$ rounds. Using $Q_v$ as defined in \Cref{sec:full-information-gathering}, each surviving node $v$ reconstructs and outputs $H_v:=G[Q_v]$. This is well-defined because $v$ knows the labeled star of every included node.

\begin{lemma}[Consistent neighborhoods on graphs of large girth]
\label{lem:linear-radius-cn}
Let $R\geq1$ be an integer, and let $G$ have girth greater than $2R$. Run \Cref{alg:full-information-gathering} for $R$ rounds, and let every surviving node $v$ output $H_v=G[Q_v]$, where $Q_v$ is the set of sources of its collected stars. This algorithm solves $\CN(R)$ deterministically in $R$ rounds in the synchronous \local model under standard crash faults, for any number of crashes.
\end{lemma}

\begin{proof}
The algorithm runs for $R$ rounds and then reconstructs $H_v$ locally. By \Cref{lem:full-information-gathering} with $t=R$, we have $v\in Q_v\subseteq B_G(v,R)$, so each output is a labeled radius-$R$ view centered at $v$. Adjacent surviving nodes exchange their own stars in the first round, so both views contain their common edge. Without crashes, $Q_v=B_G(v,R)$ and hence $H_v=G[B_G(v,R)]$.

It remains to prove consistency. The girth condition implies that there is at most one simple path of length at most $R$ between any two nodes: two distinct such paths would contain a cycle of length at most $2R$.

Fix a surviving edge $\{u,v\}$ and $x\in S_u(v)$. Choose a simple path $P$ in $H_u$ from $x$ to one of $u,v$, of length at most $R-1$. Since $x\in Q_u$, \Cref{lem:full-information-gathering} gives a simple first-reception path of length at most $R$ along which $\sigma_x$ reached $u$ in successive rounds. If $u$ lies on $P$, then $\sigma_x$ reaches $u$ along the $x$--$u$ subpath of $P$, by uniqueness. Thus $u$ knows $\sigma_x$ by round~$R-1$ and sends it to $v$ by round~$R$. Otherwise, $P$ ends at $v$, and appending $\{u,v\}$ gives a simple $x$--$u$ path of length at most $R$. By uniqueness, this is the first-reception path, so $v$ knows $\sigma_x$ by round~$R-1$. In both cases, $x\in Q_v$.

The preceding argument shows that $S_u(v) \subseteq Q_v$. Thus every node of $P$ belongs to $Q_v$. Since \(H_v=G[Q_v]\), the same path occurs in \(H_v\), giving \(x\in S_v(u)\). Thus \(S_u(v)\subseteq S_v(u)\), and symmetry gives equality. Both views induce the same labeled subgraph of \(G\) on this set, proving consistency.
\end{proof}

\begin{remark}[The strict girth condition]
For $R\geq2$, the $R$-round gathering algorithm of \Cref{lem:linear-radius-cn} can fail on a graph of girth $2R$, even when $\Delta=2$. The following example establishes this limitation of the gathering algorithm; it does not establish a lower bound for arbitrary algorithms for $\CN(R)$. Let $G$ be the cycle with nodes $p_0,\ldots,p_{2R-1}$ in cyclic order. Let $x=p_0$ crash in round~$1$, sending $\sigma_x$ only to $p_1$, while every other node survives. The star $\sigma_x$ first reaches $u=p_R$ in round~$R$, whereas $v=p_{R+1}$ has not received it by that round. All other stars reach both nodes by round~$R$, so the outputs are $H_u=G$ and $H_v=G[V\setminus\{x\}]$. Since $\dist_{H_u}(v,x)=R-1$ and $x\notin V(H_v)$, we have $x\in S_u(v)\setminus S_v(u)$, violating the consistency requirement of \Cref{def:cn} across the surviving edge $\{u,v\}$.
\end{remark}

\Needspace{8\baselineskip}
\subsection{Putting It All Together} \label{sec:cn-algorithms-end}
We now combine the results of this section to prove \Cref{thm:upper-bound}.

\begin{proof}[Proof of \Cref{thm:upper-bound}]
Part~(i) follows from \Cref{lem:linear-cn-from-propagation}, which establishes that \Cref{alg:linear-cn-from-propagation} solves $\CN(R)$ in $n-1=O(n)$ rounds for every $R\geq1$ and any number of crashes.

For every $R\geq1$, choose $T$ from \eqref{eq:parameter-cn-rounds} under the degree and crash assumptions of the chosen case. By \Cref{lem:parameter-cn-validity,lem:parameter-cn-consistency,lem:parameter-cn-completeness,lem:parameter-cn-termination}, \Cref{alg:parameter-cn} satisfies all requirements of \Cref{def:cn}. Applying \Cref{lem:parameter-cn-round-complexity} for the three choices of $T$ in \eqref{eq:parameter-cn-rounds} gives part~(ii), part~(iii), and the $\Delta=2$ case of part~(iv), respectively.

The large-girth case of part~(iv) follows from \Cref{lem:linear-radius-cn}, which solves $\CN(R)$ in $R$ rounds for every $R\geq1$ on graphs of girth greater than $2R$, under any number of crashes. Forests have infinite girth and are therefore included.
\end{proof}

\section{Lower Bounds for Consistent Neighborhoods}
\label{sec:lower-bound}

In this section, we prove the lower bounds of \Cref{thm:lower-bound}.
All lower bounds use one common construction principle, which builds on the path-versus-cycle argument behind the fault-free lower bound in \Cref{lem:fault-free-cn}. We describe this principle in \Cref{sec:lb-framework}. We then instantiate it to obtain the linear lower bound with one or two crashes in \Cref{sec:lb-linear}, the exponential lower bound in the radius in \Cref{sec:lb-degree}, and the lower bound depending on the crash bound and the radius in \Cref{sec:lb-fault-budget}. Finally, we combine these lower bounds into a proof of \Cref{thm:lower-bound} in \Cref{sec:lb-together}.

\subsection{Common Indistinguishability Framework}
\label{sec:lb-framework}

We assume that a deterministic algorithm $\mathcal{A}$ solves $\CN(R)$ within $T$ rounds on two graphs $G^-$ and $G^+$, which we define below, with the same globally known bounds $n$, $\Delta$, and $f$. The structure of the two graphs will enable us to reach a contradiction. We may assume that every surviving node outputs at the end of round~$T$: a node that computes its output earlier can retain it and report it at the end of round~$T$, while preserving its original communication behavior.

\paragraph{Common structure of $G^-$ and $G^+$.}
Each of our lower bounds uses its own pair of graphs $G^-$ and $G^+$. As in the lower-bound proof of \Cref{lem:fault-free-cn}, the two graphs differ only in the edge joining the endpoints of a path: this edge is present in $G^+$ and absent in $G^-$.

Both graphs have the same node set $V$, partitioned into three sets $C$, $F$, and $S$, with no edges between $C$ and $S$, as illustrated in \Cref{fig:lb-framework}. More generally, we say that a labeled graph \emph{respects} a partition $(C,F,S)$ of its node set if it has no edges between $C$ and $S$; in particular, $G^+$ and $G^-$ respect $(C,F,S)$. We formulate some of the definitions and lemmas below in this more general setting so that they can also be used in \Cref{sec:lcl-lower-bound}.
The induced graph $G^-[S]$ is a path with endpoints $a$ and $b$, and $G^+$ is obtained from $G^-$ by adding the edge $\{a,b\}$. Both graphs use the same identifiers and input labels on $V$. If port numbers are used, each common edge has the same local port number at each endpoint in both graphs, and the closing edge is assigned unused port numbers at $a$ and $b$. Hence, the nodes' initial states in the two graphs differ only at $a$ and $b$, whose initial stars reveal the extra edge.

The nodes of $F$ are the only nodes that may crash. The nodes of $C$ and $S$ always survive. The actual crash set $\mathcal{F}(\calE)\subseteq F$ depends on the execution. In the original graph, $C\cup F$ provides short routes between distant nodes of $S$, so that every node of $S$ is within distance $R-1$ of every node of the graph. Once all nodes of $F$ crash, these shortcuts disappear.

For each lower bound, we specify the structure of $C\cup F$ and how the nodes of $F$ attach to $S$ in \Cref{sec:lb-linear,sec:lb-degree,sec:lb-fault-budget}. There, we make the path $G^-[S]$ as long as possible under each setting's constraints, as its length determines how large a round complexity $T$ we can rule out.

\paragraph{Proof plan.}
We note that, in a fault-free execution of $\mathcal{A}$ on $G^+$, fault-free completeness forces every node of $S$ to output the whole graph, since every node of $S$ is within distance $R-1$ of every node of the graph. In particular, these outputs include the closing edge $\{a,b\}$. In contrast, by the termination and radius requirement, every surviving node's output in an execution on $G^-$ must be a labeled subgraph of $G^-$, which cannot contain $\{a,b\}$.

We also make the following observation: if all nodes of $F$ crash before sending any message, $S$ is cut off from the rest of the graph and forms a long path or cycle. Then, nodes of $S$ that are far from $a$ and $b$ cannot tell within $T$ rounds whether they are in $G^+$ or in $G^-$.

Together, these observations allow us to show that some node of $S$ outputs $G^+$ even in an execution on $G^-$, which contradicts that $\mathcal{A}$ solves $\CN(R)$. To this end, we define a \emph{chain} of \emph{similar} executions, in which each pair of consecutive executions is indistinguishable at some node of $S$. We gradually move the crashes of the nodes of $F$ earlier, until they crash before sending any message, and use consistency to show that every node of $S$ keeps outputting $G^+$. Finally, we switch from $G^+$ to $G^-$: a node of $S$ far from $a$ and $b$ still outputs $G^+$, although $G^+$ is not a subgraph of $G^-$.

\paragraph{Indistinguishable executions and similarity chains.}
To make this plan precise, we first formally define when a node cannot tell two executions apart.

For $t\in\{0,\dots,T\}$, the \emph{view} of a node $v$ at the end of round $t$ consists of the initial state of $v$ and, for every round $t'\in\{1,\dots,t\}$ and every neighbor $u$ of $v$, the message that $v$ receives from $u$ in round $t'$, or $\bot$ if no message is received. Here, a view records the local history of a node: it is distinct from the output graph $H_v$ required by $\CN(R)$.

We say that two executions are \emph{indistinguishable} at $v$ up to round $t$ if $v$ has the same view at the end of round $t$ in both. Since $\mathcal{A}$ is deterministic, a node that survives in both executions and has the same view through round~$T$ produces the same output in both.

If $\mathcal A$ is correct, every node of $S$ produces different outputs in the fault-free execution on $G^+$ and in the final execution on $G^-$, so these executions cannot be indistinguishable up to round $T$ at any node of $S$. We therefore connect them through a chain in which consecutive executions are indistinguishable at some node of $S$. Such chains are a standard tool for crash-fault lower bounds, most prominently for consensus \cite{LamportFischer1982, Lynch1996}. We formalize them for our setting as follows.

\begin{definition}[Similarity chain]\label{def:similarity-chain}
A \emph{similarity chain} for a $T$-round algorithm $\mathcal{A}$ is a finite sequence $\calE_0,\dots,\calE_K$ of executions of $\mathcal{A}$, each on a labeled graph respecting $(C,F,S)$, satisfying the following:
\begin{itemize}
\item \emph{(Admissibility)} in every $\calE_i$, at most $f$ nodes crash, and all of them lie in $F$, so every node of $C\cup S$ survives;
\item \emph{(Indistinguishability)} for every $i<K$, there is a node $w_i\in S$ at which $\calE_i$ and $\calE_{i+1}$ are indistinguishable up to round $T$.
Such a node is called a \emph{witness} of the step from $\calE_i$ to $\calE_{i+1}$.
\end{itemize}
\end{definition}

Returning to our plan, our similarity chain starts with the fault-free execution $\calE_0$ on $G^+$ and ends with an execution $\calE_K$ on $G^-$, with all intermediate executions on $G^+$. The following lemma shows that such a chain forces $\mathcal A$ to violate the requirements of $\CN(R)$ in at least one execution, provided that every node of $S$ has sufficiently small eccentricity.
\begin{lemma}[From similarity chains to lower bounds]\label{lem:lb-chain-contradiction}
Assume that every node of $S$ has eccentricity at most $R-1$ in $G^+$. In addition, let $\calE_0,\dots,\calE_K$ be a similarity chain for $\mathcal A$ such that $\calE_0$ is the fault-free execution on $G^+$, the executions $\calE_1,\dots,\calE_{K-1}$ run on $G^+$, and $\calE_K$ runs on $G^-$. Then, $\mathcal A$ does not satisfy the requirements of $\CN(R)$ in at least one execution of this chain.
\end{lemma}

To prove \Cref{lem:lb-chain-contradiction}, we carry the complete output $G^+$ from $\calE_0$ along the assumed similarity chain, using the result below.

\begin{lemma}[Propagation of a complete view]\label{lem:lb-propagation}
Let $G$ be a graph, and let $S\subseteq V(G)$ be a set of nodes such that $G[S]$ is connected and every node of $S$ has eccentricity at most $R-1$ in $G$. Suppose that the outputs $H_v$ of the surviving nodes $v$, including all nodes of $S$, satisfy the termination and radius, edge validity, and consistency requirements of $\CN(R)$. If some node $p\in S$ outputs $H_p=G$, then every node of $S$ outputs $G$.
\end{lemma}
\begin{proof}
If $S=\{p\}$, there is nothing to prove. Otherwise, since $G[S]$ is connected, node $p$ has at least one neighbor in $S$.

We show that every neighbor $q\in S$ of $p$ also outputs $H_q=G$. As $p$ and $q$ are adjacent surviving nodes, edge validity ensures that both $H_p$ and $H_q$ contain the edge $\{p,q\}$. Consistency therefore gives $H_p[S_p(q)]=H_q[S_q(p)]$. Since $H_p=G$ and $p$ has eccentricity at most $R-1$ in $G$, we have $B_{H_p}(p,R-1)=V(G)$, and hence $S_p(q)=V(G)$ and $H_p[S_p(q)]=G$. Thus, $H_q[S_q(p)]=G$, so $H_q$ contains every node and edge of $G$. By termination and radius, $H_q$ is a labeled subgraph of $G$, so $H_q=G$.

Thus, every neighbor in $S$ of a node that outputs $G$ also outputs $G$. Since $G[S]$ is connected, applying this argument along paths from $p$ shows that every node of $S$ outputs $G$.
\end{proof}

We may now present the proof of \Cref{lem:lb-chain-contradiction}.
\begin{proof}[Proof of \Cref{lem:lb-chain-contradiction}]
Suppose, for a contradiction, that $\mathcal A$ satisfies the requirements of $\CN(R)$ in every execution of the chain.

We first note how \Cref{lem:lb-propagation} applies to the executions on $G^+$. Every node of $S$ survives, $G^+[S]$ is connected, and every node of $S$ has eccentricity at most $R-1$ in $G^+$. Thus, if one node of $S$ outputs $G^+$ in such an execution, then every node of $S$ outputs $G^+$.

We now show by induction that every node of $S$ outputs $G^+$ in each execution $\calE_0,\dots,\calE_{K-1}$. In $\calE_0$, fault-free completeness requires each node of $S$ to output its complete radius-$R$ neighborhood. By the eccentricity assumption, this neighborhood is all of $G^+$.

For the induction step, suppose that every node of $S$ outputs $G^+$ in $\calE_i$, where $i<K-1$. The witness $w_i$ cannot distinguish $\calE_i$ from $\calE_{i+1}$, so it also outputs $G^+$ in $\calE_{i+1}$. Since $\calE_{i+1}$ runs on $G^+$, the propagation argument above implies that every node of $S$ outputs $G^+$ in $\calE_{i+1}$.

Finally, the witness $w_{K-1}$ outputs $G^+$ in $\calE_{K-1}$ and cannot distinguish this execution from $\calE_K$. It therefore also outputs $G^+$ in $\calE_K$. However, $\calE_K$ runs on $G^-$, so termination and radius require this output to be a labeled subgraph of $G^-$. This is impossible because $\{a,b\}$ belongs to $G^+$ but not to $G^-$.
\end{proof}

\paragraph{Constructing our similarity chains.}
\Cref{lem:lb-chain-contradiction} reduces each lower bound of \Cref{thm:lower-bound} to constructing a suitable similarity chain from the fault-free execution on $G^+$ to an execution on $G^-$. We do so first for a single standard crash (\Cref{lem:lb-standard-chain}) and then for atomic crashes (\Cref{lem:lb-atomic-chain}).

In both crash models, each step of the chain changes the messages of a single crashing node or, in the last step, the initial states of $a$ and $b$. We find a witness for each step by bounding how far such a change can spread.

For two executions $\calE$ and $\calE'$ in which every node of $C\cup S$ survives, and for $t \in \{0, \dots, T\}$, let $D_t\subseteq C\cup S$ be the set of nodes whose local histories through round~$t$ differ. Then, $D_0$ consists of the nodes in $C\cup S$ with different initial states, and every node of $S\setminus D_T$ is a witness. The following lemma shows that, once the nodes of $F$ stop sending messages, differences spread only inside $G[S]$, by at most one hop per round.
\begin{lemma}[Spreading of differences]\label{lem:lb-spreading}
Let $\calE$ and $\calE'$ be executions of $\mathcal A$ on labeled graphs respecting $(C,F,S)$, in which all nodes of $C\cup S$ survive, and let $G$ be a graph that contains the graphs of both executions as subgraphs. Let $0\leq t_0\leq T$, and suppose that no node of $F$ sends a message in any round $t\in\{t_0+1,\dots,T\}$ in either execution. Then, $D_T\cap S\subseteq B_{G[S]}(D_{t_0}\cap S, T-t_0)$.
\end{lemma}
\begin{proof}
We show by induction on $t$ that $D_t\cap S\subseteq B_{G[S]}(D_{t_0}\cap S,t-t_0)$ for every $t\in\{t_0,\dots,T\}$. The lemma then follows by setting $t=T$.

The base case $t=t_0$ is immediate. For the induction step, let $t\in\{t_0+1,\dots,T\}$ and assume that the claim holds for $t-1$. It will suffice to show that a difference can spread by at most one hop in round $t$, i.e., that $D_t\cap S\subseteq B_{G[S]}(D_{t-1}\cap S,1)$. Together with the induction hypothesis and the triangle inequality, this implies our claim:
$D_t\cap S\subseteq B_{G[S]}\bigl(B_{G[S]}(D_{t_0}\cap S,t-1-t_0),1\bigr)\subseteq B_{G[S]}(D_{t_0}\cap S,t-t_0)$.

To prove the one-hop inclusion, fix a node $v\in S$ with $v\notin B_{G[S]}(D_{t-1}\cap S,1)$. We show that $v\notin D_t$. Since $v\notin D_{t-1}$, its local histories through round $t-1$ coincide in the two executions. In particular, its initial star is identical in both. This star specifies all neighbors of $v$, so $v$ has the same neighbors in both executions.

In round $t$, $v$ receives messages only from its neighbors in $S$: the nodes of $F$ send nothing in round $t$, and there are no edges between $C$ and $S$. Let $u$ be such a neighbor. Since $\{u,v\}$ is an edge of $G[S]$ and $v$ is not within distance one of $D_{t-1}\cap S$, we have $u\notin D_{t-1}$, so $u$ has the same view at the end of round $t-1$ in both executions. Moreover, $u$ survives, and since $\mathcal A$ is deterministic, its round-$t$ message to $v$ is determined by this view. Hence, $v$ receives the same messages in round $t$ in both executions, and therefore has the same view at the end of round $t$. Consequently, $v\notin D_t$.
\end{proof}

We next compare executions on $G^+$ and $G^-$ in which no node of $F$ sends any message. Since the graphs differ only in the closing edge $\{a,b\}$, we have $D_0=\{a,b\}$. Applying \Cref{lem:lb-spreading} with $G=G^+$ and $t_0=0$ gives the following corollary, which we will use to justify the final step of each similarity chain.
\begin{corollary}[Switching the graph]\label{lem:lb-switch-graph}
Let $\calE$ be an execution of $\mathcal A$ on $G^+$ and $\calE'$ an execution of $\mathcal A$ on $G^-$, such that all nodes of $C\cup S$ survive in both and no node of $F$ sends any message in either execution. Then, $D_T\cap S\subseteq B_{G^+[S]}(\{a,b\},T)$.
\end{corollary}

\paragraph{One standard crash.}
We now construct a similarity chain with $F=\{z\}$, allowing only $z$ to crash in the standard model. Starting from the fault-free execution on $G^+$, we aim to reach an execution in which $z$ sends no messages. We can then apply \Cref{lem:lb-switch-graph} for the final switch to $G^-$. In the standard crash model, $z$'s crash-round messages may be delivered to any subset of its neighbors. We therefore compare executions in which $z$ crashes in the same round, but one additional neighbor does not receive its message from $z$.

\begin{lemma}[One standard crash]\label{lem:lb-standard-chain}
Consider a pair $G^-,G^+$ with the partition and path/cycle structure specified earlier, where $F=\{z\}$, and let $T\geq0$. Assume that every node of $S$ has eccentricity at most $R-1$ in $G^+$, and that the following hold:
\begin{enumerate}[(a)]
    \item for every $w\in N_{G^+}(z)\cap S$, there is a node $p\in S$ such that $\dist_{G^+[S]}(p,w)\geq T$;
    \item there is a node $q\in S$ such that $\dist_{G^+[S]}\bigl(q,\{a,b\}\bigr)>T$.
\end{enumerate}
Then, no deterministic $T$-round algorithm solves $\CN(R)$ on both graphs while tolerating one standard crash.
\end{lemma}

For $T=0$, condition~(b) gives a node $q\in S\setminus\{a,b\}$ with the same initial state in both graphs. The two fault-free executions therefore form a similarity chain with witness $q$, and \Cref{lem:lb-chain-contradiction} rules out such an algorithm. We henceforth assume $T\geq1$.

\begin{figure}[t]
\centering
\resizebox{\linewidth}{!}{\begin{tikzpicture}[x=1cm,y=1cm]
  \def\cA{0}\def\cB{3.0}\def\cC{6.0}\def\cD{9.0}\def\cE{12.7}
  \def\yA{8.0}\def\yB{4.0}\def\yC{0}
  \node[font=\footnotesize,anchor=south west] at (-0.9,\yA+1.55)
    {$|S|=8$ surviving nodes,\quad $R=3$ (diameter $2\le R-1$),\quad
     $T=\lfloor (|S|-1)/2\rfloor-1=2$,\quad $w_j=p_{9-j}$};
  \node[scgraph,anchor=east,text=purple!75!black] at (-0.4,\yA+0.45) {$G^+$};
  \scStandardPanel{\cA}{\yA}{1}{}{}{-1}
  \scStandardPanel{\cB}{\yA}{1}{2}{}{-1}
  \scStandardPanel{\cC}{\yA}{1}{2}{7}{3}
  \scStandardPanel{\cD}{\yA}{1}{2}{7,6}{2}
  \scStandardPanel{\cE}{\yA}{1}{2}{0,1,2,3,4,5,6,7}{4}
  \scArrowR{\cA}{\cB}{\yA+0.45}
  \scArrowR{\cB}{\cC}{\yA+0.45}
  \scArrowR{\cC}{\cD}{\yA+0.45}
  \scDots{12.05}{\yA+0.45}
  \node[sccaption] at (\cA+1.26,\yA-1.05) {$\calE_0$\\fault-free};
  \node[sccaption] at (\cB+1.26,\yA-1.05) {$\calE_1=\calE(G^+,2,M_0)$\\crashes in round $T=2$,\\all delivered (silent step)};
  \node[sccaption] at (\cC+1.26,\yA-1.05) {$\calE_2=\calE(G^+,2,M_1)$\\$p_8$ dropped};
  \node[sccaption] at (\cD+1.26,\yA-1.05) {$\calE_3=\calE(G^+,2,M_2)$\\$p_8,p_7$ dropped};
  \node[sccaption] at (\cE+1.26,\yA-1.05) {$\calE_9=\calE(G^+,2,M_8)$\\nothing delivered};
  \draw[scstep,rounded corners=4pt]
    ({\cE+1.26},\yA-1.55) -- ({\cE+1.26},\yA-1.95) -- (-0.3,\yA-1.95)
    -- (-0.3,\yB+1.75) -- (\cA+1.26,\yB+1.75);
  \node[sccaption,anchor=north] at (7.0,\yA-2.02)
    {silent step: $z$ sends the same messages in $\calE_9$ and $\calE_{10}$, so every node of $S$ is a witness};
  \node[scgraph,anchor=east,text=purple!75!black] at (-0.4,\yB+0.45) {$G^+$};
  \scStandardPanel{\cA}{\yB}{1}{1}{}{-1}
  \scStandardPanel{\cB}{\yB}{1}{1}{7}{3}
  \scStandardPanel{\cC}{\yB}{1}{1}{7,6}{2}
  \scStandardPanel{\cD}{\yB}{1}{1}{7,6,5}{1}
  \scStandardPanel{\cE}{\yB}{1}{1}{0,1,2,3,4,5,6,7}{4}
  \scArrowR{\cA}{\cB}{\yB+0.45}
  \scArrowR{\cB}{\cC}{\yB+0.45}
  \scArrowR{\cC}{\cD}{\yB+0.45}
  \scDots{12.05}{\yB+0.45}
  \node[sccaption] at (\cA+1.26,\yB-1.05) {$\calE_{10}=\calE(G^+,1,M_0)$\\crashes in round 1,\\all delivered};
  \node[sccaption] at (\cB+1.26,\yB-1.05) {$\calE_{11}=\calE(G^+,1,M_1)$\\$p_8$ dropped};
  \node[sccaption] at (\cC+1.26,\yB-1.05) {$\calE_{12}=\calE(G^+,1,M_2)$\\$p_8,p_7$ dropped};
  \node[sccaption] at (\cD+1.26,\yB-1.05) {$\calE_{13}=\calE(G^+,1,M_3)$\\$p_8,p_7,p_6$ dropped};
  \node[sccaption] at (\cE+1.26,\yB-1.05) {$\calE_{18}=\calE(G^+,1,M_8)$\\crashes before round 1};
  \draw[scstep] ({\cE+1.26},\yB-1.5) -- ({\cE+1.26},\yC+1.55)
    node[midway,left=3pt,sccaption,align=right]
    {closing edge\\removed;\\$p_4$ cannot tell};
  \node[scgraph,anchor=east,text=purple!75!black] at (\cE-0.4,\yC+0.45) {$G^-$};
  \scStandardPanel{\cE}{\yC}{0}{1}{0,1,2,3,4,5,6,7}{3}
  \foreach \i [evaluate=\i as \j using int(\i+1)] in {0,...,7}
    \node[sccaption] at ({\cE*1cm+\i*\scsp},\yC-0.43) {$p_\j$};
  \node[sccaption] at (\cE+1.26,\yC-1.05) {$\calE_{19}=\calE(G^-,1,M_8)$\\same schedule on $G^-$};
\end{tikzpicture}
}
\caption{The similarity chain of \Cref{lem:lb-standard-chain} for the
one-standard-crash construction with $|S|=8$ surviving nodes, so
$T=\lfloor (|S|-1)/2\rfloor-1=2$, and radius $R=3$: both graphs have diameter
$2\le R-1$, so every fault-free output is the whole graph. A dark $z$
crashes in the indicated round; a dotted spoke is a message of that round
that is not delivered. Every execution has at most one crash. A ring marks
a witness, at which the execution is indistinguishable from the previous one.}
\label{fig:lb-standard-chain}
\end{figure}
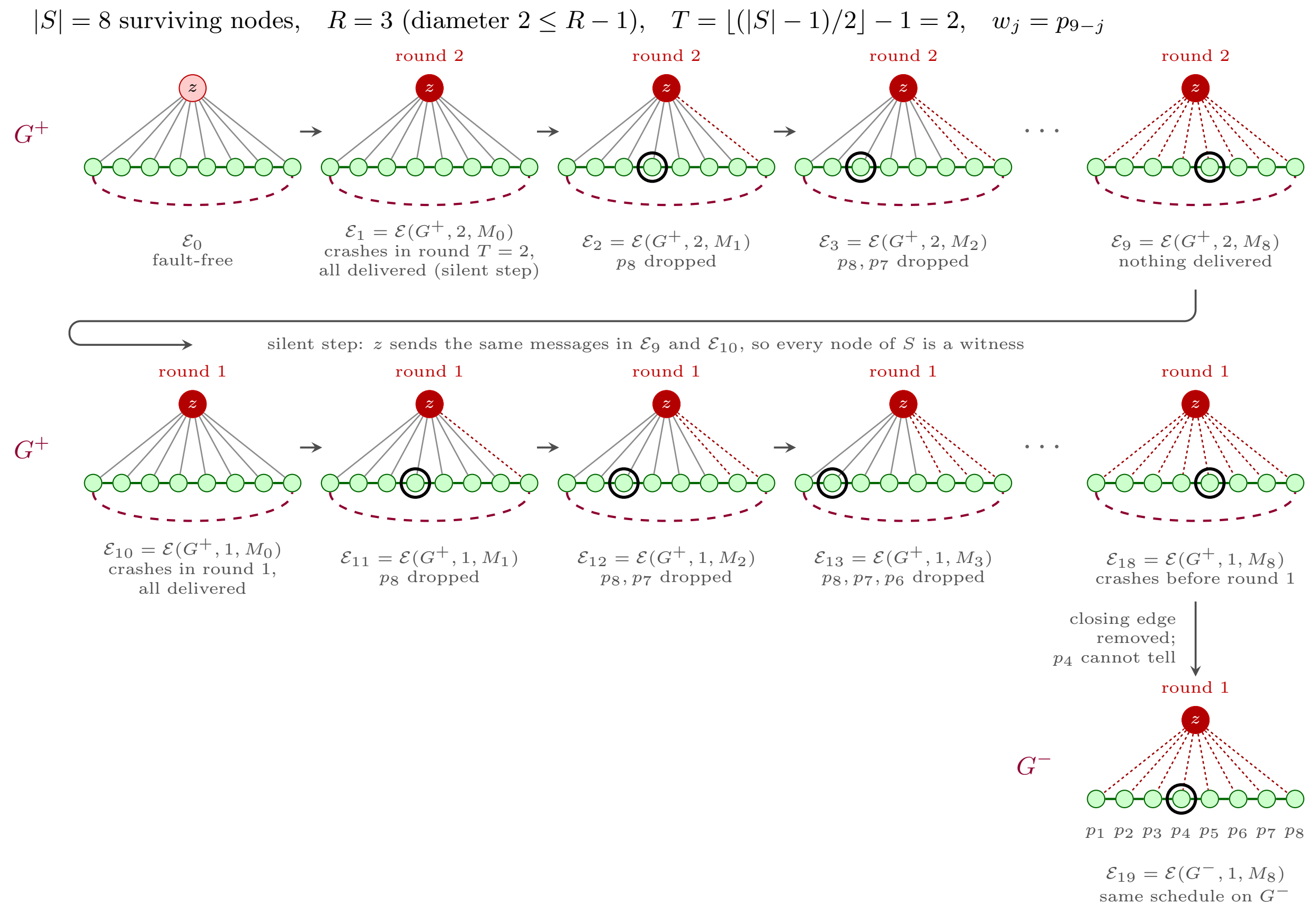

As displayed in \Cref{fig:lb-standard-chain}, for $T\ge 1$, our chain here first lets $z$ crash in round $T$ while still sending all its messages. It then removes these messages one at a time. Once $z$ sends nothing in round $T$, it behaves exactly as if it had crashed in round $T-1$ while sending all its messages, and the chain continues with round $T-1$, and so on, until $z$ crashes in round $1$ without sending any message. Finally, it switches from $G^+$ to $G^-$.

Since $z\notin\{a,b\}$ and the graphs $G^+$ and $G^-$ differ only in the edge $\{a,b\}$, the neighbors of $z$ are the same in both graphs. We write $N(z):=N_{G^+}(z)=N_{G^-}(z)$.

For the fixed deterministic algorithm $\mathcal A$, specifying the graph and the crash behavior of $z$ determines the execution. We describe this behavior by a \emph{standard crash schedule}: either $\bot$, indicating that $z$ does not crash, or a pair $(r,M)$, where $r\in\{1,\dots,T\}$ is its crash round and $M\subseteq N(z)$. Under schedule $(r,M)$, node $z$ follows $\mathcal A$ before round $r$. In round $r$, its prescribed messages to neighbors in $M$ are delivered, while those to other neighbors are not. Node $z$ remains inactive thereafter.

For $G\in\{G^+,G^-\}$, we write $\calE(G,\bot)$ for the fault-free execution on $G$ and $\calE(G,r,M)$ for the execution with crash schedule $(r,M)$. In each case, every node other than $z$ survives and follows $\mathcal A$.

To describe the schedules of our chain, we denote the $d:=|N(z)|$ neighbors of $z$ by $w_1,\dots,w_d$, in the order in which their messages are removed, and let $M_j:=N(z)\setminus\{w_1,\dots,w_j\}$ for $j\in\{0,\dots,d\}$, so that $M_0=N(z)$ and $M_d=\emptyset$. Our chain consists of the following executions, in this order:
\[
  \calE(G^+,\bot),\ \calE(G^+,T,M_0),\ \dots,\ \calE(G^+,T,M_d),\ \calE(G^+,T-1,M_0),\ \dots,\ \calE(G^+,1,M_d),\ \calE(G^-,1,M_d).
\]
Apart from the last step, which switches from $G^+$ to $G^-$, the chain has two kinds of steps: \emph{silent step} and \emph{drop step}.
In a silent step, from $\calE(G^+,\bot)$ to $\calE(G^+,T,M_0)$, or from $\calE(G^+,r+1,M_d)$ to $\calE(G^+,r,M_0)$ for some $r<T$, the node $z$ sends exactly the same messages in both executions: all of them in the first case, and all of them through round~$r$ and none afterwards in the second. Hence, every node other than $z$ has the same view in both executions.
In a drop step, from $\calE(G^+,r,M_{j-1})$ to $\calE(G^+,r,M_j)$ for some $r\in\{1,\dots,T\}$ and $j\in\{1,\dots,d\}$, only the round-$r$ message of $z$ to $w_j$ is removed, so only nodes of $S$ close to $w_j$ can notice the step, as described by the lemma below.

\begin{lemma}[Dropping one message]\label{lem:lb-drop-message}
Let $r\in\{1,\dots,T\}$ and $j\in\{1,\dots,d\}$, and consider the two executions $\calE=\calE(G^+,r,M_{j-1})$ and $\calE'=\calE(G^+,r,M_j)$. Then, $D_T\cap S\subseteq B_{G^+[S]}\bigl(\{w_j\}\cap S,\;T-r\bigr)$.
\end{lemma}
\begin{proof}
In both executions, $z$ is the only node that crashes, and it does so in round $r$. Hence, the two executions coincide through round $r-1$, i.e., $D_{r-1}=\emptyset$. In round $r$, they differ only in the message from $z$ to $w_j$, so $D_r\subseteq\{w_j\}$. Since $z$ sends no message after round $r$ in either execution, \Cref{lem:lb-spreading} with $t_0=r$ gives $D_T\cap S\subseteq B_{G^+[S]}(D_r\cap S,T-r)\subseteq B_{G^+[S]}\bigl(\{w_j\}\cap S,T-r\bigr)$.
\end{proof}

We are now ready to prove the lower bound for standard crashes.

\begin{proof}[Proof of \Cref{lem:lb-standard-chain}]
Fix a deterministic $T$-round algorithm $\mathcal A$, and consider the chain described above. Every execution in this chain has at most one standard crash, namely of $z$, so the chain is admissible. It remains to verify the indistinguishability property of \Cref{def:similarity-chain}. Recall that a node of $S$ outside $D_T$ is a witness, so for each step it suffices to find a node of $S$ that cannot notice it.

In a silent step, as shown above, $z$ sends the same messages in both executions, so every node other than $z$ has the same view in both. Hence, every node of $S$ is a witness.

Next, consider a drop step, from $\calE(G^+,r,M_{j-1})$ to $\calE(G^+,r,M_j)$. By \Cref{lem:lb-drop-message}, every node of $D_T\cap S$ is within distance $T-r<T$ of $\{w_j\}\cap S$ in $G^+[S]$. If $w_j\notin S$, then $D_T\cap S=\emptyset$, and every node of $S$ is a witness. Otherwise, condition~(a) provides a node $p\in S$ at distance at least $T$ from $w_j$, so $p\notin D_T$, and $p$ is a witness.

Finally, consider the last step, from $\calE(G^+,1,M_d)$ to $\calE(G^-,1,M_d)$. Since $M_d=\emptyset$, the node $z$ sends no message in either execution. Hence, by \Cref{lem:lb-switch-graph}, every node of $D_T\cap S$ is within distance $T$ of $\{a,b\}$ in $G^+[S]$. Condition~(b) provides a node $q\in S$ at distance more than $T$ from $\{a,b\}$, so $q\notin D_T$, and $q$ is a witness.

Consequently, the chain is a similarity chain satisfying the assumptions of \Cref{lem:lb-chain-contradiction}, so $\mathcal A$ violates the requirements of $\CN(R)$ in at least one of its executions. Since each of these executions has at most one standard crash, $\mathcal A$ does not solve $\CN(R)$ on both graphs while tolerating one standard crash.
\end{proof}

\paragraph{Atomic crashes.}
We now turn to atomic crashes, where $F$ may contain several nodes. Since a node that crashes atomically sends either all or none of its messages in a round, consecutive executions now differ in the crash round of an entire node of $F$. Such a step changes the messages to all its neighbors at once, so condition~(a) below asks for a node of $S$ far from all of them.
\begin{lemma}[Atomic crashes]\label{lem:lb-atomic-chain}
Consider a pair $G^-,G^+$ with the partition and path/cycle structure specified above, and let $T\geq0$. Assume that every node of $S$ has eccentricity at most $R-1$ in $G^+$, and that the following hold:
\begin{enumerate}[(a)]
    \item for every $x\in F$, there is a node $p\in S$ such that $\dist_{G^+[S]}\bigl(p,N_{G^+}(x)\cap S\bigr)\geq T$;
    \item there is a node $q\in S$ such that $\dist_{G^+[S]}\bigl(q,\{a,b\}\bigr)>T$.
\end{enumerate}
Here the distance to an empty set is infinity. Then, for every globally known crash bound $f\geq|F|$, no deterministic $T$-round algorithm solves $\CN(R)$ on both graphs under atomic crashes.
\end{lemma}

We first handle $T=0$. Condition~(b) gives a node $q\in S\setminus\{a,b\}$. Since the graphs differ only in the edge $\{a,b\}$, node $q$ has the same initial state in both graphs. The two fault-free executions therefore form a similarity chain with witness $q$. By \Cref{lem:lb-chain-contradiction}, no deterministic zero-round algorithm solves $\CN(R)$ on both graphs. We henceforth assume $T\geq1$.

\begin{figure}[hbt]
\centering
\begin{tikzpicture}[x=1cm,y=1cm]
  \def\cA{0}\def\cB{3.0}\def\cC{6.0}\def\cD{9.0}\def\cE{12.0}
  \def\yTop{4.2}\def\yBot{0}
  \node[font=\footnotesize,anchor=south west] at (-0.9,\yTop+1.55)
    {$|S|=8$ surviving nodes,\quad $R=4$ (diameter $3\le R-1$),\quad
     $T=\lfloor |S|/4\rfloor=2$};
  \node[scgraph,anchor=east,text=purple!75!black] at (-0.4,\yTop+0.45) {$G^+$};
  \scAtomicPanel{\cA}{\yTop}{1}{}{}{-1}
  \scAtomicPanel{\cB}{\yTop}{1}{2}{}{6}
  \scAtomicPanel{\cC}{\yTop}{1}{2}{2}{1}
  \scAtomicPanel{\cD}{\yTop}{1}{1}{2}{6}
  \scAtomicPanel{\cE}{\yTop}{1}{1}{1}{1}
  \scArrowR{\cA}{\cB}{\yTop+0.45}
  \scArrowR{\cB}{\cC}{\yTop+0.45}
  \scArrowR{\cC}{\cD}{\yTop+0.45}
  \scArrowR{\cD}{\cE}{\yTop+0.45}
  \draw[scstep] ({\cE+1.26},\yTop-1.5) -- ({\cE+1.26},\yBot+1.55)
    node[midway,left=3pt,sccaption,align=right]
    {closing edge\\removed;\\$p_4$ cannot tell};
  \node[scgraph,anchor=east,text=purple!75!black] at (\cE-0.4,\yBot+0.45) {$G^-$};
  \scAtomicPanel{\cE}{\yBot}{0}{1}{1}{3}
  \foreach \i [evaluate=\i as \j using int(\i+1)] in {0,...,7}
    \node[sccaption] at ({\cE*1cm+\i*\scsp},\yBot-0.43) {$p_\j$};
  \node[sccaption] at (\cA+1.26,\yTop-1.05) {$\calE_0$\\fault-free};
  \node[sccaption] at (\cB+1.26,\yTop-1.05) {$\calE_1$\\$x_1$ crashes in round $T=2$};
  \node[sccaption] at (\cC+1.26,\yTop-1.05) {$\calE_2$\\$x_2$ crashes in round $2$};
  \node[sccaption] at (\cD+1.26,\yTop-1.05) {$\calE_3$\\$x_1$ moved to round $1$};
  \node[sccaption] at (\cE+1.26,\yTop-1.05) {$\calE_4$\\$x_2$ moved to round $1$};
  \node[sccaption] at (\cE+1.26,\yBot-1.05) {$\calE_5$\\same schedule on $G^-$};
\end{tikzpicture}
\caption{The similarity chain of \Cref{lem:lb-atomic-chain} for the
two-atomic-crash construction with $|S|=8$ surviving nodes, so
$T=\lfloor |S|/4\rfloor=2$, and radius $R=4$: both graphs have diameter
$3\le R-1$, so every fault-free output is the whole graph. Here $x_1$ is
adjacent to $p_1,\dots,p_4$ and $x_2$ to $p_5,\dots,p_8$. Each panel is an
execution: a dark node crashes atomically in the indicated round, a light
one never crashes. A ring marks a witness, at which the execution is
indistinguishable from the previous one.}
\label{fig:lb-atomic-chain}
\end{figure}
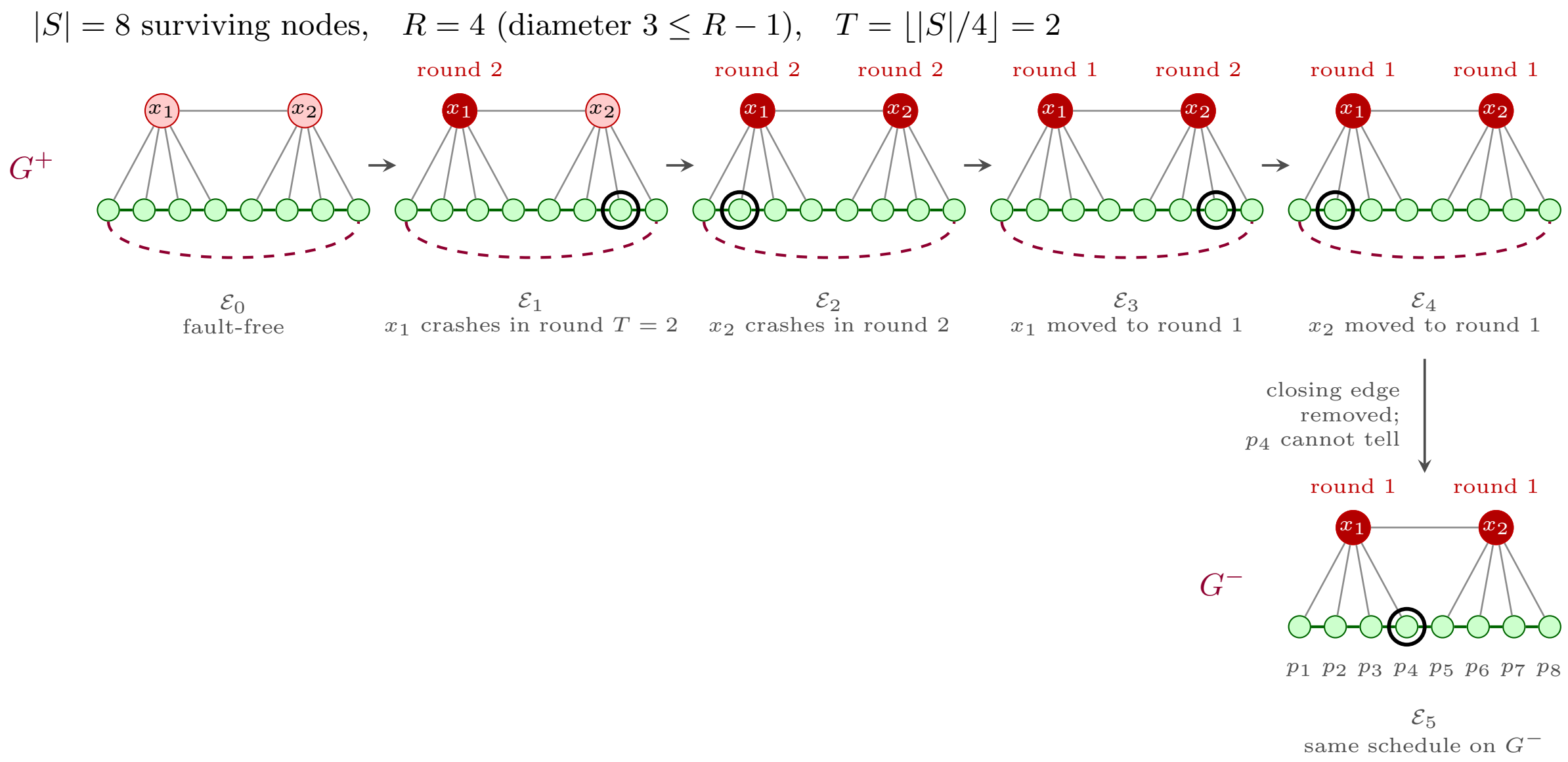

As displayed in \Cref{fig:lb-atomic-chain}, for $T\geq1$, our chain starts from the fault-free execution. Each successive execution makes one additional node of $F$ crash at the beginning of round $T$, until all nodes of $F$ crash in that round. We then move the crashes earlier, one node and one round at a time, until they all occur at the beginning of round~$1$. Finally, we switch from $G^+$ to $G^-$.

An \emph{atomic crash schedule} is a map $c\colon F\to\{1,\dots,T\}\cup\{\infty\}$. If $c(x)=r<\infty$, node $x$ follows $\mathcal A$ through round $r-1$, crashes before sending any message in round $r$, and remains inactive thereafter. If $c(x)=\infty$, node $x$ survives. We write $c_\infty$ for the schedule with no crashes.

For a labeled graph $G$ respecting $(C,F,S)$, let $\calE(G,c)$ denote the execution determined by $\mathcal A$, $G$, and $c$. Every node of $C\cup S$ survives, so this execution has at most $|F|$ atomic crashes, all in $F$.

To describe the schedules of our chain, we denote the $k:=|F|$ nodes of $F$ by $x_1,\dots,x_k$. For $r\in\{1,\dots,T\}$ and $s\in\{0,\dots,k\}$, we denote by $c_{r,s}$ the schedule in which $x_1,\dots,x_s$ crash at the beginning of round $r$, and $x_{s+1},\dots,x_k$ crash at the beginning of round $r+1$ if $r<T$, or not at all if $r=T$. That is, the index $s$ marks the cut between the nodes that crash in round $r$ and those that crash one round later:
\[
  c_{r,s}(x_i):=
  \begin{cases}
    r   & \text{if } i\le s,\\
    r+1 & \text{if } i>s \text{ and } r<T,\\
    \infty & \text{if } i>s \text{ and } r=T.
  \end{cases}
\]
Our chain consists of the following executions, in this order:
\[
\calE(G^+,c_{T,0}),\calE(G^+,c_{T,1}),\dots,\calE(G^+,c_{T,k}),\calE(G^+,c_{T-1,1}),\dots,\calE(G^+,c_{1,k}),\calE(G^-,c_{1,k}),
\]
where $c_{T,0}=c_\infty$. We omit $c_{r-1,0}$, since it equals $c_{r,k}$: in both, all nodes of $F$ crash at the beginning of round $r$. Hence, apart from the last step, which switches from $G^+$ to $G^-$, every step, from $\calE(G^+,c_{r,s-1})$ to $\calE(G^+,c_{r,s})$ for some $r\in\{1,\dots,T\}$ and $s\in\{1,\dots,k\}$, changes only the crash time of $x_s$: from no crash to round $T$ if $r = T$, and from round $r + 1$ to $r$ otherwise. Only nodes of $S$ close to the neighbors of $x_s$ can notice such a step, as described by the lemma below.

The only restriction on the graph in the next lemma is that there are no edges between $C$ and $S$; it does not require $G[S]$ to be a path or cycle. We will reuse this generality in \Cref{sec:lcl-crash-bounds}.

\begin{lemma}[Moving one atomic crash]\label{lem:lb-move-atomic}
Let $G$ be a labeled graph respecting $(C,F,S)$, let $r\in\{1,\dots,T\}$ and $s\in\{1,\dots,k\}$, and consider the two executions $\calE=\calE(G,c_{r,s-1})$ and $\calE'=\calE(G,c_{r,s})$. Then, $D_T\cap S\subseteq B_{G[S]}\bigl(N_G(x_s)\cap S,\;T-r\bigr)$.
\end{lemma}
\begin{proof}
The two executions run on the same labeled graph, and no node crashes before round $r$. Their states and messages therefore coincide through round $r-1$, so $D_{r-1}=\emptyset$.

In round $r$, every node other than $x_s$ has the same state and crash behavior in both executions, so it sends the same messages in both or no messages in either. Node $x_s$ sends its prescribed messages in $\calE$ but crashes before sending them in $\calE'$. Thus, only its neighbors can receive different messages, giving $D_r\cap S\subseteq N_G(x_s)\cap S$.

If $r<T$, every node of $F$ crashes by the beginning of round $r+1$ in both executions; if $r=T$, there are no remaining rounds. Hence, \Cref{lem:lb-spreading} applies with $t_0=r$ and gives
$$
D_T\cap S
\subseteq B_{G[S]}(D_r\cap S,T-r)
\subseteq B_{G[S]}\bigl(N_G(x_s)\cap S,T-r\bigr).
$$
\end{proof}

We are now ready to prove the lower bound for atomic crashes.
\begin{proof}[Proof of \Cref{lem:lb-atomic-chain}]
Fix a crash bound $f\ge|F|$ and a deterministic $T$-round algorithm $\mathcal A$, and consider the chain described above. Every execution in this chain has at most $|F|\le f$ atomic crashes, all of them in $F$, so the chain is admissible. It remains to verify the indistinguishability property of \Cref{def:similarity-chain}. As in the proof of \Cref{lem:lb-standard-chain}, it suffices to find, for each step, a node of $S$ outside $D_T$.

First, consider a step from $\calE(G^+,c_{r,s-1})$ to $\calE(G^+,c_{r,s})$. By \Cref{lem:lb-move-atomic}, every node of $D_T\cap S$ is within distance $T-r<T$ of $N_{G^+}(x_s)\cap S$ in $G^+[S]$. Condition~(a) provides a node $p\in S$ at distance at least $T$ from $N_{G^+}(x_s)\cap S$, so $p\notin D_T$, and $p$ is a witness.

Finally, consider the last step, from $\calE(G^+,c_{1,k})$ to $\calE(G^-,c_{1,k})$. Under $c_{1,k}$, every node of $F$ crashes at the beginning of round~1, so no node of $F$ sends a message in either execution. Hence, by \Cref{lem:lb-switch-graph}, every node of $D_T\cap S$ is within distance $T$ of $\{a,b\}$ in $G^+[S]$. Condition~(b) provides a node $q\in S$ at distance more than $T$ from $\{a,b\}$, so $q\notin D_T$, and $q$ is a witness.

Consequently, the chain is a similarity chain satisfying the assumptions of \Cref{lem:lb-chain-contradiction}, so $\mathcal A$ violates the requirements of $\CN(R)$ in at least one of its executions. Since each of these executions has at most $f$ atomic crashes, $\mathcal A$ does not solve $\CN(R)$ on both graphs with crash bound $f$ under atomic crashes.
\end{proof}

\subsection{Linear Time with One or Two Crashes}
\label{sec:lb-linear}

We now focus on \Cref{thm:lower-bound}(i), which concerns general graphs: a single standard crash can force every algorithm for $\CN(R)$ to take $\Omega(n)$ rounds for every $R\geq3$, and under atomic crashes, two crashes suffice for every $R\geq4$. The following lemma states these bounds precisely.

\begin{lemma}[Linear lower bounds]\label{lem:lb-linear}
The following statements hold for deterministic algorithms in the synchronous LOCAL model, with the globally known size bound $n$.
\begin{enumerate}[(a)]
    \item Let $R\geq3$ and $n\geq9$. Every algorithm that solves $\CN(R)$ on all graphs with at most $n$ nodes while tolerating one standard crash needs more than $\lfloor n/2\rfloor-2$ rounds.
    \item Let $R\geq4$, $n\geq10$, and $f\geq2$. Every algorithm that solves $\CN(R)$ on all graphs with at most $n$ nodes under at most $f$ atomic crashes needs more than $\lfloor(n-2)/4\rfloor$ rounds.
\end{enumerate}
\end{lemma}

\begin{figure}[ht]
\centering
\begin{tikzpicture}[x=1cm,y=1cm,
  fail/.style={circle,draw=red!75!black,fill=red!20,
    minimum size=6mm,inner sep=1pt},
  survivor/.style={circle,draw=green!40!black,fill=green!20,
    minimum size=4mm,inner sep=0pt},
  path/.style={green!40!black,line width=.9pt},
  spoke/.style={black!45,line width=.6pt}]
  \node at (2.2,2.55) {One standard crash};
  \node[font=\small] at (2.2,2.15) {$R\geq3$};
  \node[fail] (z) at (2.2,1.45) {$z$};
  \foreach \i in {0,...,5} {
    \node[survivor] (p\i) at ({.88*\i},0) {};
    \draw[spoke] (z)--(p\i);
  }
  \foreach \i/\j in {0/1,1/2,2/3,3/4,4/5}
    \draw[path] (p\i)--(p\j);
  \draw[purple!75!black,dashed,line width=.9pt]
    (p0) .. controls (0,-.9) and (4.4,-.9) .. (p5);
  \begin{scope}[xshift=7cm]
    \node at (2.2,2.55) {Two atomic crashes};
    \node[font=\small] at (2.2,2.15) {$R\geq4$};
    \node[fail] (x1) at (.88,1.45) {$x_1$};
    \node[fail] (x2) at (3.52,1.45) {$x_2$};
    \draw[spoke] (x1)--(x2);
    \foreach \i in {0,...,5}
      \node[survivor] (q\i) at ({.88*\i},0) {};
    \foreach \i in {0,1,2} \draw[spoke] (x1)--(q\i);
    \foreach \i in {3,4,5} \draw[spoke] (x2)--(q\i);
    \foreach \i/\j in {0/1,1/2,2/3,3/4,4/5}
      \draw[path] (q\i)--(q\j);
    \draw[purple!75!black,dashed,line width=.9pt]
      (q0) .. controls (0,-.9) and (4.4,-.9) .. (q5);
  \end{scope}
\end{tikzpicture}
\caption{The linear lower bounds in \Cref{thm:lower-bound}(i).
Red nodes form $F$, green nodes
form $S$, and $C$ is empty. Dashed closing edges occur only in $G^+$.
The original graphs have diameter two on the left and three on the right,
while removing $F$ leaves a long path or cycle. The drawings show six
surviving nodes; the constructions extend to arbitrarily many.}
\label{fig:lb-linear}
\end{figure}
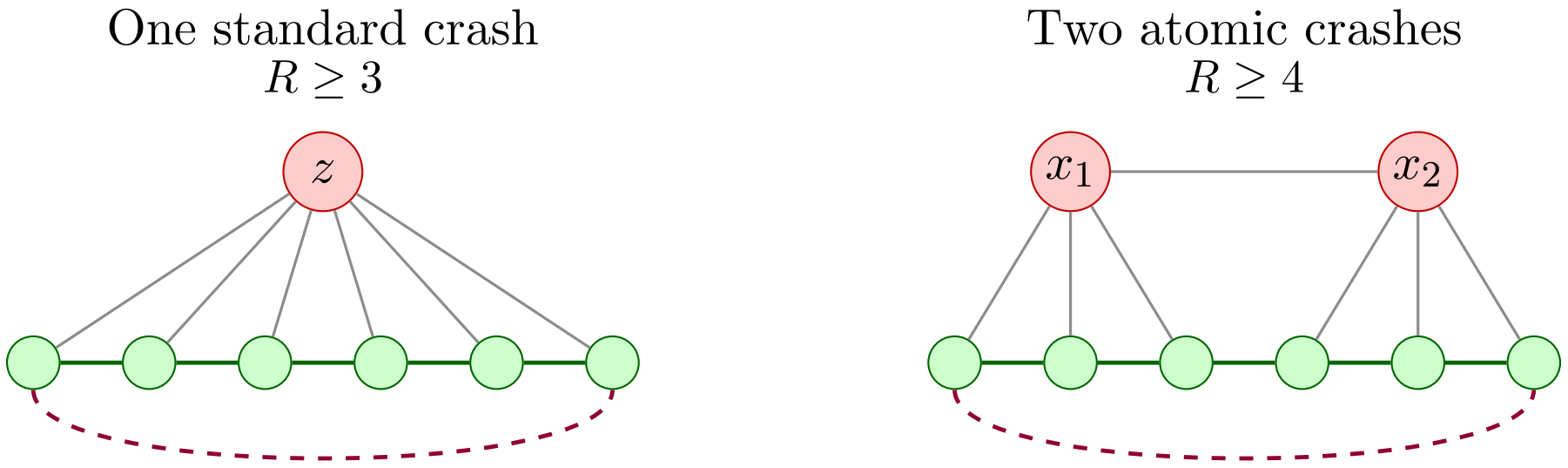

We prove both parts by instantiating the graphs $G^+$ and $G^-$ of the framework of \Cref{sec:lb-framework} with $C = \emptyset$, as illustrated in \Cref{fig:lb-linear}. The set $F$ consists of a single node in the standard crash model and of two nodes in the atomic crash model, which together are adjacent to all of $S$. Through $F$, every node of $S$ is within constant distance of every other node, so the eccentricity condition of \Cref{lem:lb-standard-chain,lem:lb-atomic-chain} holds already for small $R$. Without $F$, however, $S$ induces a path or cycle on almost all $n$ nodes, in which distances are linear in $n$. Since conditions~(a) and~(b) of both lemmas only concern distances in $G^+[S]$, they hold with $T=\Theta(n)$.

\paragraph{One standard crash.}
Here, a single crashing node $z$ serves as a shortcut between all nodes of $S$. Let $C=\emptyset$, $F=\{z\}$, and $S=\{p_1,\ldots,p_{n-1}\}$, so that $|S|=n-1\geq8$. We join $z$ to every node of $S$, and obtain $G^-$ by adding the path $p_1,\dots,p_{|S|}$, and $G^+$ by additionally adding the closing edge $\{a,b\}:=\{p_1,p_{|S|}\}$. We now show that these graphs satisfy the requirements of \Cref{lem:lb-standard-chain} with $T=\lfloor (|S|-1)/2\rfloor-1$.

\begin{proof}[Proof of \Cref{lem:lb-linear}(a)]
The construction has $n$ nodes, and both graphs have diameter at most two, since any two nodes are joined through $z$. Hence, every node of $S$ has eccentricity at most $2\leq R-1$ in $G^+$. We apply \Cref{lem:lb-standard-chain} with $T:=\lfloor (|S|-1)/2\rfloor-1=\lfloor n/2\rfloor-2$. For condition~(a), every $w\in S$ has a node at distance $\lfloor |S|/2\rfloor\geq T$ from it in the cycle $G^+[S]$. For condition~(b), the endpoints $p_1$ and $p_{|S|}$ are adjacent in this cycle, so some node has distance $\lfloor(|S|-1)/2\rfloor>T$ from both of them. Hence, by \Cref{lem:lb-standard-chain}, no $T$-round algorithm tolerating one standard crash solves $\CN(R)$ on both graphs.
\end{proof}

\paragraph{Two atomic crashes.}
Under atomic crashes, a single node adjacent to all of $S$ does not satisfy condition~(a) of \Cref{lem:lb-atomic-chain}, which requires, for every crashing node, a node of $S$ far from all of its neighbors. We therefore split $z$ into two adjacent nodes, each adjacent to one half of $S$. Let $C=\emptyset$, $F=\{x_1,x_2\}$, and $S=\{p_1,\ldots,p_{n-2}\}$, so that $|S|=n-2\geq8$. We join $x_1$ and $x_2$ by an edge, join $x_1$ to $p_1,\ldots,p_{\lfloor |S|/2\rfloor}$, and join $x_2$ to the remaining nodes of $S$. As before, we obtain $G^-$ by adding the path $p_1,\dots,p_{|S|}$, and $G^+$ by additionally adding the closing edge $\{a,b\}:=\{p_1,p_{|S|}\}$. The edge $\{x_1,x_2\}$ keeps the threshold at $R\geq4$: replacing it by a path through an additional node in $C$ would increase the diameter to four, and the eccentricity condition would then require $R\geq5$. We now show that these graphs satisfy the requirements of \Cref{lem:lb-atomic-chain} with $T=\lfloor |S|/4\rfloor$.

\begin{proof}[Proof of \Cref{lem:lb-linear}(b)]
The construction has $n$ nodes, and both graphs have diameter at most three, since any two nodes of $S$ are joined through $x_1$, $x_2$, or both. Hence, every node of $S$ has eccentricity at most $3\leq R-1$ in $G^+$. We apply \Cref{lem:lb-atomic-chain} with $T:=\lfloor |S|/4\rfloor$ and crash bound $f\geq2=|F|$. For condition~(a), each set $N_{G^+}(x_i)\cap S$ is a contiguous block of the cycle $G^+[S]$, and the complementary block contains at least $\lfloor |S|/2\rfloor$ nodes. Its middle node therefore has distance at least $\lceil\lfloor |S|/2\rfloor/2\rceil\geq T$ from $N_{G^+}(x_i)\cap S$. For condition~(b), as before, some node has distance $\lfloor(|S|-1)/2\rfloor>T$ from $p_1$ and $p_{|S|}$. Hence, by \Cref{lem:lb-atomic-chain}, no $T$-round algorithm solves $\CN(R)$ on both graphs with crash bound $f$ under atomic crashes.
\end{proof}

\subsection{An Exponential Lower Bound in the Radius}
\label{sec:lb-degree}

We next focus on \Cref{thm:lower-bound}(ii), which concerns graphs of maximum degree at most $\Delta$: on such graphs, any algorithm for $\CN(R)$ may need a number of rounds exponential in $R$. We prove the bounds under atomic crashes whenever $f\geq2$. The additional case $R=4$ and $f=1$ uses the standard crash-fault model.

The constructions in \Cref{sec:lb-linear} use crashing nodes whose degrees grow linearly with $n$. We now seek lower bounds for graphs of maximum degree at most $\Delta$, where $\Delta$ may be fixed independently of $n$. The eccentricity condition requires every node of $S$ to be within distance $R-1$ of every node of the graph. For $\Delta\geq3$, the degree bound therefore limits the graph to $O((\Delta-1)^R)$ nodes. Our goal is to make the cycle $G^+[S]$ as long as possible subject to these degree and eccentricity constraints.

\begin{figure}[t]
\centering
\begin{tikzpicture}[
    x=.85cm,y=1.05cm,
    cncenter/.style={circle,draw=blue!65!black,fill=blue!18,
      minimum size=5.5mm,inner sep=0pt},
    cnfailure/.style={circle,draw=red!75!black,fill=red!20,
      minimum size=5.5mm,inner sep=0pt},
    cnpath/.style={circle,draw=green!45!black,fill=green!20,
      minimum size=4.7mm,inner sep=0pt},
    treeedge/.style={draw=black!55,line width=.65pt},
    pathedge/.style={draw=green!40!black,line width=1pt}
]
\node[cncenter] (c0) at (0,2.8) {};
\node[cncenter] (c1) at (-3.2,1.9) {};
\node[cncenter] (c2) at (3.2,1.9) {};
\foreach \i/\x in {1/-4.8,2/-1.6,3/1.6,4/4.8}
  \node[cnfailure] (f\i) at (\x,1) {};
\foreach \i/\x in {1/-5.6,2/-4,3/-2.4,4/-.8,5/.8,6/2.4,7/4,8/5.6}
  \node[cnpath] (p\i) at (\x,0) {};
\draw[treeedge] (c0)--(c1) (c0)--(c2)
  (c1)--(f1) (c1)--(f2) (c2)--(f3) (c2)--(f4)
  (f1)--(p1) (f1)--(p2) (f2)--(p3) (f2)--(p4)
  (f3)--(p5) (f3)--(p6) (f4)--(p7) (f4)--(p8);
\foreach \i/\j in {1/2,2/3,3/4,4/5,5/6,6/7,7/8}
  \draw[pathedge] (p\i)--(p\j);
\draw[purple!75!black,line width=1pt,dashed,rounded corners=5pt]
  (p1.south)--(-5.6,-.65)--(5.6,-.65)--(p8.south);
\node[anchor=west,text=blue!65!black] at (5.95,2.25) {$C$};
\node[anchor=west,text=red!75!black] at (5.95,1) {$F$};
\node[anchor=west,text=green!40!black] at (5.95,0) {$S$};
\node[font=\small,fill=white,inner sep=1pt] at (0,-.65)
  {closing edge};
\end{tikzpicture}
\caption{The tree construction for the $\Omega((\Delta-1)^h)$ bound in \Cref{thm:lower-bound}(ii), illustrated with $\Delta=3$ and $h=3$. The blue nodes form $C$, the red nodes form $F$ and are the only ones that may crash, and the $2^h=8$ green leaves form $S$. The dashed closing edge is present only in $G^+$. Each node of $F$ is adjacent to a consecutive block of $\Delta-1$ nodes of $S$, which is short compared to the surviving path or cycle.}
\label{fig:lb-degree-tree}
\end{figure}
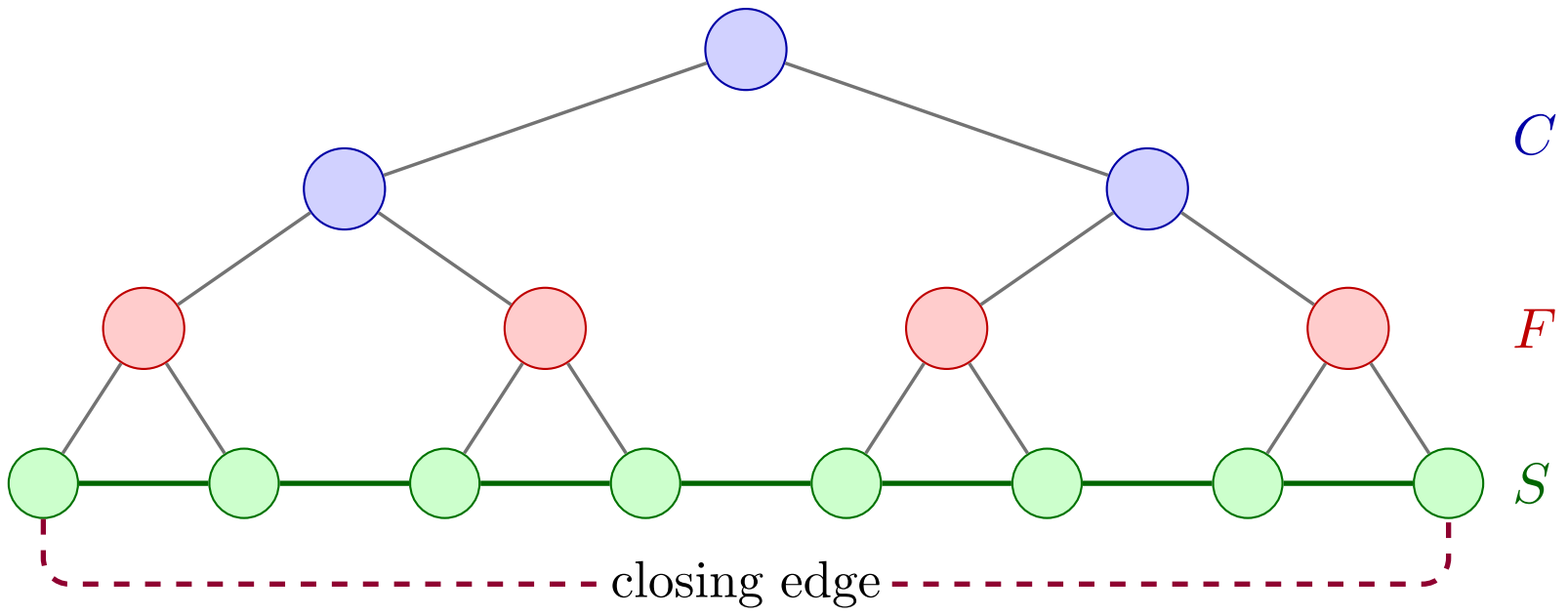

We give two constructions satisfying the degree and eccentricity constraints. The first uses a tree whose leaves form $S$: the tree provides short paths between these nodes before crashes, while the leaves form a long cycle in $G^+$ (\Cref{fig:lb-degree-tree}). This construction yields a lower bound of $\Omega((\Delta-1)^{\lfloor(R-1)/2\rfloor})$ rounds.

To obtain a stronger bound, we use a graph of small diameter as a \emph{core}. We insert crashing nodes into selected edges of the core and connect each of them to a distinct node of $S$ (\Cref{fig:lb-sparse-core}). Choosing these edges far apart limits the increase in distances caused by the insertions. This allows a longer cycle while preserving the eccentricity condition, improving the exponent from $\lfloor(R-1)/2\rfloor$ to $R-O_\Delta(\sqrt R)$.

\subsubsection{The tree construction}

\begin{lemma}[Tree bound]\label{lem:lb-tree}
Let $\Delta\geq3$ and $R\geq4$, let $h:=\lfloor(R-1)/2\rfloor$, and let $n\geq4\cdot(\Delta-1)^h$ and $f\geq(\Delta-1)^{h-1}$. Every deterministic algorithm that solves $\CN(R)$ on all graphs with at most $n$ nodes and maximum degree at most $\Delta$ under at most $f$ standard crashes needs at least $(\Delta-1)^h/8$ rounds. The same lower bound holds under atomic crashes if $f\geq2$.
\end{lemma}

We start with a complete rooted $(\Delta-1)$-ary tree of height $h$. Let $S$ be its leaves, let $F$ be the parents of the leaves, and let $C$ consist of all remaining nodes.

For $R\geq5$, we have $h\geq2$, so $F$ contains at least two nodes. We treat the case $R=4$ separately below. As depicted in \Cref{fig:lb-degree-tree}, we obtain $G^-$ by adding an edge between every two consecutive leaves in left-to-right order, so that $G^-[S]$ is a path, and $G^+$ by additionally adding the closing edge between the first and the last leaf. The root has degree $\Delta-1$, every other internal node of the tree has degree $\Delta$, and every leaf has degree at most three, so both $G^-$ and $G^+$ have maximum degree at most $\Delta$. The construction has $|S|=(\Delta-1)^h$, $|F|=(\Delta-1)^{h-1}$, and, by the geometric-series formula, $|V|=((\Delta-1)^{h+1}-1)/(\Delta-2)\leq2\cdot (\Delta-1)^h$ nodes. Moreover, both graphs contain the original tree, in which any two nodes are joined by a path of length at most $2h\leq R-1$. The additional edges cannot increase distances, so every node of $S$ has eccentricity at most $R-1$ in both $G^-$ and $G^+$.

For $R=4$, we have $h=1$, and the tree is a single parent adjacent to all leaves, which does not suffice under atomic crashes, as discussed in \Cref{sec:lb-linear}. We therefore use the two-crash construction of \Cref{sec:lb-linear} with $|S|=2(\Delta-1)$: we have $F=\{x_1,x_2\}$, $C=\emptyset$, and $S=\{p_1,\ldots,p_{2(\Delta-1)}\}$, where $x_1$ is adjacent to $p_1,\ldots,p_{\Delta-1}$ and $x_2$ to $p_{\Delta},\ldots,p_{2(\Delta-1)}$. The nodes $x_1$ and $x_2$ have degree $\Delta$, every node of $S$ has degree at most three, and the construction has $2\Delta$ nodes. Any two leaves are joined by a path of length at most three through $x_1$, $x_2$, or both, so every node of $S$ has eccentricity at most $3=R-1$ in $G^+$.

In both atomic-crash constructions, every node of $F$ has exactly $\Delta-1$ neighbors in $S$, which form a consecutive block of the path $G^-[S]$, and $|F|\geq2$. We first verify the requirements of \Cref{lem:lb-atomic-chain} for these graphs, then treat $f=1$ in the standard crash-fault model.

\begin{proof}[Proof of \Cref{lem:lb-tree}]
First, suppose $f\geq2$, and consider the constructions above. We verify that $G^-$ and $G^+$ satisfy the lemma statement's requirements, i.e., that they have maximum degree at most $\Delta$ and at most $n$ nodes. The degree bound holds by construction. For the number of nodes, we have $|V|\leq2\cdot (\Delta-1)^h\leq n$ if $R\geq5$, and $|V|=2\Delta\leq4(\Delta-1)\leq n$ if $R=4$, by the assumption on $n$.

Next, we verify the assumptions of \Cref{lem:lb-atomic-chain}. The crash bound is sufficient because $|F|=(\Delta-1)^{h-1}\leq f$ if $R\geq5$, and $|F|=2\leq f$ if $R=4$. Moreover, as shown above, every node of $S$ has eccentricity at most $R-1$ in $G^+$.

Both constructions have $|S|\geq4$, so we can set $T:=\lfloor|S|/4\rfloor-1\geq0$. We subtract one because condition~(b) requires a distance strictly greater than $T$.

For condition~(a), the sets $N_{G^+}(x)\cap S$, for $x\in F$, partition $S$ into consecutive blocks of size $\Delta-1$. Since $|F|\geq2$, each block contains at most half the nodes of the cycle $G^+[S]$. Its complementary block therefore contains at least $|S|/2$ nodes, and a middle node of this complement is at distance at least $\lceil|S|/4\rceil\geq T$ from the original block in $G^+[S]$. Thus, condition~(a) holds for every $x\in F$.

For condition~(b), consider the endpoints of the closing edge. In $G^+[S]$, a middle node of the path $G^-[S]$ is at distance at least $\lfloor(|S|-1)/2\rfloor$ from both endpoints. This distance is at least $\lfloor|S|/4\rfloor=T+1>T$, as required.

All assumptions of \Cref{lem:lb-atomic-chain} now hold, so no deterministic $T$-round algorithm solves $\CN(R)$ on both graphs under atomic crashes. Hence, the worst-case round complexity is at least $T+1=\lfloor|S|/4\rfloor$ rounds. Since $|S|\geq4$, rounding $|S|/4$ down loses at most a factor of two. Together with $|S|\geq(\Delta-1)^h$, this gives $\lfloor|S|/4\rfloor\geq|S|/8\geq(\Delta-1)^h/8.$ Atomic crashes are a special case of standard crashes, so this proves the bound in both models for $f\geq2$.

It remains to consider $f=1$. Since $(\Delta-1)^{h-1}\leq f$ and $\Delta\geq3$, we have $h=1$, and hence $R=4$. We use the construction for a single crash from \Cref{sec:lb-linear} with $|S|=\Delta$: the $\Delta$ nodes of $S$ form a path in $G^-$ and a cycle in $G^+$, and $z$ is adjacent to every node of $S$. Here, $C=\emptyset$ and $F=\{z\}$. Both graphs have maximum degree at most $\Delta$ and $\Delta+1\leq4(\Delta-1)\leq n$ nodes. Every node of $S$ has eccentricity at most two in $G^+$, so the eccentricity condition holds for $R=4$.

Set $T:=\lfloor(\Delta-1)/4\rfloor$. For condition~(a) of \Cref{lem:lb-standard-chain}, every $w\in N_{G^+}(z)\cap S$ has a node at distance $\lfloor\Delta/2\rfloor\geq T$ from it in the cycle $G^+[S]$. For condition~(b), a middle node of the path $G^-[S]$ is at distance at least $\lfloor(\Delta-1)/2\rfloor>T$ from each endpoint of the closing edge, since $\Delta\geq3$. Thus, \Cref{lem:lb-standard-chain} applies, including when $T=0$. The worst-case round complexity is therefore at least $T+1>(\Delta-1)/4\geq(\Delta-1)^h/8$ in the standard crash-fault model. This completes the proof.
\end{proof}

\subsubsection{The sparse-core construction}
We now turn to the stronger bound. In the tree construction, $S$ consists of the leaves of the tree itself, and this loses a factor of two in the exponent: we ensure the eccentricity condition by choosing the tree height so that $2h \le R - 1$, so it has only $|S|=(\Delta-1)^h$ leaves with $h=\lfloor(R-1)/2\rfloor$. The following lemma shows that this factor can be avoided.

\begin{lemma}[Sparse-core bound]\label{lem:lb-sparse-core}
For every integer $\Delta\geq3$, there exists an integer $R_0(\Delta)\geq4$ such that for every $R\geq R_0(\Delta)$, every $n\geq(\Delta-1)^R$, and every $f\geq\lceil(\Delta-1)^R/\Delta\rceil$, the following holds: every deterministic algorithm that solves $\CN(R)$ on all graphs with at most $n$ nodes and maximum degree at most $\Delta$ under at most $f$ atomic crashes needs at least $(\Delta-1)^{R-3\lceil\sqrt R\rceil}$ rounds.
\end{lemma}

To obtain a larger set $S$, we build our construction around a $\Delta$-regular graph $H$ of diameter at most $D$, which we call the \emph{core}. By the following lemma, such a graph exists with $N=\Omega_\Delta((\Delta-1)^D/D)$ nodes.

\begin{lemma}[Small-diameter regular cores]\label{lem:lb-regular-core}
For every fixed integer $\Delta\geq3$, there are constants $c_\Delta>0$ and $D_0(\Delta)$ such that, for every integer $D\geq D_0(\Delta)$, there is a simple $\Delta$-regular graph $H$ of diameter at most $D$ with $N$ nodes, where $c_\Delta\cdot(\Delta-1)^D/D\leq N\leq(\Delta-1)^D$.
\end{lemma}
\begin{proof}
We use the diameter theorem of Bollob\'as and Fernandez de la Vega~\cite{BollobasFernandez1982}, taking $\varepsilon=1$ in \cite[Theorem~10.14]{Bollobas2001RandomGraphs}. For fixed $\Delta\geq3$, this theorem implies that, for all sufficiently large $N$ with $\Delta\cdot N$ even, a simple $\Delta$-regular graph with $N$ nodes and diameter at most $D$ exists whenever $(\Delta-1)^{D-1}\geq3\cdot\Delta\cdot N\cdot\ln N$. Let $N$ be the largest even integer not exceeding $A:=(\Delta-1)^{D-1}/(6\cdot\Delta\cdot D\cdot\ln(\Delta-1))$. For sufficiently large $D$, we have $A/2\leq N\leq A\leq (\Delta-1)^D$, and hence $\ln N\leq D\cdot\ln(\Delta-1)$. Therefore, $3\cdot\Delta\cdot N\cdot\ln N\leq3\cdot\Delta\cdot A\cdot D\cdot\ln(\Delta-1)=(\Delta-1)^{D-1}/2$, so the diameter theorem applies. Moreover, $N\geq A/2=(\Delta-1)^D/(12\cdot\Delta\cdot(\Delta-1)\cdot\ln(\Delta-1)\cdot D)$, which gives the claim with $c_\Delta:=1/(12\cdot\Delta\cdot(\Delta-1)\cdot\ln(\Delta-1))$.
\end{proof}

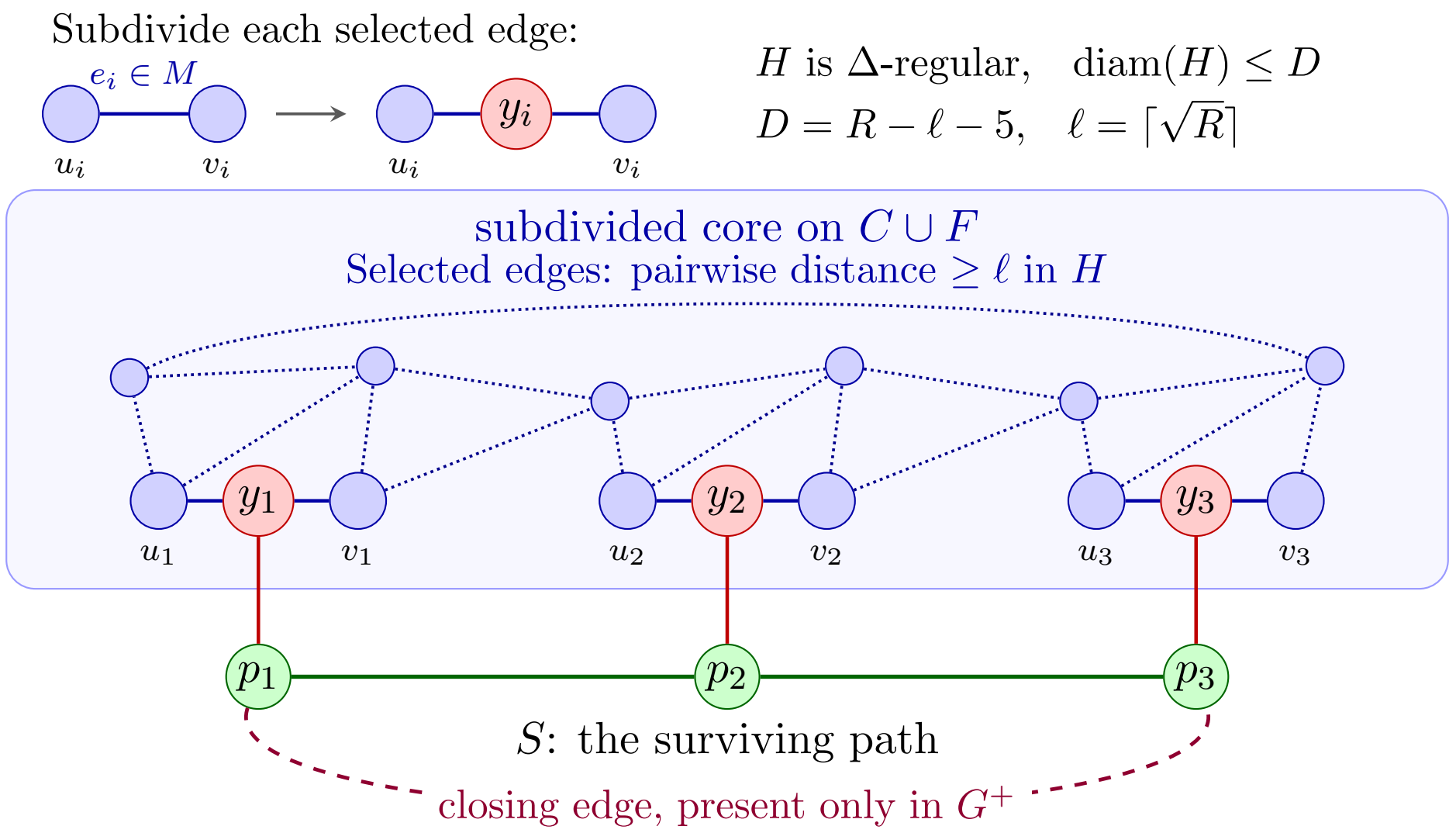
\begin{figure}[t]
\centering
\begin{tikzpicture}[x=1cm,y=1cm,
  center/.style={circle,draw=blue!65!black,fill=blue!18,
    minimum size=4.8mm,inner sep=0pt},
  interior/.style={center,minimum size=3.2mm},
  failing/.style={circle,draw=red!75!black,fill=red!20,
    minimum size=6mm,inner sep=1pt},
  survivor/.style={circle,draw=green!40!black,fill=green!20,
    minimum size=5.5mm,inner sep=1pt},
  coreedge/.style={blue!65!black,line width=.85pt},
  coreroute/.style={blue!55!black,densely dotted,line width=.7pt}]
  \node[font=\small,anchor=west] at (0,5.5)
    {Subdivide each selected edge:};
  \node[center,label={[font=\scriptsize]below:$u_i$}] (oldU) at (.3,4.8) {};
  \node[center,label={[font=\scriptsize]below:$v_i$}] (oldV) at (1.55,4.8) {};
  \draw[coreedge] (oldU)--(oldV)
    node[midway,above=2pt,font=\scriptsize] {$e_i\in M$};
  \draw[->,>=stealth,black!65,line width=.7pt] (2.05,4.8)--(2.65,4.8);
  \node[center,label={[font=\scriptsize]below:$u_i$}] (newU) at (3.15,4.8) {};
  \node[failing] (newF) at (4.1,4.8) {$y_i$};
  \node[center,label={[font=\scriptsize]below:$v_i$}] (newV) at (5.05,4.8) {};
  \draw[coreedge] (newU)--(newF)--(newV);
  \node[font=\small,anchor=west,align=left] at (6.0,4.95)
    {$H$ is $\Delta$-regular,\quad $\diam(H)\leq D$\\[3pt]
     $D=R-\ell-5$,\quad $\ell=\lceil\sqrt R\rceil$};

  \draw[blue!40,fill=blue!3,rounded corners=7pt]
    (-.25,.75) rectangle (12.05,4.15);
  \node[text=blue!65!black] at (5.9,3.85)
    {subdivided core on $C\cup F$};
  \node[font=\small,text=blue!65!black] at (5.9,3.45)
    {Selected edges: pairwise distance $\geq\ell$ in $H$};
  \foreach \i/\x/\y in {1/.8/2.55,2/2.9/2.65,3/4.9/2.35,
                         4/6.9/2.65,5/8.9/2.35,6/11/2.65}
    \node[interior] (h\i) at (\x,\y) {};
  \draw[coreroute]
    (h1)--(h2)--(h3)--(h4)--(h5)--(h6)
    (h1) .. controls (2.3,3.35) and (9.5,3.35) .. (h6);
  \foreach \i/\x in {1/1.9,2/5.9,3/9.9} {
    \node[center,label={[font=\scriptsize]below:$u_{\i}$}]
      (u\i) at ({\x-.85},1.5) {};
    \node[center,label={[font=\scriptsize]below:$v_{\i}$}]
      (v\i) at ({\x+.85},1.5) {};
    \node[failing] (f\i) at (\x,1.5) {$y_{\i}$};
    \draw[coreedge] (u\i)--(f\i)--(v\i);
    \node[survivor] (p\i) at (\x,0) {$p_{\i}$};
    \draw[red!75!black,line width=.85pt] (f\i)--(p\i);
  }
  \draw[coreroute]
    (u1)--(h1) (u1)--(h2) (v1)--(h2) (v1)--(h3)
    (u2)--(h3) (u2)--(h4) (v2)--(h4) (v2)--(h5)
    (u3)--(h5) (u3)--(h6) (v3)--(h6);
  \draw[green!40!black,line width=1pt] (p1)--(p2)--(p3);
  \node[below] at (5.9,-.25) {$S$: the surviving path};
  \draw[purple!75!black,dashed,line width=.9pt]
    (p1) .. controls (1.4,-1.45) and (10.4,-1.45) .. (p3);
  \node[fill=white,text=purple!75!black,font=\small] at (5.9,-1.1)
    {closing edge, present only in $G^+$};
\end{tikzpicture}
\caption{The sparse-core construction for the stronger
$(\Delta-1)^{R-3\lceil\sqrt R\rceil}$ bound in \Cref{thm:lower-bound}(ii).
The inset shows how a selected edge $e_i=\{u_i,v_i\}$ of the
$\Delta$-regular core $H$ is replaced by $u_i,y_i,v_i$; the original edge
is removed. Each red $y_i\in F$ also receives one green neighbor
$p_i\in S$. Blue nodes form $C$ and retain degree $\Delta$.
Dotted lines represent routes with omitted core nodes and edges;
neither their drawn lengths nor the depicted degrees are literal.
The selected edges are pairwise at distance at least $\ell$ in $H$,
so shortest core paths encounter few subdivisions. Three of the
$|F|=|S|=k$ attachment sites are shown. Only red nodes may crash;
after they do, $S$ remains a path, or a cycle with the dashed closing edge.}
\label{fig:lb-sparse-core}
\end{figure}

The nodes of $H$ form the set $C$, and we select a set $M$ of edges of $H$ as \emph{attachment sites}. Since $H$ is $\Delta$-regular, we cannot connect the nodes of $F$ to $C$ by additional edges, so we subdivide the selected edges instead. More precisely, for each edge $e=\{u,v\}\in M$, we add a new node $y_e$ to $F$ and a new node $p_e$ to $S$, replace $e$ by the path $u,y_e,v$, and add the edge $\{y_e,p_e\}$, as shown in \Cref{fig:lb-sparse-core}. We obtain $G^-$ by adding a path through all nodes of $S$, in an arbitrary order, and $G^+$ by additionally adding the closing edge between its endpoints. This way, every selected edge contributes exactly one node of $F$ and one node of $S$, so $|S|=|M|$. Moreover, there are no edges between $C$ and $S$.

It remains to decide which edges of $H$ to select. Each subdivision makes paths through that edge one step longer: if we subdivided every edge of $H$, all distances in $H$ would double. Bounding eccentricities using paths through the subdivided core would then require choosing $D$ to be about $R/2$. Instead, we select edges of $H$ that are pairwise at distance at least $\ell\approx\sqrt R$. Then, a shortest path in $H$ passes through only $O(D/\ell)=O(\sqrt R)$ selected edges, so distances grow by only $O(\sqrt R)$, and we can choose $D=R-O(\sqrt R)$.

We select edges greedily, keeping distance at least $\ell$ between them. Once no further edge can be selected, every node of $H$ lies within distance $\ell$ of some selected edge. Each selected edge has only $O((\Delta-1)^\ell)$ nodes this close to it. Thus, covering all $N$ nodes requires $\Omega(N/(\Delta-1)^\ell)$ selected edges.

Since $\ell=\lceil\sqrt R\rceil$, this separation reduces the exponent by only $O(\sqrt R)$. Choosing the smaller core diameter $D=R-O(\sqrt R)$ causes another reduction of the same order. Together, these give $|S|=(\Delta-1)^{R-O_\Delta(\sqrt R)}$.

\paragraph{Choosing the parameters and attachment sites.}
We now make this precise. Fix $\Delta\geq3$, and assume that $R$ is sufficiently large, depending on $\Delta$. We set $\ell:=\lceil\sqrt R\rceil$, the minimum distance between attachment sites, and $D:=R-\ell-5$, the upper bound on the core's diameter. This choice of $D$ leaves room for the extra distance caused by the subdivisions and by the nodes of $S$, so that every node of $S$ has eccentricity at most $R-1$, as we show below. Since $R$ is large, so is $D$, and \Cref{lem:lb-regular-core} provides a $\Delta$-regular graph $H$ of diameter at most $D$ with $N$ nodes, where $c_\Delta\cdot (\Delta-1)^D/D\leq N\leq (\Delta-1)^D$.

To select the attachment sites, we define the distance between edges $e,e'$ of $H$ as $\dist_H(e,e'):=\min\{\dist_H(u,v):u\in e,\ v\in e'\}$. Then, we choose an inclusion-maximal set $M$ of edges of $H$ with pairwise distance at least $\ell$, and write $k:=|M|=|F|=|S|$. Since two edges sharing an endpoint have distance $0<\ell$, the edges of $M$ form a matching. By maximality, every node of $H$ lies within distance $\ell$ of an endpoint of some edge of $M$: every node has an incident edge, and every edge outside $M$ has an endpoint at distance less than $\ell$ from an endpoint of an edge in $M$. Moreover, at most $2\cdot\sum_{j=0}^{\ell}(\Delta-1)^j\leq4\cdot (\Delta-1)^\ell$ nodes lie within distance $\ell$ of the endpoints of a single edge: in a breadth-first search from the two endpoints, each endpoint has at most $\Delta-1$ neighbors other than the other endpoint, and each later node has at most $\Delta-1$ new neighbors. Consequently,
\begin{equation}\label{eq:lb-sparse-packing}
    k\geq\frac{N}{4\cdot (\Delta-1)^\ell}
      \geq\frac{c_\Delta\cdot (\Delta-1)^{D-\ell}}{4D}.
\end{equation}

We apply the construction described above with this set $M$ of attachment sites. We now show that the resulting graphs $G^-$ and $G^+$ satisfy the requirements of \Cref{lem:lb-atomic-chain} with $T=\Omega(k)$, hence proving \Cref{lem:lb-sparse-core}.

\begin{proof}[Proof of \Cref{lem:lb-sparse-core}]
Fix an integer $\Delta\geq3$. Choose an integer $R_0(\Delta)\geq4$ such that, for every $R\geq R_0(\Delta)$, the core construction above is available, the lower bound on $k$ in \eqref{eq:lb-sparse-packing} is at least eight, and
\[
    \frac{c_\Delta}{32}\cdot(\Delta-1)^{\lceil\sqrt R\rceil-5}\geq R.
\]
Such a choice is possible by \Cref{lem:lb-regular-core}, \eqref{eq:lb-sparse-packing}, and the fact that $(\Delta-1)^{\lceil\sqrt R\rceil}/R\to\infty$ as $R\to\infty$. The threshold depends only on $\Delta$. Fix $R\geq R_0(\Delta)$, and let $n$ and $f$ satisfy the assumptions of the lemma.

We first check that $G^-$ and $G^+$ have maximum degree at most $\Delta$ and at most $n$ nodes. The nodes of $C$ keep degree $\Delta$, every node of $F$ has degree three, and every node of $S$ has degree at most three. Since $M$ is a matching, $k\leq N/2$, so the construction has $N+2k\leq2N\leq2\cdot (\Delta-1)^D\leq (\Delta-1)^R\leq n$ nodes. The bound $2(\Delta-1)^D\leq(\Delta-1)^R$ follows from $R-D=\ell+5$ and $(\Delta-1)^{\ell+5}\geq2$. The final inequality uses the assumption $n\geq(\Delta-1)^R$.

Next, we verify the requirements of \Cref{lem:lb-atomic-chain}. First, the construction respects the crash bound, since $|F|=k\leq(\Delta-1)^D/2\leq(\Delta-1)^R/\Delta\leq f.$ The first inequality follows from $k\leq N/2$ and $N\leq(\Delta-1)^D$. The second uses $R-D=\ell+5$ and $(\Delta-1)^{\ell+5}\geq\Delta/2$, while the last follows from the assumption on $f$.

For the eccentricity condition, we first bound distances between nodes of $C$. Consider a shortest path in $H$ of length $\lambda\leq D$, containing $\kappa$ selected edges. Subdividing these edges increases its length to $\lambda+\kappa$. If $\kappa\geq1$, the separation between consecutive selected edges gives $\lambda\geq\kappa+(\kappa-1)\ell$. Thus, $\kappa\leq\lceil\lambda/(\ell+1)\rceil$, which also holds when $\kappa=0$. Consequently, any two nodes of $C$ have distance at most $D+\lceil D/(\ell+1)\rceil\leq D+\ell$ in $G^+$, where the last inequality follows from $D\leq R\leq\ell^2$.

Finally, every node of the construction is within distance two of $C$. We can therefore connect any node of $S$ to any other node by reaching $C$, following a path between two nodes of $C$, and leaving $C$. The resulting distance is at most $2+(D+\ell)+2=R-1$. Hence, every node of $S$ has eccentricity at most $R-1$ in $G^+$.

It remains to verify conditions~(a) and~(b) of \Cref{lem:lb-atomic-chain}. The graph $G^+[S]$ is a cycle on $k$ nodes, and each $y_e$ has exactly one neighbor $p_e$ in $S$. The distances are therefore as in the one-standard-crash construction of \Cref{sec:lb-linear}: for~(a), $p_e$ has a node at distance $\lfloor k/2\rfloor$ in $G^+[S]$, and for~(b), some node has distance $\lfloor(k-1)/2\rfloor$ from the endpoints of the closing edge. By the choice of $R_0(\Delta)$, we have $k\geq8$, so both conditions hold with $T:=\lfloor k/4\rfloor$. By \Cref{lem:lb-atomic-chain}, no deterministic $T$-round algorithm solves $\CN(R)$ on both graphs under atomic crashes, so every algorithm needs more than $\lfloor k/4\rfloor\geq k/8$ rounds.

Finally, we express this bound in terms of $R$. Substituting $D=R-\ell-5$ into \eqref{eq:lb-sparse-packing} and using $D\leq R$, every algorithm needs at least $(c_\Delta/32)\cdot (\Delta-1)^{R-2\lceil\sqrt R\rceil-5}/R$ rounds. The inequality imposed when choosing $R_0(\Delta)$ makes this at least $(\Delta-1)^{R-3\lceil\sqrt R\rceil}$ rounds.
\end{proof}

\subsection{Dependence on the Crash Bound and Radius}
\label{sec:lb-fault-budget}

We now focus on \Cref{thm:lower-bound}(iii), which shows how the round complexity depends on the crash bound $f$: on graphs of maximum degree at most $\Delta$, any algorithm for $\CN(R)$ may need $\Omega\bigl((R-2\log_{\Delta-1}f)\cdot f\cdot\Delta\bigr)$ rounds.
The following lemma gives bounds for atomic crashes with $f\geq2$ and $f = 1$ for standard crashes.

\begin{lemma}[Lower bounds in terms of $f$, $R$, and $\Delta$]\label{lem:lb-budget}
Let $\Delta\geq3$. The following hold for deterministic algorithms that solve $\CN(R)$ on all graphs with at most $n$ nodes and maximum degree at most $\Delta$:
\begin{enumerate}[(a)]
  \item  Let $f\geq2$, $R\geq2\lceil\log_{\Delta-1}f\rceil+4$, and $n\geq R\cdot f\cdot\Delta$. Every such algorithm that tolerates at most $f$ atomic crashes needs at least $(\Delta-1)\cdot f\cdot\bigl(R-2\lceil\log_{\Delta-1}f\rceil-3\bigr)/4$ rounds.
  \item Let $R\geq4$ and $n\geq R\cdot\Delta$. Every such algorithm that tolerates one standard crash needs more than $\lfloor(\Delta-1)\cdot(R-3)/4\rfloor$ rounds.
\end{enumerate}
\end{lemma}

\begin{figure}[ht]
\centering
\begin{tikzpicture}[x=1cm,y=0.9cm,
  core/.style={circle,draw=blue!65!black,fill=blue!18,inner sep=2.8pt},
  failing/.style={circle,draw=red!75!black,fill=red!20,inner sep=3pt},
  site/.style={circle,draw=green!40!black,fill=green!20,inner sep=2pt}]
  \node[core] (root) at (5.7,3.6) {};
  \node[above] at (5.7,3.8) {core tree: internal nodes $C$, leaves $F$};
  \node[core] (left) at (2.7,2.7) {};
  \node[core] (right) at (8.7,2.7) {};
  \draw[blue!65!black,thick] (root)--(left) (root)--(right);
  \foreach \j/\x in {1/1.2,2/4.2,3/7.2,4/10.2} {
    \node[failing,label=above:{$y_{\j}$}] (f\j) at (\x,1.45) {};
  }
  \draw[blue!65!black,thick]
    (left)--(f1) (left)--(f2) (right)--(f3) (right)--(f4);
  \foreach \i in {0,...,11} {
    \node[site] (a\i) at ({0.2+\i},0) {};
  }
  \foreach \i in {0,...,10} {
    \pgfmathtruncatemacro{\j}{\i+1}
    \draw[green!40!black,thick] (a\i)--(a\j);
  }
  \foreach \j/\first/\last in {1/0/2,2/3/5,3/6/8,4/9/11} {
    \foreach \i in {\first,...,\last} {
      \draw[red!75!black] (f\j)--(a\i);
    }
  }
  \draw[<->,black!65,thin] (2.2,0.23)--(3.2,0.23);
  \node[above,font=\small,fill=white,inner sep=1pt] at (2.7,0.29)
    {$L$ edges};
  \node[below] at (5.7,-0.15) {$S$: the surviving path};
  \node[left] at (0.05,0) {$a_1$};
  \node[right] at (11.35,0) {$a_{f(\Delta-1)}$};
  \draw[purple!75!black,thick,dashed]
    (a0) .. controls (-0.4,-1.75) and (11.8,-1.75) .. (a11);
  \node[fill=white,text=purple!75!black] at (5.7,-1.31)
    {closing edge, present only in $G^+$};
\end{tikzpicture}
\caption{The construction for \Cref{thm:lower-bound}(iii), illustrated with
$\Delta=4$ and $f=4$. Every horizontal link represents a path of $L$ edges;
its omitted internal nodes also belong to $S$. The dashed closing
edge is a single edge, present only in $G^+$. Each
failing node is adjacent to a consecutive block of $\Delta-1$ attachment
sites. After $F$ crashes, the core is disconnected from the surviving path or
cycle, whose length is $\Theta(L\cdot f\cdot \Delta)$.}
\label{fig:lb-spaced-cycle}
\end{figure}
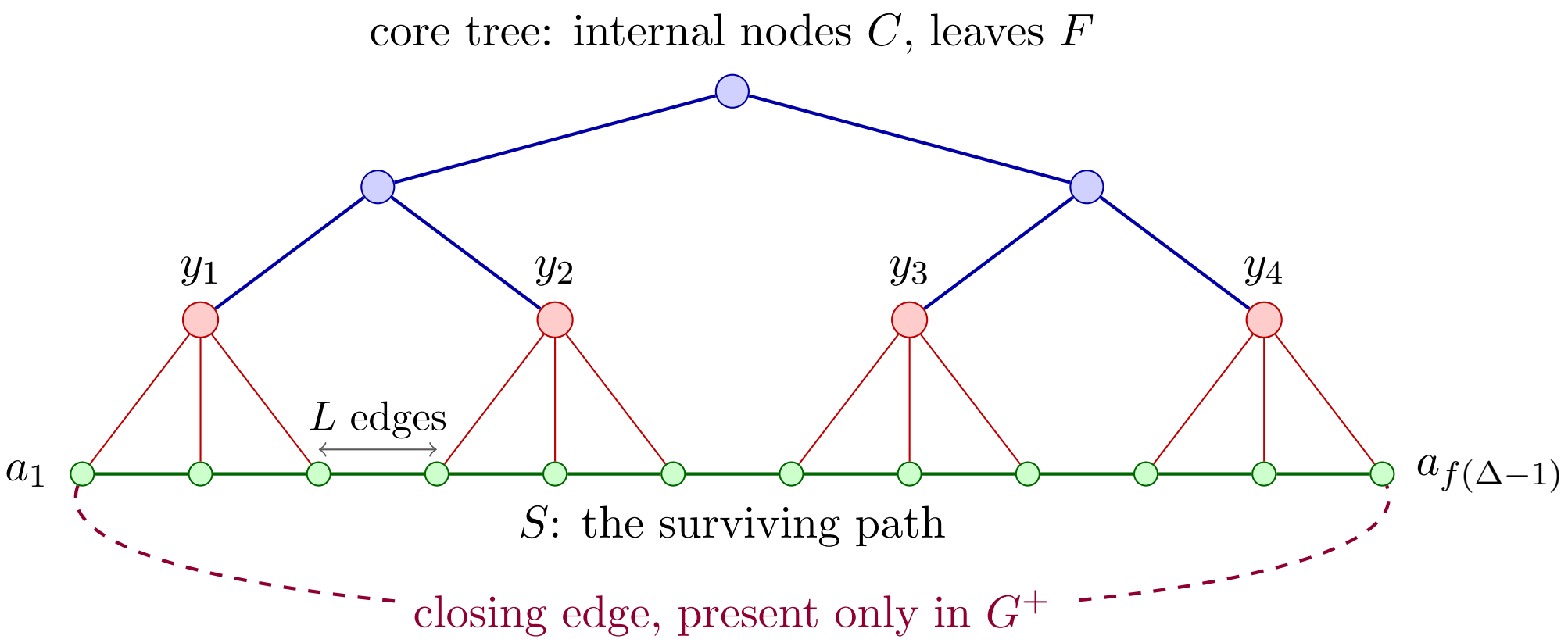

The main constraint is the bound on the number of crashes: since at most $f$ nodes crash and each has degree at most $\Delta$, at most $f\cdot\Delta$ nodes of $S$, which we call \emph{attachment sites}, can be adjacent to $F$. If $S$ consisted only of these sites, the cycle $G^+[S]$ would have length $O(f\cdot\Delta)$, so conditions~(a) and~(b) of \Cref{lem:lb-atomic-chain,lem:lb-standard-chain} could hold only for $T=O(f\cdot\Delta)$, and the resulting lower bound would not grow with $R$.
We therefore place about $R-2\log_{\Delta-1}f$ additional nodes of $S$ between any two consecutive attachment sites, as illustrated in \Cref{fig:lb-spaced-cycle}. Along each subpath between consecutive attachment sites, every node is at distance at most half the subpath's length from an attachment site. Each attachment site is adjacent to a node of $F$, and these nodes are leaves of a tree of diameter at most $2\lceil\log_{\Delta-1}f\rceil$. We choose the subpath lengths so that every node of $S$ has eccentricity at most $R-1$ in $G^+$. This increases the cycle length by a factor of $\Theta(R-2\log_{\Delta-1}f)$.

\subsubsection{Atomic crashes with $f\geq2$}
We first prove the bound for atomic crashes in \Cref{lem:lb-budget}, following the idea above: we connect the $f$ crashing nodes by a small tree, and separate their attachment sites by long subpaths of $S$. To this end, let $h:=\lceil\log_{\Delta-1} f\rceil$, so that the tree has height at most $h$, and let $L:=R-2h-3$ be the number of edges between consecutive attachment sites. The assumption on $R$ in \Cref{lem:lb-budget}(a) gives $R\geq2h+4$, and hence $L\geq1$.

We take a rooted tree of height at most $h$ with exactly $f$ leaves, in which every internal node has at least two and at most $\Delta-1$ children. We first show that such a tree exists: start with the complete $(\Delta-1)$-ary tree of height $h$, keep only the paths from the root to $f$ of its leaves. Repeatedly remove any non-root node with exactly one child and connect its parent directly to that child. If the root has only one child, remove it and make its child the new root. We let $F=\{y_1,\ldots,y_f\}$ be its leaves and $C$ its internal nodes. Since every internal node has at least two children, $|C|\leq f-1$. Furthermore, the tree has height at most $h$, which implies that any two nodes of the tree have distance at most $2h$.

Next, we form a path with $f\cdot (\Delta-1)$ attachment sites $a_1,\ldots,a_{f(\Delta-1)}$ in this order, with exactly $L$ edges between consecutive sites. Let $S$ be the node set of this path, including the nodes between the sites. For each $j\in\{1,\ldots,f\}$, we join $y_j$ to the $\Delta-1$ consecutive sites $a_{(j-1)(\Delta-1)+1},\ldots,a_{j(\Delta-1)}$.
This defines $G^-$, and adding the closing edge $\{a_1,a_{f(\Delta-1)}\}$ gives $G^+$, as illustrated in \Cref{fig:lb-spaced-cycle}. Thus, $G^-[S]$ is a path, $G^+[S]$ is a cycle, and the nodes of $C$ have no neighbors in $S$.

We now prove the bound for atomic crashes in \Cref{lem:lb-budget}, using \Cref{lem:lb-atomic-chain} with $T=\Omega(L\cdot f\cdot\Delta)$.

\begin{proof}[Proof of \Cref{lem:lb-budget}(a)]
We first check that $G^-$ and $G^+$ satisfy the requirements of the lemma statement, i.e., that they have maximum degree at most $\Delta$ and at most $n$ nodes. Every node of $C$ has at most one parent and $\Delta-1$ children, every $y_j$ has one parent and $\Delta-1$ attachment sites as neighbors, and every node of $S$ has degree at most three, so the maximum degree is at most $\Delta$. Since $|S|=(f\cdot (\Delta-1)-1)\cdot L+1$, the number of nodes is $|S|+|F|+|C|\leq L\cdot f\cdot (\Delta-1)+2f\leq R\cdot f\cdot\Delta\leq n$.

Next, we check the requirements of \Cref{lem:lb-atomic-chain}. The lemma applies with crash bound $f=|F|$. For the eccentricity condition, note that every node of $S$ is within distance $\lfloor L/2\rfloor$ of an attachment site. Passing through the tree, any $p,q\in S$ therefore satisfy $\dist_{G^+}(p,q)\leq2\lfloor L/2\rfloor+2+2h\leq L+2h+2=R-1$. Similarly, for every $p\in S$ and $z\in C\cup F$, we have $\dist_{G^+}(p,z)\leq\lfloor L/2\rfloor+1+2h\leq R-1$. This implies that every node of $S$ has eccentricity at most $R-1$ in $G^+$.

It remains to verify conditions~(a) and~(b) of \Cref{lem:lb-atomic-chain}. All distances in this verification are measured in the cycle $G^+[S]$.

For condition~(a), fix $y_j\in F$. Its $\Delta-1$ attachment sites occur consecutively along the cycle and occupy a path with $(\Delta-2)\cdot L$ edges. The other path between the first and last of these sites has $|S|-(\Delta-2)\cdot L=(f-1)\cdot(\Delta-1)\cdot L+1$ edges and has no neighbor of $y_j$ in its interior. Choose a middle node $p$ of this other path. The nearest attachment site of $y_j$ is therefore one of the two endpoints of this other path. Hence,
$$
\dist_{G^+[S]}\bigl(p,N_{G^+}(y_j)\cap S\bigr)=\left\lfloor\frac{(f-1)\cdot(\Delta-1)\cdot L+1}{2}\right\rfloor\geq\frac{f\cdot(\Delta-1)\cdot L}{4},
$$
where the inequality uses $f\geq2$.

For condition~(b), choose a middle node $q$ of the path from $a_1$ to $a_{f(\Delta-1)}$ that avoids the closing edge. This path has $(f\cdot(\Delta-1)-1)\cdot L$ edges, so
$$
\dist_{G^+[S]}\bigl(q,\{a_1,a_{f(\Delta-1)}\}\bigr)=\left\lfloor\frac{(f\cdot(\Delta-1)-1)\cdot L}{2}\right\rfloor\geq\frac{f\cdot(\Delta-1)\cdot L}{4},
$$
where the inequality uses $f\cdot(\Delta-1)\geq4$ and $L\geq1$.

Taking $T:=\lceil f\cdot(\Delta-1)\cdot L/4\rceil-1$ gives $T<f\cdot(\Delta-1)\cdot L/4$, so both conditions hold.

By \Cref{lem:lb-atomic-chain}, no deterministic $T$-round algorithm solves $\CN(R)$ on both graphs under at most $f$ atomic crashes. Hence, the worst-case round complexity is at least $T+1\geq(\Delta-1)\cdot f\cdot L/4$.
\end{proof}

\subsubsection{One standard crash}
We now prove \Cref{lem:lb-budget}(b) in the standard crash-fault model, with $z$ as the only node that may crash.

In the preceding construction for atomic crashes, the tree connects the $f$ nodes that may crash. Here, only $z$ may crash, so no tree is needed. We instead adapt the construction for a single crash from \Cref{sec:lb-linear}: $z$ is adjacent to $\Delta$ attachment sites, with additional nodes placed between consecutive sites along the path.

More precisely, let $C=\emptyset$ and $F=\{z\}$, and let $L:=R-3$ be the number of edges between consecutive attachment sites; the assumption $R\geq4$ gives $L\geq1$. We take $\Delta$ attachment sites on a path with $L$ edges between consecutive sites, let $S$ be the node set of this path, and join $z$ to every attachment site. This defines $G^-$, and adding the closing edge between the first and the last site gives $G^+$.

We now present the proof for one standard crash in \Cref{lem:lb-budget}, using \Cref{lem:lb-standard-chain} with $T=\Omega(R\cdot\Delta)$.

\begin{proof}[Proof of \Cref{lem:lb-budget}(b)]
We first check that $G^-$ and $G^+$ satisfy the requirements in the lemma statement, i.e., that they have maximum degree at most $\Delta$ and at most $n$ nodes. The node $z$ has degree $\Delta$, and every node of $S$ has degree at most three. Since $|S|=(\Delta-1)\cdot L+1$, the construction has $|S|+1\leq R\cdot\Delta\leq n$ nodes.

Next, we check the requirements of \Cref{lem:lb-standard-chain}. Every node of $S$ can reach $z$ by following at most $\lfloor L/2\rfloor$ edges to an attachment site and then its edge to $z$. Thus, any two nodes of $S$ have distance at most $2\lfloor L/2\rfloor+2\leq L+2=R-1$ in $G^+$. Since $z$ is the only node outside $S$, every node of $S$ has eccentricity at most $R-1$ in $G^+$.

For conditions~(a) and~(b), set $T:=\lfloor(|S|-1)/4\rfloor$. All distances in these two conditions are measured in the cycle $G^+[S]$. For condition~(a), fix $w\in N_{G^+}(z)\cap S$ and choose a node $p$ farthest from $w$ along the cycle. Their distance is $\lfloor|S|/2\rfloor\geq T$, as required.

For condition~(b), consider the path between the endpoints of the closing edge that avoids this edge. It has $|S|-1$ edges. Choose a middle node $q$ of this path. Its distance in the cycle to the nearer endpoint is $\lfloor(|S|-1)/2\rfloor>T$, where the strict inequality follows from $|S|\geq3$. Thus, both endpoints are at distance greater than $T$ from $q$.

By \Cref{lem:lb-standard-chain}, no deterministic $T$-round algorithm solves $\CN(R)$ on both graphs when $z$ may crash in the standard crash-fault model. Since $|S|-1=(\Delta-1)\cdot(R-3)$, the worst-case round complexity is greater than $\lfloor(\Delta-1)\cdot(R-3)/4\rfloor$.
\end{proof}
\subsection{Putting It All Together}
\label{sec:lb-together}

We now combine the lemmas of \Cref{sec:lb-linear,sec:lb-degree,sec:lb-fault-budget} into a proof of \Cref{thm:lower-bound}. Recall that atomic crashes are a special case of standard crashes, so every lower bound under atomic crashes also holds in the standard crash-fault model.

\begin{proof}[Proof of \Cref{thm:lower-bound}]
Part~(i) follows from \Cref{lem:lb-linear}. Under a single standard crash, i.e., for $f=1$, part~(a) of the lemma gives a lower bound of more than $\lfloor n/2\rfloor-2=\Omega(n)$ rounds for every $R\geq3$. In the atomic crash-fault model, part~(b) of the lemma gives more than $\lfloor(n-2)/4\rfloor=\Omega(n)$ rounds for every $R\geq4$ and $f=2$. The fixed lower thresholds on $n$ do not affect the asymptotic bounds.

Part~(ii) follows from \Cref{lem:lb-tree,lem:lb-sparse-core}. Under the assumptions on $\Delta$, $R$, $n$, and $f$ in the theorem, \Cref{lem:lb-tree} gives a lower bound of $(\Delta-1)^{\lfloor(R-1)/2\rfloor}/8=\Omega\bigl((\Delta-1)^{\lfloor(R-1)/2\rfloor}\bigr)$ rounds in the standard crash-fault model. The same lemma proves this bound under atomic crashes with the additional assumption $f\geq2$, as required by the theorem.

For each integer $\Delta\geq3$, take $R_0(\Delta)$ from \Cref{lem:lb-sparse-core}. For every $R\geq R_0(\Delta)$, that lemma gives the stronger bound $(\Delta-1)^{R-3\lceil\sqrt R\rceil}$, provided that $n\geq(\Delta-1)^R$ and $f\geq\lceil(\Delta-1)^R/\Delta\rceil$. The threshold is independent of $n$ and $f$. This bound holds under atomic crashes, and therefore also under standard crashes, proving the stronger claim in part~(ii).

Part~(iii) follows from \Cref{lem:lb-budget}. First, suppose $f\geq2$, and let $L:=R-2\lceil\log_{\Delta-1}f\rceil-3\geq1$. By \Cref{lem:lb-budget}(a), the round complexity is at least $(\Delta-1)\cdot f\cdot L/4$ under atomic crashes, and therefore also under standard crashes. To compare this with the bound in the theorem, note that $\lceil\log_{\Delta-1}f\rceil<\log_{\Delta-1}f+1$ gives $R-2\log_{\Delta-1}f<L+5\leq6L$. Together with $\Delta-1\geq2\Delta/3$, this yields the following:
$$
\frac{(\Delta-1)\cdot f\cdot L}{4}\geq\frac{\Delta\cdot f\cdot L}{6}\geq\frac{(R-2\log_{\Delta-1}f)\cdot f\cdot\Delta}{36}.
$$

For $f=1$, \Cref{lem:lb-budget}(b) requires more than $\lfloor(\Delta-1)\cdot(R-3)/4\rfloor$ rounds in the standard crash-fault model. Since the round complexity is an integer, it is at least
$$
\left\lfloor\frac{(\Delta-1)\cdot(R-3)}{4}\right\rfloor+1>\frac{(\Delta-1)\cdot(R-3)}{4}\geq\frac{R\cdot\Delta}{24},
$$
where the last inequality uses $R-3\geq R/4$ for $R\geq4$ and $\Delta-1\geq2\Delta/3$ for $\Delta\geq3$. Since $\log_{\Delta-1}1=0$, this establishes the claimed bound for $f=1$.
\end{proof}

\section{Crash Tolerance for Locally Checkable Labelings}
\label{sec:lcl-complexity}

We prove both parts of \Cref{thm:lcl-crash-complexity}. \Cref{sec:lcl-crash-complexity} focuses on part~(i) and shows that fault-free $O(\log^* n)$ complexity is preserved under crash faults. Afterwards, \Cref{sec:lcl-lower-bound}  constructs an LCL with fault-free complexity $\Theta(\log n)$ and crash-tolerant complexity $\Theta(n)$, proving part~(ii).

\subsection{Preserving Log-Star Complexity}\label{sec:lcl-crash-complexity}

To prove \Cref{thm:lcl-crash-complexity}(i), fix an integer degree bound $\Delta\geq2$ and an LCL problem $\Pi$ with checking radius $\ell\geq1$ that admits a deterministic $O(\log^* n)$-round fault-free algorithm on all graphs of maximum degree at most $\Delta$. We show that $\Pi$ also admits a deterministic algorithm with the same asymptotic round complexity that tolerates any number of crashes.

On graphs of bounded degree, every LCL problem with the assumed deterministic fault-free round complexity, including $\Pi$, admits a deterministic fault-free algorithm in the following two-stage \emph{normal form}~\cite{ChangKopelowitzPettie2019} (see also~\cite[proof of Theorem~7]{BalliuEtAl2021}): it first computes, in $O(\log^* n)$ rounds, a coloring with a constant number of colors such that any two distinct nodes within a fixed constant number of hops receive distinct colors. It then runs a constant-round algorithm that uses this coloring.

More precisely, the normal form provides constants $t,q\geq1$, depending only on $\Pi$ and $\Delta$, and a deterministic \local algorithm $\mathcal B$. The constant $t$ specifies the coloring distance and bounds the running time of $\mathcal B$. We fix a normal form whose first stage is the Linial reduction schedule described below, and let $q$ be its terminal palette bound. The candidate families and stopping rule are fixed as part of this choice.

We express the coloring distance using graph powers. For a graph $G$ of maximum degree at most $\Delta$, let $G^t$ denote its \emph{$t$-th power}: it has the same node set as $G$, and two distinct nodes are adjacent if and only if their distance in $G$ is at most $t$. The first stage computes a proper coloring $c\colon V(G)\to\{1,\ldots,q\}$ of $G^t$, i.e., any two distinct nodes within distance $t$ in $G$ receive distinct colors.
Given any such coloring on any graph $G$ of maximum degree at most $\Delta$, algorithm $\mathcal B$ solves $\Pi$ within $t$ fault-free rounds when each node $v$ receives its color $c(v)$ as an additional input. This guarantee applies independently of how the coloring was computed.

Our crash-tolerant algorithm for $\Pi$ follows these two stages. First, we show that the color reduction still produces the required $q$ colors in $O(\log^* n)$ rounds when nodes crash. Then, we construct consistent views of constant radius and use them to simulate $\mathcal B$ locally.

\paragraph{Linial's color reduction.}
We recall Linial's color reduction without crashes~\cite{Linial1992}, applied to $G^t$. Let  $\Delta_{G^t}$ be a known upper bound on the maximum degree of $G^t$. Since $\Delta$ and $t$ are constants, $\Delta_{G^t}$ is constant as well.

We call the set of allowed colors the \emph{palette}, and its size the \emph{palette bound}. Let $M$ denote the current palette bound. Initially, each node uses its identifier as its color, giving a proper coloring with the known palette bound $M=\max\{2,n\}^{c_{\mathrm{ID}}}$.
The algorithm then shrinks the palette in a sequence of reduction steps: each step maps a proper coloring with palette $\{1,\ldots,M\}$ to a proper coloring with a smaller palette $\{1,\ldots,K\}$, which becomes the palette for the next step.

In each reduction step, all nodes use the same globally known family of candidate sets $F_1,\ldots,F_M\subseteq\{1,\ldots,K\}$, where $F_i$ contains the possible new colors associated with old color $i$. The family is chosen so that every $F_i$ contains at least one color outside the union of any at most $\Delta_{G^t}$ candidate sets associated with other old colors. Such a family exists with $K=O(\Delta_{G^t}^2\log M)$. Writing $c(v)$ for the current color of $v$, each node learns its neighbors' current colors in $G^t$ and chooses the smallest candidate in $F_{c(v)}$ that belongs to none of their candidate sets. Its new color is therefore $c'(v)=\min\left(F_{c(v)}\setminus\bigcup_{u\in N_{G^t}(v)}F_{c(u)}\right)$.

It remains to verify that every node has an available new color and that the new coloring is proper. First, each node $v$ has at most $\Delta_{G^t}$ neighbors in $G^t$. Since the current coloring is proper, all these neighbors have colors different from $c(v)$. The defining property of the candidate sets therefore ensures that $F_{c(v)}$ contains at least one color outside all of their candidate sets, so $v$ can choose a new color. Second, consider adjacent nodes $u,v$ in $G^t$. Each chooses its new color from its own candidate set while excluding the other node's candidate set: $c'(u)\in F_{c(u)}\setminus F_{c(v)}$ and $c'(v)\in F_{c(v)}\setminus F_{c(u)}$. These two sets are disjoint, so $c'(u)\neq c'(v)$. Thus, the new coloring is proper.

After the step, all colors lie in $\{1,\ldots,K\}$. The next step uses a new candidate family based on this palette bound and the same degree bound $\Delta_{G^t}$. For constant $\Delta_{G^t}$, the palette bound decreases from $n^{O(1)}$ to $O(\log n)$, then to $O(\log\log n)$, and eventually to at most $q$ after $O(\log^* n)$ steps. Each step on $G^t$ is simulated by forwarding current colors for $t = O(1)$ rounds in $G$, so the total communication complexity is $O(\log^* n)$.

\paragraph{Linial's color reduction with crashes.}
We use the same initial colors, palette bounds, and globally known families of candidate sets as in the fault-free procedure. In each reduction step, nodes send and forward current colors for $t$ rounds in $G$, keeping their own current colors fixed during this exchange. Each message contains the source identifier and the reduction step number. At the end of the step, each node $v$ that has not crashed forms the set $J_v$ of distinct colors received from other nodes for this step, discarding any value equal to $c(v)$. It then chooses the smallest candidate in $F_{c(v)}$ that belongs to none of the candidate sets associated with colors in $J_v$. Its new color is therefore $c'(v)=\min\left(F_{c(v)}\setminus\bigcup_{j\in J_v}F_j\right)$. A received color is included in $J_v$ even if its source crashes during the step; if no color is received from a node, its candidate set is omitted for this step. The next step uses a new candidate family based on the new palette bound $K$ and the same degree bound $\Delta_{G^t}$.

\begin{lemma}[Linial's color reduction under crashes]\label{lem:linial-crashes}
Fix integers $\Delta\geq2$ and $t\geq1$, independent of $n$. Let $G=(V,E)$ be a graph with at most $n$ nodes and maximum degree at most $\Delta$. Assume that nodes have identifiers from the known universe specified in \Cref{sec:model} and globally known bounds $n$ and $\Delta$. Consider the color-reduction procedure above in the standard crash-fault model, and let $A$ be the set of nodes that complete the coloring phase without crashing. These nodes may crash in later phases. The procedure has the following guarantees:
\begin{itemize}
    \item \textbf{Existence of an available color.} In every reduction step, each node that has not crashed has at least one available color to choose.
    \item \textbf{Palette bound.} Each reduction step uses the same palette bound as its fault-free version. The fixed terminal palette bound $q$ depends only on $\Delta$ and $t$, and every node in $A$ outputs a color in $\{1,\ldots,q\}$.
    \item \textbf{Proper coloring.} The resulting coloring $c:A\to\{1,\ldots,q\}$ is proper on $(G[A])^t$: $c(u)\ne c(v)$ for all distinct $u,v\in A$ with $\dist_{G[A]}(u,v)\leq t$.
    \item \textbf{Round complexity.} The phase uses the same predetermined reduction steps and takes the same number of rounds as the fault-free procedure, namely $t$ rounds per reduction step and $O(t\log^* n)=O(\log^* n)$ rounds in total for fixed $\Delta$ and $t$.
\end{itemize}
\end{lemma}
\begin{proof}
We first prove that, in every reduction step, an available color exists. Fix a reduction step and a node $v$ that has not crashed by the end of that step. Every color received from another node during the $t$ communication rounds originates at a node of $N_{G^t}(v)$. Hence $|J_v|\leq \Delta_{G^t}$, where $\Delta_{G^t}$ is the fixed degree bound used in the procedure.
Moreover, $c(v)\notin J_v$, since $v$ discards any received color equal to $c(v)$ when forming $J_v$.
The defining property of the candidate sets therefore gives $F_{c(v)}\setminus\bigcup_{j\in J_v}F_j\neq\emptyset$, so $v$ can choose an available color. This holds even if some received colors originate at nodes that crash during the step.

Next, we show that crashes do not affect the palette bounds or the number of rounds. In every reduction step, each node $v$ chooses its new color from its own candidate set $F_{c(v)}\subseteq\{1,\ldots,K\}$. Crashes can only reduce the set $J_v$ of excluded candidate sets, so the new color still lies in $\{1,\ldots,K\}$. Moreover, the degree bound $\Delta_{G^t}$, the candidate families, and the sequence of palette bounds are fixed in advance and coincide with those of the fault-free procedure. Since $\Delta_{G^t}$ depends only on $\Delta$ and $t$, this sequence reaches its fixed terminal palette bound $q$ after $O(\log^* n)$ reduction steps. Each step takes $t$ communication rounds, so the procedure takes $O(t\log^* n)=O(\log^* n)$ rounds in total.

We prove properness on $(G[A])^t$ by induction over the reduction steps. Initially, the coloring is proper because identifiers are unique. Suppose it is proper at the beginning of a step, and let $u,v\in A$ be distinct nodes with $\dist_{G[A]}(u,v)\leq t$. A path of length at most $t$ in $G[A]$ connects them, and every node on this path completes all $t$ communication rounds of the step without crashing. Their current colors therefore reach each other before either endpoint updates its color. By the induction hypothesis, $c(u)\neq c(v)$, so neither received color is discarded: $c(u)\in J_v$ and $c(v)\in J_u$. Consequently, $v$ chooses $c'(v)\notin F_{c(u)}$, whereas $u$ chooses $c'(u)\in F_{c(u)}$. Thus $c'(u)\neq c'(v)$, completing the induction.
\end{proof}

\paragraph{Simulating $\mathcal B$ on consistent views.}
Our crash-tolerant algorithm first runs the color reduction above, using the same palette schedule as the fault-free color reduction of the normal form. It then simulates $\mathcal B$ on consistent views of radius $R:=t+\ell+1$. To establish this, we run the $\CN(R)$ algorithm of \Cref{alg:parameter-cn} with fresh initial stars, each carrying its source's original input label and its color after color reduction. For the analysis only, assign color $1$ to all nodes outside $A$; they send no fresh stars or messages in this phase, so these assigned colors do not occur in any view. Each surviving node $v$ then simulates $\mathcal B$ on its view $H_v$ and adopts its own simulated output.

\begin{proof}[Proof of \Cref{thm:lcl-crash-complexity}(i)]
By \Cref{lem:linial-crashes}, every node in $A$ obtains a color in the normal form's palette $\{1,\ldots,q\}$, and this coloring is proper on $(G[A])^t$.

Next, we show that $\mathcal B$ solves $\Pi$ on every view. In the second stage, we treat the color of each node as part of its input label. Since $R\geq3$, the proof of \Cref{thm:upper-bound} shows that \Cref{alg:parameter-cn} solves $\CN(R)$ under any number of crashes, where nodes that crashed during the first stage count as crashed from the beginning.
Therefore, by termination and radius, every surviving node $v$ obtains a view $H_v$ that is a labeled subgraph of $G$. Since colors are part of the input labels, every node of $H_v$ appears in $H_v$ with its color from the first stage, even if it crashes during the second stage. Moreover, $H_v=G[W_v]$, and every token in $W_v$ was initiated by a node that completed the gathering phase of \Cref{alg:parameter-cn}, and hence the earlier color reduction. Thus, $W_v\subseteq A$ and $H_v\subseteq G[A]$. Since distances in $H_v$ are at least those in $G[A]$, the coloring is proper on $H_v^t$, so $\mathcal B$ solves $\Pi$ on $H_v$ within $t$ rounds.

We next verify local satisfaction. Fix a surviving node $v$. If any node in $B_G(v,\ell)$ crashes during the execution, then $v$ is locally satisfied by definition. Otherwise, every node in this neighborhood completes color reduction and survives the construction of the views. Thus, $B_G(v,\ell)\subseteq A$. As shown above, $\mathcal B$ produces a valid solution on each view within $t$ rounds. Since the views satisfy $\CN(R)$ with $R=t+\ell+1$, the simulation argument in the proof of \Cref{thm:cn-to-lcl}(ii) shows that $v$'s checking neighborhood, including the output labels, is identical in the actual execution and in the simulation on $H_v$. Hence, the actual outputs satisfy the constraints of $\Pi$ at $v$, so $v$ is locally satisfied.

Finally, color reduction takes $O(\log^* n)$ rounds by \Cref{lem:linial-crashes}. Since $\Delta$ and $R=t+\ell+1$ are constants, constructing the consistent views takes $O((\Delta-1)^R)=O(1)$ rounds by \Cref{thm:upper-bound}(ii) if $\Delta\geq3$, and $O(R)=O(1)$ rounds by \Cref{thm:upper-bound}(iv) if $\Delta=2$. Simulating $\mathcal B$ on these views requires no further communication, so the algorithm takes $O(\log^* n)$ rounds in total.
\end{proof}

\subsection{An LCL with a Linear Crash-Fault Lower Bound}
\label{sec:lcl-lower-bound}
To prove \Cref{thm:lcl-crash-complexity}(ii), we first construct an LCL problem with checking radius $\ell=2$ on graphs of maximum degree at most five. We then show that it has deterministic fault-free round complexity $\Theta(\log n)$ and deterministic crash-tolerant round complexity $\Theta(n)$.

Our problem, denoted $\Pi_{\mathrm{bit}}$, is built around a complete rooted ordered binary tree: every internal node has a left and a right child, and all leaves have the same depth. We join consecutive nodes in left-to-right order on each level by an edge so that each level forms a path. This graph family also appears in earlier LCL constructions~\cite{BalliuBrandtOlivettiSuomela2020,BalliuEtAl2021}. The input labels describe this tree-based structure and assign an input bit to the root. On a correctly labeled tree-based graph (without crashes), $\Pi_{\mathrm{bit}}$ requires every node to output the root's bit. To define $\Pi_{\mathrm{bit}}$ on arbitrary input-labeled graphs, we also allow error outputs that certify structural defects or point toward them, subject to local checks.

For the lower bound, we allow crashes only in the layer three edges above the leaves:  crashing this entire layer before communication cuts the leaves off from the root, while every leaf's radius-two checking neighborhood survives. We then use a similarity chain, as in \Cref{sec:lower-bound}, to compare executions with two inputs on the same graph that differ only in the root's bit.  In the executions of this chain, all leaves must agree on whether they output $0$, output $1$, or use an error output. Hence, if a leaf witnesses a step of the chain, then all leaves make the same choice in both executions of that step. When the entire crashing layer fails before communication, the leaves cannot distinguish the two root bits. The chain therefore forces all leaves to output $0$ even in the fault-free execution with root bit $1$, a contradiction.

\subsubsection{A locally verifiable ordered binary tree}
\label{sec:lcl-tree-structure}
This subsection formally defines the tree-based graphs $G_h$ and the input labels that describe their structure. We give five tests, similar to local structural checks in~\cite{BalliuBrandtOlivettiSuomela2020,BalliuEtAl2021}, that each node can evaluate within radius two. We show that a component in which every node passes all five tests is a correctly labeled copy of $G_h$ for some $h\geq0$. If a component contains a node that fails a test, then every node in that component is within distance $O(\log n)$ of a node that fails a test (\Cref{lem:lcl-tree-defects}).

For an integer $h\geq0$, define $G_h$ by $V(G_h):=\{(d,k)\in\mathbb Z^2:0\leq d\leq h,\ 0\leq k<2^d\}.$ The node $(d,k)$ is at depth $d$ and position $k$ in its level; the root is $(0,0)$. Every node $(d,k)$ with $d<h$ has left child $(d+1,2k)$ and right child $(d+1,2k+1)$. The edges of $G_h$ are the parent--child edges and the horizontal edges joining $(d,k)$ to $(d,k+1)$ whenever $k+1<2^d$.

The number of nodes is $n_h:=|V(G_h)|=2^{h+1}-1$, the maximum degree is at most five, and $\diam(G_h)\leq2h$.

We now define the structural input labels. To refer to specific neighbors while keeping the input alphabet fixed and finite, we use local port names. Consider an arbitrary graph of maximum degree at most five. At each node, assign distinct local port names from $\{1,\ldots,5\}$ to its incident edges. Unused names are allowed, and the two endpoints of an edge may use different names.

The structural part of the input label at $v$ is $\tau(v)=(P(v),L(v),R(v),j(v),t(v)).$ The fields $P(v),L(v),R(v)\in\{\bot,1,\ldots,5\}$ designate ports toward the parent, the left horizontal neighbor, and the right horizontal neighbor, respectively; $\bot$ denotes an absent reference. The field $j(v)\in\{\bot,\leftchild,\rightchild\}$ specifies whether the node is a left or right child, with $\bot$ for the root, and $t(v)\in\{\mathsf{internal},\mathsf{leaf}\}$ specifies the node type. The fields may take arbitrary values in their specified domains. An edge incident to $v$ is called a child edge at $v$ if its local port at $v$ is not among $P(v),L(v),R(v)$; its other endpoint is called a child of $v$.

A copy of $G_h$ is called \emph{correctly labeled} if its input labels specify the parents, horizontal neighbors, child indices, and node types prescribed by its coordinates.

An arbitrary input labeling might not follow these conventions consistently: the tests below check whether it does. Any required reference to a missing port or any other undefined expression causes the corresponding test to fail. We define the predicate $\mathsf{Good}(v)$ to hold if and only if all five tests pass at node $v$.

\begin{enumerate}[(1)]
\item The non-$\bot$ values among $P(v),L(v),R(v)$ are pairwise distinct and name existing ports. For every child $u$ of $v$, the edge $\{u,v\}$ has local port name $P(u)$ at $u$. If $v$ has parent $p$, then $t(p)=\mathsf{internal}$ and $\{p,v\}$ is a child edge at $p$. For every right horizontal neighbor $u$ of $v$, the edge $\{u,v\}$ has local port name $L(u)$ at $u$; for every left horizontal neighbor $u$ of $v$, it has local port name $R(u)$ at $u$.

\item If $P(v)=\bot$, then $j(v)=L(v)=R(v)=\bot$; otherwise $j(v)\in\{\leftchild,\rightchild\}$. If $t(v)=\mathsf{internal}$, then $v$ has exactly two children, one of each index. If $t(v)=\mathsf{leaf}$, then $v$ has no children.

\item If $R(v)\neq\bot$, then the right horizontal neighbor $u$ of $v$ satisfies $t(u)=t(v)$.

\item Suppose $P(v)\neq\bot$, and let $p$ be the parent of $v$. If $j(v)=\leftchild$, then $v$ has a right horizontal neighbor $u$ with $j(u)=\rightchild$, and the edge $\{u,p\}$ exists and has local port name $P(u)$ at $u$. If $j(v)=\rightchild$, then $R(v)=\bot$ if and only if $R(p)=\bot$. When $j(v)=\rightchild$ and both entries are non-$\bot$, let $u$ and $q$ be the right horizontal neighbors of $v$ and $p$, respectively. Then $j(u)=\leftchild$, and the edge $\{u,q\}$ exists and has local port name $P(u)$ at $u$.

\item Suppose $P(v)\neq\bot$, and let $p$ be the parent of $v$. If $j(v)=\rightchild$, then $v$ has a left horizontal neighbor $u$ with $j(u)=\leftchild$, and the edge $\{u,p\}$ exists and has local port name $P(u)$ at $u$. If $j(v)=\leftchild$, then $L(v)=\bot$ if and only if $L(p)=\bot$. When $j(v)=\leftchild$ and both entries are non-$\bot$, let $u$ and $q$ be the left horizontal neighbors of $v$ and $p$, respectively. Then $j(u)=\rightchild$, and the edge $\{u,q\}$ exists and has local port name $P(u)$ at $u$.
\end{enumerate}

Each test can be evaluated from the labeled radius-two neighborhood of $v$: tests (1), (2), (3) only involve $v$ and its neighbors, while tests (4) and (5) compare the nodes reached by walks of length at most two. All input fields range over fixed finite sets, independently of $n$ and $h$.

Our error outputs will either certify a structural defect or point toward one. For an arbitrary assignment $\tau$ of structural input labels, we define the \emph{defect set} $D_\tau:=\{v:\neg\mathsf{Good}(v)\}$, consisting of the nodes at which at least one structural test fails. We define $J_\tau$ as the spanning subgraph of $G$ consisting of mutually consistent parent--child edges. Error pointers may use only these edges. Specifically, its edge set consists of all $\{p,c\}\in E(G)$ such that the local port name of $\{p,c\}$ at $c$ is $P(c)$, $j(c)\in\{\leftchild,\rightchild\}$, $t(p)=\mathsf{internal}$, and $\{p,c\}$ is a child edge at $p$. Membership in $J_\tau$ requires only these conditions; other structural tests may fail. If $\mathsf{Good}(v)$ holds, every parent or child edge incident to $v$ belongs to $J_\tau$.

The next lemma shows that the tests characterize correctly labeled copies of $G_h$. It also guarantees that, in every other component, each node can reach a structural defect through $J_\tau$ within $O(\log n)$ steps.

\begin{lemma}[Valid structures and nearby defects]
\label{lem:lcl-tree-defects}
Let $G$ be a graph with at most $n$ nodes and maximum degree at most five, equipped with the structural inputs above, and let $K:=\lceil\log_2 n\rceil+1$. Let $X$ be a connected component of $G$.
\begin{itemize}
\item If every node of $X$ satisfies $\mathsf{Good}$, then $X$ is a correctly labeled copy of $G_h$ for some $h\geq0$.
\item Otherwise, every node $v$ of $X$ satisfies $\dist_{J_\tau}(v,D_\tau)\leq4K$.
\end{itemize}
\end{lemma}

\begin{proof}
We first prove the second claim. Suppose that some node of $X$ fails a structural test. Since every correctly labeled copy of $G_h$ passes all five tests, $X$ is not such a copy. We prove the distance bound by contradiction.

Fix $v\in X$, and assume that no node of $D_\tau$ lies within $J_\tau$-distance $4K$ of $v$.
Every node $u$ within this distance therefore satisfies $\mathsf{Good}(u)$. We show that $X$ is a correctly labeled copy of some $G_h$, contradicting the assumption on $X$. We first show that following parent edges from $v$ reaches a root $r$ after fewer than $K$ steps. Otherwise, the structural tests would force $2^K>n$ distinct nodes below the node reached after $K$ steps, a contradiction. Afterwards, we show that the levels below $r$ end in a leaf level and form a correctly labeled copy of some $G_h$, which contains $v$ and is therefore all of $X$.

Assume that there is a walk of length $K$ from $v$ along parent edges, and let $u$ be its endpoint. Reversing this walk gives a walk of length $K$ from $u$ to $v$ in $J_\tau$. Each node except the last is the parent of the next node, so test~(1) requires it to be internal.

We construct levels below $u$, starting with level zero consisting of $u$. To form the next level, we list each parent's left and then right child, preserving the parents' order. We prove that, for every $d\leq K$, level $d$ exists and contains $2^d$ distinct nodes joined in this order by their right horizontal edges. Distinctness is needed only within each level; different levels need not yet be disjoint. Before proving the claim, we note that all nodes considered satisfy $\mathsf{Good}$: a node at level $d\leq K$ is reachable from $u$ by $d$ parent--child edges in $J_\tau$, so its $J_\tau$-distance from $v$ is at most $K+d\leq2K<4K$.

The claim holds at level zero. Suppose it holds at level $d<K$. The walk from $u$ to $v$ contains an internal node at this level, so test~(3) makes the whole level internal. By test~(2), each node has two children, one of each index. These $2^{d+1}$ children are distinct because no node can have two designated parents. Tests~(1), (4), and~(5) connect each parent's two children and connect the right child of each parent to the next parent's left child. Thus the next level forms a horizontal path in the listed order, completing the induction.

Level $K$ would contain $2^K>n$ distinct nodes, a contradiction. Hence, no walk of length $K$ from $v$ along parent edges exists. Following parent edges from $v$ therefore stops after fewer than $K$ steps, and by test~(1), this happens only at a node $r$ with $P(r)=\bot$, i.e., at a root.

The same construction below $r$ produces levels that form horizontal paths and, by test~(3), consist either entirely of internal nodes or entirely of leaves. Since $2^K>n$, a leaf level occurs at some depth $h<K$. Every node considered is within $J_\tau$-distance less than $2K$ of $v$, so its tests pass. Different levels are disjoint because each node has a unique parent and $r$ has none.

The root conditions and tests~(4)--(5) give exactly the horizontal edges of $G_h$, while tests~(1)--(2) permit no additional parent or child edges. Hence no edge leaves the constructed copy of $G_h$. Since it contains $v$, it is the whole connected component of $v$.
This contradicts our assumption on $X$, so $\dist_{J_\tau}(v,D_\tau)\leq4K$, proving the second claim.

For the first claim, suppose that every node of $X$ satisfies $\mathsf{Good}$, but $X$ is not a correctly labeled copy of any $G_h$. The preceding argument uses only the assumption that $X$ is not such a copy. Hence, for any $v\in X$, it gives a node $w\in D_\tau$ with $\dist_{J_\tau}(v,w)\leq4K$. Since $J_\tau\subseteq G$, the nodes $v$ and $w$ belong to the same connected component of $G$, so $w\in X$. This contradicts the assumption that every node of $X$ satisfies $\mathsf{Good}$.
\end{proof}

\subsubsection{The LCL and its fault-free complexity}
\label{sec:lcl-promise-free}

Using the structural tests of \Cref{sec:lcl-tree-structure}, we now define our LCL problem and show that its fault-free complexity is $\Theta(\log n)$.

\paragraph{Problem definition.}
In addition to its structural label $\tau(v)$, a node $v$ receives an input bit $x(v)\in\{0,1\}$ exactly when $P(v)=\bot$. Thus, on a correctly labeled component $G_h$, only the root receives an input bit.
As defined below, every node of a correctly labeled component $G_h$ must output the input bit of the root $(0,0)$. In components with structural defects, nodes may instead output $\mathsf E$ to certify a defect, or $\mathsf{ptr}(p)$ to point toward one along the incident edge with local port name $p$.

\begin{definition}[The problem $\Pi_{\mathrm{bit}}$]\label{def:pi-bit}
$\Pi_{\mathrm{bit}}$ is the locally checkable labeling problem with checking radius two on graphs of maximum degree at most five with local port names in $\{1,\dots,5\}$, specified as follows.
\begin{itemize}
\item \emph{Inputs.} Every node $v$ receives its structural label $\tau(v)=(P(v),L(v),R(v),j(v),t(v))$. If $P(v)=\bot$, its input label is $(\tau(v),x(v))$, where $x(v)\in\{0,1\}$ is an additional input bit; otherwise, its input label is $\tau(v)$. The input alphabet $\Sigma_{\mathrm{in}}$ consists of all labels of these two forms, with the structural fields ranging over the finite sets specified in \Cref{sec:lcl-tree-structure}.

\item \emph{Outputs.} Every node $v$ outputs a label from $\Sigma_{\mathrm{out}}:=\{0,1,\mathsf E\}\cup\{\mathsf{ptr}(p):1\leq p\leq5\}$. The labels $\mathsf E$ and $\mathsf{ptr}(p)$ are called \emph{error outputs}.

\item \emph{Verifier.} A node $v$ accepts an output labeling if and only if the following checks pass in its labeled radius-two neighborhood:
  \begin{enumerate}[(a)]
  \item either $v$ and all its neighbors output the same bit, or they all use error outputs;
  \item if node $v$ outputs a bit $b\in\{0,1\}$, then $\mathsf{Good}(v)$ holds, and if, in addition, $P(v)=\bot$, then $b=x(v)$;
  \item if $v$ outputs $\mathsf E$, then $v\in D_\tau$;
  \item if $v$ outputs $\mathsf{ptr}(p)$, then port $p$ names an edge $\{v,u\}$ of $J_\tau$. If $u$ also outputs a pointer, it must point to a node other than $v$.
  \end{enumerate}
\end{itemize}
An output labeling is a solution of $\Pi_{\mathrm{bit}}$ if and only if every node accepts.
\end{definition}

On a correctly labeled $G_h$, the root is the unique node $v$ with $P(v)=\bot$, so check~(b) refers to it through its structural label, without using its coordinates. The proof of \Cref{lem:lcl-bit-fault-free} shows that checks~(a), (c), and~(d) together rule out error outputs on such a component. Check~(a) then forces all nodes to output the same bit, which must equal the root's input bit by check~(b).

On components with structural defects, check~(a) allows different error outputs at adjacent nodes. For example, a node outputting $\mathsf E$ may have a neighbor outputting $\mathsf{ptr}(p)$. In our constructions, pointers follow shortest paths in $J_\tau$ to a defect, which outputs $\mathsf E$.

We now verify that $\Pi_{\mathrm{bit}}$ is an LCL. Each check is determined by the labeled radius-two neighborhood of $v$. The structural tests, and hence $\mathsf{Good}(v)$ and membership in $D_\tau$, can be evaluated within radius two (\Cref{sec:lcl-tree-structure}). Whether an incident edge belongs to $J_\tau$ and whether a neighbor's pointer points back to $v$ are determined by the input and output labels at the endpoints. Since both alphabets are fixed finite sets and the maximum degree is at most five, $\Pi_{\mathrm{bit}}$ is an LCL on all such input-labeled graphs, including those that are not correctly labeled copies of any $G_h$.

\paragraph{Fault-Free Complexity.} We may now establish the fault-free part of \Cref{thm:lcl-crash-complexity}(ii). For the upper bound, \Cref{lem:lcl-tree-defects} ensures that a node can gather enough information in $O(\log n)$ rounds to either find a defect reachable through $J_\tau$ and certify or point toward it, or reconstruct its correctly labeled component and output the root's bit. For the lower bound, a leaf at distance $h=\Theta(\log n)$ from the root cannot distinguish opposite root bits in fewer than $h$ rounds.

\begin{lemma}[Fault-free round complexity]
\label{lem:lcl-bit-fault-free}
On input-labeled graphs of maximum degree at most five, $\Pi_{\mathrm{bit}}$ has deterministic fault-free \local round complexity $\Theta(\log n)$.
\end{lemma}

\begin{proof}
We first prove the upper bound. Let $K:=\lceil\log_2 n\rceil+1$.
Our algorithm runs \Cref{alg:full-information-gathering} for $4K+2$ fault-free rounds and then computes its outputs locally, so it takes $O(\log n)$ rounds.

After these rounds, every node $v$ knows its labeled radius-$(4K+2)$ neighborhood. It can therefore evaluate the structural tests at all nodes within distance $4K$ and determine whether a defect is reachable by a path of at most $4K$ edges in $J_\tau$. If so, it can compute its $J_\tau$-distance to $D_\tau$.

If $v\in D_\tau$, it outputs $\mathsf E$. Otherwise, if its $J_\tau$-distance to $D_\tau$ is at most $4K$, it outputs $\mathsf{ptr}(p)$ along an edge of $J_\tau$ to a neighbor whose distance to $D_\tau$ is one less, breaking ties by local port name. If no defect is reachable within this distance, it outputs the input bit of the root of its component.
The last case is well defined: if no node of $D_\tau$ is within $J_\tau$-distance $4K$ of $v$, then \Cref{lem:lcl-tree-defects} implies that the component of $v$ is a correctly labeled $G_h$. Since $n_h\leq n$ gives $h<K$, this component has diameter at most $2h<2K$, so $v$ sees the entire component, including its root.

We now show that all checks pass at every node. By \Cref{lem:lcl-tree-defects}, every component is either a correctly labeled copy of some $G_h$, or every node of the component has $J_\tau$-distance at most $4K$ to $D_\tau$.

Consider first a correctly labeled $G_h$ with root $r$. It contains no node of $D_\tau$, so every node outputs $x(r)$. Every node satisfies $\mathsf{Good}$, and $r$ is the only node with $P(r)=\bot$, so checks~(a) and~(b) pass. Since no node outputs $\mathsf E$ or a pointer, checks~(c) and~(d) pass as well.

We now consider a component with a structural defect. Every node outputs $\mathsf E$ or a pointer, so check~(a) passes, and check~(b) does not apply. Check~(c) passes because only nodes of $D_\tau$ output $\mathsf E$. For check~(d), each pointer uses an edge of $J_\tau$ and decreases the $J_\tau$-distance to $D_\tau$ by one. Thus, if $v$ points to $u$, any pointer at $u$ must lead to a node closer to $D_\tau$ than $v$, so it cannot point back to $v$.

It remains to prove the fault-free lower bound of $\Omega(\log n)$ rounds. We first show that, in every solution on a correctly labeled $G_h$ with root $r$, every node outputs $x(r)$. Since check~(a) passes throughout this connected component, either all nodes output the same bit or they all use error outputs. Suppose that all nodes use error outputs. Check~(c) rules out $\mathsf E$, since the component contains no defect, so every node outputs a pointer along an edge of the parent--child tree $J_\tau$. Following pointers from any node eventually revisits a node. Since $J_\tau$ is a tree, the resulting cycle consists of two nodes pointing to each other, contrary to check~(d). Hence, all nodes output the same bit, which is $x(r)$ by check~(b) at $r$.

Now let $n$ be a sufficiently large size bound, let $h$ be the largest integer with $n_h\leq n$, and add $n-n_h$ isolated, correctly labeled one-node trees to $G_h$. Since $n<n_{h+1}$, we have $h=\Theta(\log n)$. Assume that a deterministic algorithm $\mathcal A$ solves $\Pi_{\mathrm{bit}}$ within $T<h$ rounds. Consider two fault-free executions of $\mathcal A$ on this graph with identical identifiers, port names, and inputs except for opposite bits at the root of $G_h$. Horizontal edges preserve depth and vertical edges change it by one, so every leaf is at distance $h>T$ from the root. Hence, a leaf has the same view through round $T$ in both executions and produces the same output in both. However, by the claim above, it outputs the root bit, which differs between the two executions, a contradiction. Consequently, every deterministic algorithm for $\Pi_{\mathrm{bit}}$ needs at least $h=\Omega(\log n)$ rounds.
\end{proof}

\subsubsection{Crash-tolerant round complexity} \label{sec:lcl-crash-bounds}

In the following, we show that the crash-tolerant round complexity of $\Pi_{\mathrm{bit}}$ is $\Theta(n)$, which completes the proof of \Cref{thm:lcl-crash-complexity}(ii). The lower bound is established by the lemma below.

\begin{lemma}[Linear crash-tolerant lower bound]
\label{lem:lcl-crash-lower-bound}
For every prescribed crash budget $g(n)=\Theta(n)$, every deterministic algorithm for $\Pi_{\mathrm{bit}}$ that tolerates up to $g(n)$ atomic crashes has worst-case round complexity $\Omega(n)$, with correctness as in \Cref{def:lcl-away-from-crashes}.
\end{lemma}

To prove \Cref{lem:lcl-crash-lower-bound}, we fix a correctly labeled $G_h$ with $\Theta(n)$ nodes, in which only the nodes three levels above the leaves may crash. As in \Cref{sec:lower-bound}, we construct a similarity chain, here from the fault-free execution on $G_h$ with root bit~$0$ to the fault-free execution on $G_h$ with root bit~$1$, in which each step is witnessed by a leaf (\Cref{lem:lcl-chain}). In every execution, either all leaves output the same bit or they all use error outputs. The unchanged output of each witness then forces all leaves to output $0$ throughout the chain, contradicting the required output $1$ at the other endpoint.

\paragraph{Setup.}
Let $n\geq n_4$ be the globally known size bound, and let $h$ be the largest integer with $n_h\leq n$. Then $h\geq4$ and $n<n_{h+1}=2n_h+1$, so $n_h\geq n/2$. Let $G$ be the graph obtained by adding $n-n_h$ isolated, correctly labeled one-node trees to $G_h$. Thus, $G$ has exactly $n$ nodes. The added components exchange no messages with $G_h$, and their input bits remain fixed throughout the argument. The crashes will occur in the level at depth $h-3$, which contains $f:=2^{h-3}=(n_h+1)/16=\Theta(n)$ nodes.

We equip $G_h$ with its correct structural inputs and fix its identifiers, port names, and the remaining global bounds. We also fix the input bits at the roots of the isolated components. Only the input bit $x((0,0))$ of the root of $G_h$, which we call the \emph{root bit}, will vary between our executions.

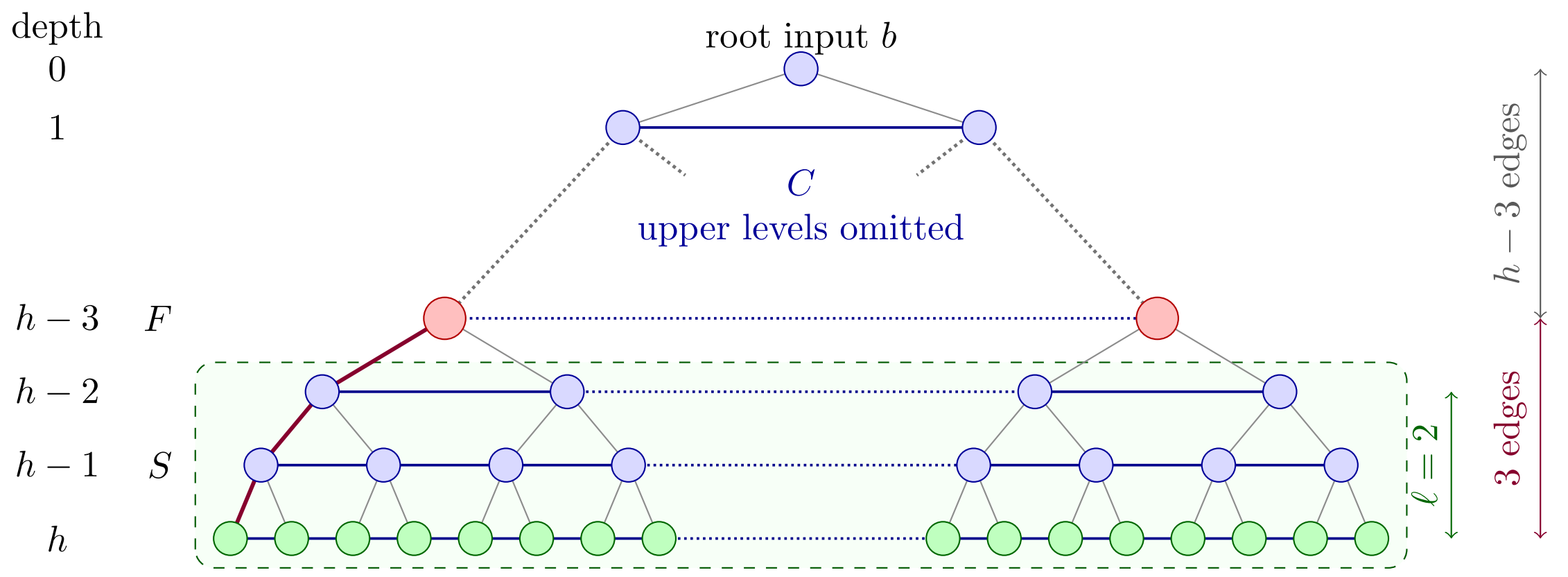
\begin{figure}[!h]
\centering
\begin{tikzpicture}[x=.85cm,y=.7cm,
  treevertex/.style={circle,draw=blue!60!black,fill=blue!15,
    minimum size=3.2mm,inner sep=0pt},
  failvertex/.style={circle,draw=red!70!black,fill=red!25,
    minimum size=4mm,inner sep=0pt},
  leafvertex/.style={circle,draw=green!40!black,fill=green!25,
    minimum size=3.2mm,inner sep=0pt}]
  \draw[green!35!black,dashed,fill=green!3,rounded corners=5pt]
    (-.05,-.4) rectangle (13.55,2.4);
  \coordinate (root) at (6.75,6.4);
  \coordinate (upperleft) at (4.75,5.6);
  \coordinate (upperright) at (8.75,5.6);
  \draw[black!45] (upperleft)--(root)--(upperright);
  \draw[blue!55!black,line width=.7pt] (upperleft)--(upperright);
  \draw[black!55,densely dotted,line width=.9pt]
    (upperleft)--(2.75,3) (upperright)--(10.75,3)
    (upperleft)--(5.45,4.95) (upperright)--(8.05,4.95);
  \node[treevertex] at (root) {};
  \node[treevertex] at (upperleft) {};
  \node[treevertex] at (upperright) {};
  \node[above,font=\small] at (root) {root input $b$};
  \node[font=\small,align=center,text=blue!60!black] at (6.75,4.5)
    {$C$\\upper levels omitted};

  \foreach \depth in {0,...,3} {
    \pgfmathtruncatemacro{\blocksize}{2^\depth}
    \pgfmathtruncatemacro{\last}{2*\blocksize-1}
    \foreach \pos in {0,...,\last} {
      \pgfmathtruncatemacro{\block}{floor(\pos/\blocksize)}
      \pgfmathtruncatemacro{\localpos}{mod(\pos,\blocksize)}
      \coordinate (v-\depth-\pos) at
        ({8*\block+5.5*(\localpos+.5)/\blocksize},{3-\depth});
    }
  }
  \foreach \depth in {0,...,3} {
    \pgfmathtruncatemacro{\blocksize}{2^\depth}
    \pgfmathtruncatemacro{\last}{2*\blocksize-1}
    \pgfmathtruncatemacro{\prevdepth}{\depth-1}
    \foreach \pos in {0,...,\last} {
      \ifnum\depth>0
        \pgfmathtruncatemacro{\parentpos}{floor(\pos/2)}
        \draw[black!45] (v-\depth-\pos)--(v-\prevdepth-\parentpos);
      \fi
      \ifnum\pos>0
        \pgfmathtruncatemacro{\prevpos}{\pos-1}
        \ifnum\pos=\blocksize
          \draw[blue!55!black,densely dotted,line width=.7pt]
            (v-\depth-\prevpos)--(v-\depth-\pos);
        \else
          \draw[blue!55!black,line width=.7pt]
            (v-\depth-\prevpos)--(v-\depth-\pos);
        \fi
      \fi
    }
  }
  \draw[purple!70!black,line width=1.1pt]
    (v-0-0)--(v-1-0)--(v-2-0)--(v-3-0);
  \foreach \depth in {0,...,3} {
    \pgfmathtruncatemacro{\last}{2^(\depth+1)-1}
    \foreach \pos in {0,...,\last} {
      \ifnum\depth=0
        \node[failvertex] at (v-\depth-\pos) {};
      \else\ifnum\depth=3
        \node[leafvertex] at (v-\depth-\pos) {};
      \else
        \node[treevertex] at (v-\depth-\pos) {};
      \fi\fi
    }
  }
  \node[left,font=\small] at (-.15,3) {$F$};
  \node[left,font=\small] at (-.15,1) {$S$};
  \node[font=\small] at (-1.6,6.95) {depth};
  \foreach \height/\depthlabel in {6.4/0,5.6/1,3/{h-3},2/{h-2},1/{h-1},0/h}
    \node[font=\small] at (-1.6,\height) {$\depthlabel$};
  \draw[<->,black!65] (15.05,3)--(15.05,6.4)
    node[midway,right,rotate=90,anchor=south,font=\small] {$h-3$ edges};
  \draw[<->,purple!70!black] (15.05,0)--(15.05,3)
    node[midway,right,rotate=90,anchor=south,font=\small] {$3$ edges};
  \draw[<->,green!40!black] (14.05,0)--(14.05,2)
    node[midway,right,rotate=90,anchor=south,font=\small] {$\ell=2$};
\end{tikzpicture}
\caption{The graph $G_h$ and the crash layer $F$ at depth $h-3$. The dashed box marks $S$, the bottom three levels, and the leaves are shown in green. Only the leftmost and rightmost subtrees below $F$ are expanded. Dotted branches and horizontal segments indicate omitted nodes. Every leaf is three edges from $F$, so its radius-two checking neighborhood is contained in the surviving subgraph $G_h[S]$.}
\label{fig:lcl-crash-layer}
\end{figure}

Let $F$ be the set of nodes of $G_h$ at depth $h-3$, let $S$ consist of its bottom three levels, at depths $h-2,h-1,h$, and let $C:=V(G)\setminus(F\cup S)$. In particular, $C$ includes the added isolated nodes, which are omitted from \Cref{fig:lcl-crash-layer}.

We use three properties of this partition. First, every edge of $G_h$ joins nodes at equal or adjacent depths, and the added nodes are isolated. Hence, $G$ has no edges between $C$ and $S$ and respects $(C,F,S)$ for both root bits. Second, every leaf of $G_h$ is three edges from $F$, so $B_G(v,2)\subseteq S$ for every such leaf $v$. Third, the leaves of $G_h$ induce a horizontal path.

Let $T:=f$, and fix a deterministic $T$-round algorithm $\mathcal A$. Since $G$ respects $(C,F,S)$, we can use the atomic crash schedules, views, indistinguishability, sets $D_t$, and similarity chains of \Cref{sec:lb-framework} and \Cref{def:similarity-chain}. The spreading lemmas used below apply to every deterministic algorithm on such a graph; they do not require $\mathcal A$ to solve $\CN(R)$. In the rest of this argument, a \emph{leaf} means a leaf of $G_h$, rather than one of the added isolated nodes.

For $b\in\{0,1\}$ and an atomic crash schedule $c\colon F\to\{1,\dots,T\}\cup\{\infty\}$, we denote by $\calE(b,c)$ the execution of $\mathcal A$ on $G$, with bit $b$ at the root of $G_h$, in which every $z\in F$ with $c(z)<\infty$ crashes at the beginning of round $c(z)$ and every other node runs $\mathcal A$ for all $T$ rounds. Every such execution has at most $|F|=f$ atomic crashes, all in $F$, and all nodes of $C\cup S$ survive in it.

Most steps of our chain change the crash time of a single node of $F$. To witness such a step, we need a leaf far from that node's neighbors in $S$, with distances measured in $G_h[S]$. The following observation supplies this leaf and plays the role of condition~(a) of \Cref{lem:lb-atomic-chain}.

\begin{observation}[Distant leaves]\label{obs:lcl-distant-leaf}
Each node $z_i=(h-3,i)\in F$, $0\leq i<f$, has a leaf $p_i$ with $\dist_{G_h[S]}\bigl(p_i,\,N_{G_h}(z_i)\cap S\bigr)\geq f$.
\end{observation}
\begin{proof}
The top level of $S$, at depth $h-2$, has $2f$ nodes at positions $0,\dots,2f-1$. For $w\in S$, let $\pi(w)$ be the position of its ancestor at depth $h-2$. A parent--child edge in $G_h[S]$ preserves $\pi$, and a horizontal edge changes it by at most one. Consequently, $\dist_{G_h[S]}(u,v)\geq|\pi(u)-\pi(v)|$ for all $u,v\in S$.

The neighbors of $z_i$ in $S$ are its two children, whose $\pi$-values are $2i$ and $2i+1$. The leftmost leaf has $\pi$-value $0$, so its distance to this neighbor set is at least $2i$. The rightmost leaf has $\pi$-value $2f-1$, giving a lower bound of $2f-2i-2$. Choose $p_i$ to be the extreme leaf with the larger of these two bounds.

The two bounds are even and sum to $2f-2$, so their maximum is at least $f-1$. Since $f=2^{h-3}$ is even for $h\geq4$, an even integer at least $f-1$ is at least $f$. Thus, $\dist_{G_h[S]}\bigl(p_i,N_{G_h}(z_i)\cap S\bigr)\geq f$, as required.
\end{proof}

We now construct the chain. As in the proof of \Cref{lem:lb-atomic-chain}, it moves the crashes of the nodes of $F$ earlier, one node and one round at a time, until all of $F$ crashes before communication. At that point, $S$ is cut off from the root, so we can switch the root bit without any leaf noticing. Finally, we undo the crashes with root bit~$1$.

\tikzset{
  lcfail/.style={circle,draw=red!75!black,fill=red!20,minimum size=3.2mm,inner sep=0pt,font=\tiny},
  lccrashed/.style={fill=red!70!black,text=white},
  lcroot/.style={circle,draw=blue!60!black,fill=blue!15,minimum size=3.2mm,inner sep=0pt,font=\tiny},
  lcleaf/.style={circle,draw=green!40!black,fill=green!20,minimum size=2.0mm,inner sep=0pt},
  lcchild/.style={circle,draw=blue!60!black,fill=blue!15,minimum size=2.0mm,inner sep=0pt},
  lcband/.style={draw=green!35!black,dashed,fill=green!6,rounded corners=2pt},
  lcpath/.style={green!40!black,line width=.7pt},
  lcedge/.style={black!45,line width=.5pt},
  lcwitness/.style={circle,draw=black,line width=1.1pt,minimum size=3.4mm,inner sep=0pt},
  lcstep/.style={->,>=stealth,black!70,line width=.7pt},
  lccap/.style={font=\tiny,text=black!70,align=center},
}
\newlength{\lcsp}\setlength{\lcsp}{3.0mm}
% LCL panel: x/y position, root bit, two crash rounds, and witness index.
% Witness index -1 means none, 0--7 selects one leaf, and 8 marks every leaf.
\newcommand{\lcPanel}[6]{\begin{scope}[shift={(#1,#2)}]
  \fill[lcband] (-0.15,-0.2) rectangle (7*\lcsp+0.15,0.95);
  \foreach \i in {0,...,7} \node[lcleaf] (l\i) at ({\i*\lcsp},0) {};
  \foreach \i/\j in {0/1,1/2,2/3,3/4,4/5,5/6,6/7} \draw[lcpath] (l\i)--(l\j);
  \foreach \i in {0,...,3} \node[lcchild] (w\i) at ({(2*\i+0.5)*\lcsp},0.75) {};
  \foreach \i/\j in {0/1,1/2,2/3} \draw[lcedge] (w\i)--(w\j);
  \node[lcfail] (z0) at ({1.5*\lcsp},1.55) {$z_0$};
  \node[lcfail] (z1) at ({5.5*\lcsp},1.55) {$z_1$};
  \draw[lcedge] (z0)--(z1) (z0)--(w0) (z0)--(w1) (z1)--(w2) (z1)--(w3);
  \node[lcroot] (rt) at ({3.5*\lcsp},2.35) {$#3$};
  \draw[lcedge] (rt)--(z0) (rt)--(z1);
  \if\relax\detokenize{#4}\relax\else
    \node[lcfail,lccrashed] at (z0) {$z_0$};
    \node[lccap,text=red!75!black,anchor=east] at ([xshift=-1pt]z0.west) {rd.\,#4};
  \fi
  \if\relax\detokenize{#5}\relax\else
    \node[lcfail,lccrashed] at (z1) {$z_1$};
    \node[lccap,text=red!75!black,anchor=west] at ([xshift=1pt]z1.east) {rd.\,#5};
  \fi
  \ifnum#6>-1 \ifnum#6<8
    \node[lcwitness] at ({#6*\lcsp},0) {};
  \else
    \foreach \i in {0,...,7} \node[lcwitness,line width=.6pt] at ({\i*\lcsp},0) {};
  \fi\fi
\end{scope}}
\newcommand{\lcArrowR}[3]{\draw[lcstep] ({#1*1cm+2.1cm+1.6mm},#3) -- ({#2*1cm-1.6mm},#3);}
\newcommand{\lcArrowL}[3]{\draw[lcstep] ({#2*1cm-1.6mm},#3) -- ({#1*1cm+2.1cm+1.6mm},#3);}

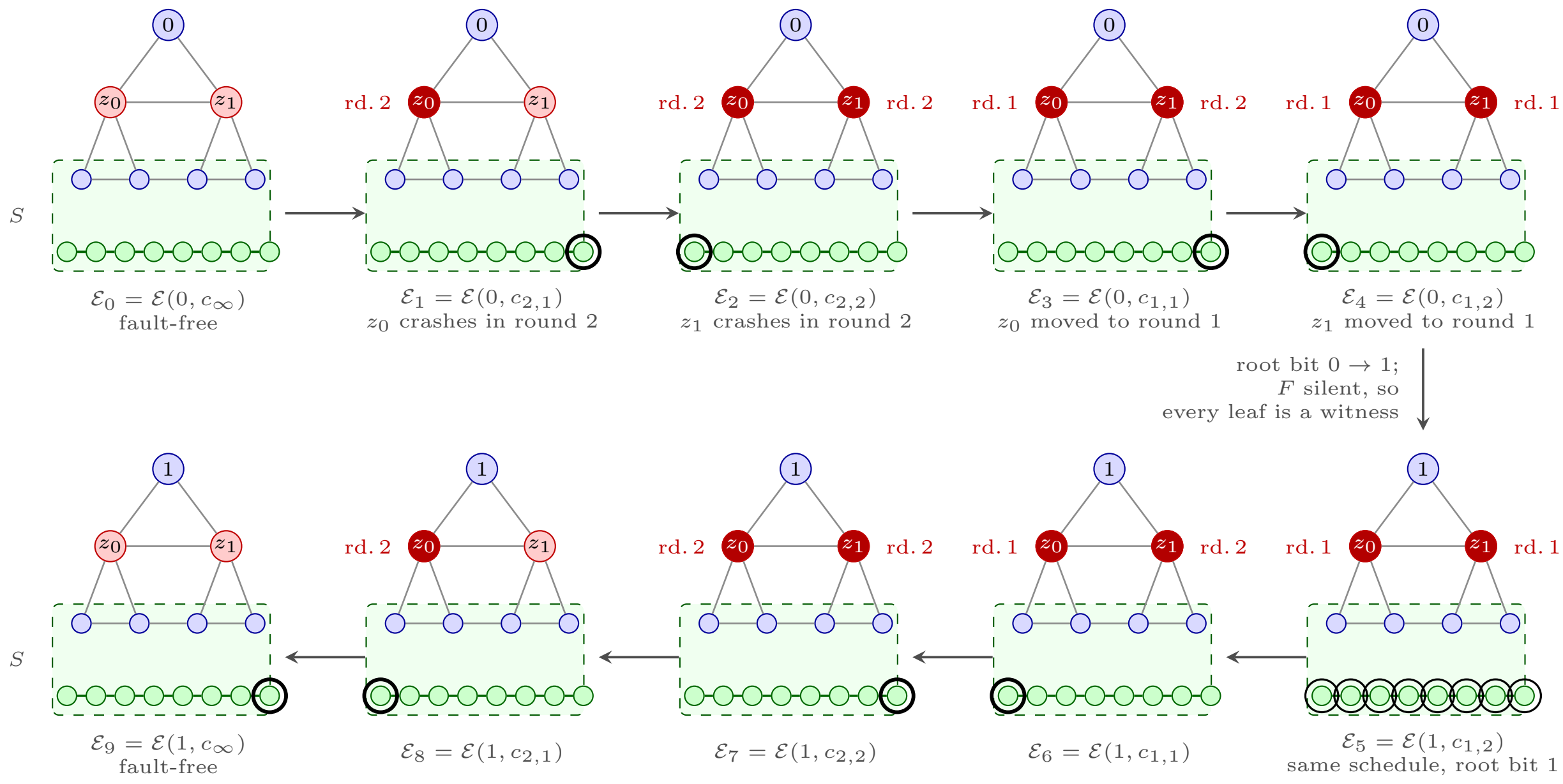
\begin{figure}[t]
\centering
\begin{tikzpicture}[x=1cm,y=1cm]
  \def\cA{0}\def\cB{3.25}\def\cC{6.5}\def\cD{9.75}\def\cE{13.0}
  \def\yTop{4.6}\def\yBot{0}
  \node[font=\footnotesize,anchor=south west] at (-0.9,\yTop+2.75)
    {$h=4$: $F=\{z_0,z_1\}$ at depth $1$, $f=T=2$; the band $S$ (depths $2$--$4$) is drawn schematically};
  \node[lccap,anchor=east] at (-0.3,\yTop+0.38) {$S$};
  \node[lccap,anchor=east] at (-0.3,\yBot+0.38) {$S$};
  \lcPanel{\cA}{\yTop}{0}{}{}{-1}
  \lcPanel{\cB}{\yTop}{0}{2}{}{7}
  \lcPanel{\cC}{\yTop}{0}{2}{2}{0}
  \lcPanel{\cD}{\yTop}{0}{1}{2}{7}
  \lcPanel{\cE}{\yTop}{0}{1}{1}{0}
  \lcArrowR{\cA}{\cB}{\yTop+0.4}\lcArrowR{\cB}{\cC}{\yTop+0.4}
  \lcArrowR{\cC}{\cD}{\yTop+0.4}\lcArrowR{\cD}{\cE}{\yTop+0.4}
  \node[lccap] at (\cA+1.05,\yTop-0.6) {$\calE_0=\calE(0,c_\infty)$\\fault-free};
  \node[lccap] at (\cB+1.05,\yTop-0.6) {$\calE_1=\calE(0,c_{2,1})$\\$z_0$ crashes in round $2$};
  \node[lccap] at (\cC+1.05,\yTop-0.6) {$\calE_2=\calE(0,c_{2,2})$\\$z_1$ crashes in round $2$};
  \node[lccap] at (\cD+1.05,\yTop-0.6) {$\calE_3=\calE(0,c_{1,1})$\\$z_0$ moved to round $1$};
  \node[lccap] at (\cE+1.05,\yTop-0.6) {$\calE_4=\calE(0,c_{1,2})$\\$z_1$ moved to round $1$};
  \draw[lcstep] ({\cE+1.05},\yTop-1.0) -- ({\cE+1.05},\yBot+2.75)
    node[midway,left=3pt,lccap,align=right] {root bit $0\to1$;\\$F$ silent, so\\every leaf is a witness};
  \lcPanel{\cA}{\yBot}{1}{}{}{7}
  \lcPanel{\cB}{\yBot}{1}{2}{}{0}
  \lcPanel{\cC}{\yBot}{1}{2}{2}{7}
  \lcPanel{\cD}{\yBot}{1}{1}{2}{0}
  \lcPanel{\cE}{\yBot}{1}{1}{1}{8}
  \lcArrowL{\cA}{\cB}{\yBot+0.4}\lcArrowL{\cB}{\cC}{\yBot+0.4}
  \lcArrowL{\cC}{\cD}{\yBot+0.4}\lcArrowL{\cD}{\cE}{\yBot+0.4}
  \node[lccap] at (\cA+1.05,\yBot-0.6) {$\calE_9=\calE(1,c_\infty)$\\fault-free};
  \node[lccap] at (\cB+1.05,\yBot-0.6) {$\calE_8=\calE(1,c_{2,1})$};
  \node[lccap] at (\cC+1.05,\yBot-0.6) {$\calE_7=\calE(1,c_{2,2})$};
  \node[lccap] at (\cD+1.05,\yBot-0.6) {$\calE_6=\calE(1,c_{1,1})$};
  \node[lccap] at (\cE+1.05,\yBot-0.6) {$\calE_5=\calE(1,c_{1,2})$\\same schedule, root bit $1$};
\end{tikzpicture}
\caption{The similarity chain of \Cref{lem:lcl-chain} for $h=4$,
where $F=\{z_0,z_1\}$ and $f=T=2$. Each panel is an execution $\calE(b,c)$:
the root carries its input bit $b$, a dark node of $F$ crashes atomically in
the indicated round, and the band is the surviving region $S$, of which only
the children of $F$ and the leaves are drawn. The top row moves the crashes
from $\infty$ to round $2$ to round $1$ with root bit $0$; the vertical step
changes the root bit while $F$ sends nothing; the bottom row traverses the
same schedules backwards with root bit $1$. A ring marks a leaf that cannot
distinguish the execution from the previous one in the chain: the rightmost
leaf for changes at $z_0$, the leftmost leaf for changes at $z_1$, and every
leaf for the bit switch. All leaves output $0$ in $\calE_0$, so they
would still do so in the fault-free execution $\calE_9$ with root bit $1$.}
\label{fig:lcl-chain}
\end{figure}

\begin{lemma}[A chain between the root bits]\label{lem:lcl-chain}
For the deterministic $T$-round algorithm $\mathcal A$ fixed above, with $T=f$, there is a similarity chain from $\calE(0,c_\infty)$ to $\calE(1,c_\infty)$ in which every step has a leaf as a witness.
\end{lemma}
\begin{proof}
We use the schedules $c_{r,s}$ of \Cref{sec:lb-framework} with $k=|F|=f$ and $x_s:=z_{s-1}$ for $s=1,\dots,f$, i.e., with the nodes of $F$ ordered by position. Recall that $c_{T,0}=c_\infty$ and $c_{r-1,0}=c_{r,f}$ for $r\geq2$. Since $G$ respects $(C,F,S)$, the lemmas of \Cref{sec:lb-framework} apply to $G$. Moreover, $G[S]=G_h[S]$ and $N_G(z_i)\cap S=N_{G_h}(z_i)\cap S$, so they give the following two facts:
\begin{enumerate}[(i)]
    \item For $\calE(b,c_{r,s-1})$ and $\calE(b,c_{r,s})$, which move the crash of $z_{s-1}$ into round $r$, \Cref{lem:lb-move-atomic} implies $D_T\cap S\subseteq B_{G_h[S]}\bigl(N_{G_h}(z_{s-1})\cap S,\;T-r\bigr)$. Since $T=f$ and $r\geq1$, we have $T-r<f$. By \Cref{obs:lcl-distant-leaf}, the leaf $p_{s-1}$ is at distance at least $f$ from this neighbor set, so $p_{s-1}\notin D_T$.
    \item For $\calE(0,c)$ and $\calE(1,c)$ with a schedule $c$ under which no node of $F$ sends any message, the initial states differ only at the root, so $D_0\cap S=\emptyset$, and \Cref{lem:lb-spreading} with $t_0=0$ gives $D_T\cap S=\emptyset$: every leaf has the same view in both executions.
\end{enumerate}
Let $\calE_0:=\calE(0,c_\infty)$ and, for $r=T,\dots,1$ and $s=1,\dots,f$, $\calE_{(T-r)\cdot f+s}:=\calE(0,c_{r,s})$.
This defines $\calE_0,\dots,\calE_{f\cdot T}$. In execution $\calE_{f\cdot T}=\calE(0,c_{1,f})$, all nodes of $F$ crash before round $1$.

Then, let $\calE_{f\cdot T+1}:=\calE(1,c_{1,f})$, and traverse the first part backwards with root bit $1$: for $j=1,\dots,f\cdot T$, let $\calE_{f\cdot T+1+j}$ use root bit $1$ and the schedule of $\calE_{f\cdot T-j}$. The sequence ends at $\calE_{2f\cdot T+1}=\calE(1,c_\infty)$; see \Cref{fig:lcl-chain}. In the following, we show that this sequence of executions is a similarity chain, i.e., that it satisfies \Cref{def:similarity-chain}.

Every execution in this sequence is of the form $\calE(b,c)$, so it has at most $f$ atomic crashes, all in $F$, and is therefore admissible. It remains to show that every step has a leaf as a witness. We consider the three parts of the chain separately.

For $i<f\cdot T$, the executions $\calE_i$ and $\calE_{i+1}$ both have root bit $0$. The identities $c_{T,0}=c_\infty$ and $c_{r-1,0}=c_{r,f}$ ensure that every such step compares $c_{r,s-1}$ with $c_{r,s}$ for some $r$ and $s$, including the first step and the transitions between rounds. By~(i), the leaf $p_{s-1}$ witnesses this step.

The step from $\calE_{f\cdot T}$ to $\calE_{f\cdot T+1}$ keeps the schedule $c_{1,f}$ and only changes the root bit. Under $c_{1,f}$, all of $F$ crashes before round $1$, so no node of $F$ sends any message. By~(ii), every leaf is a witness of this step.

Finally, consider a step from $\calE_i$ to $\calE_{i+1}$ after the root-bit switch. Both executions have root bit $1$, and this part of the chain traverses the crash schedules from the first part in reverse order. Thus, for some $r$ and $s$, the step changes the schedule from $c_{r,s}$ to $c_{r,s-1}$, reversing the earlier change from $c_{r,s-1}$ to $c_{r,s}$. Statement~(i) applies to both root bits. In particular, with root bit $1$, it shows that the leaf $p_{s-1}$ has the same view through round $T$ under these two schedules. Reversing the order of the executions does not change this equality of views. Hence, $p_{s-1}$ is also a witness of the step from $\calE_i$ to $\calE_{i+1}$.

Hence, the sequence is a similarity chain in which every step has a leaf as a witness.
\end{proof}

We may now present the proof of \Cref{lem:lcl-crash-lower-bound}.
\begin{proof}[Proof of \Cref{lem:lcl-crash-lower-bound}]
Consider a deterministic algorithm for $\Pi_{\mathrm{bit}}$ that tolerates up to $f$ atomic crashes, and suppose, for a contradiction, that it terminates within $T=f$ rounds in every such execution on $G$. Once a node produces an output, we keep that output fixed and compare its local history through round $T$. Thus, we may use this algorithm as the $T$-round algorithm $\mathcal A$ of the setup.

We first show that, in every execution $\calE(b,c)$, either all leaves output the same bit or they all use error outputs. Only nodes of $F$ crash, so every leaf survives. In addition, the radius-two neighborhood of every leaf lies in $S$, where no node crashes. Since $\mathcal A$ solves $\Pi_{\mathrm{bit}}$, \Cref{def:lcl-away-from-crashes} requires the checks of \Cref{def:pi-bit} to pass at every surviving node whose radius-two neighborhood contains no crashed node. In particular, check~(a) passes at every leaf. Thus, adjacent leaves either output the same bit or both use error outputs. Since the leaves induce a path, the claim follows.

Next, in the fault-free execution $\calE(b,c_\infty)$, all leaves output $b$. In this execution, no node crashes, so all checks pass at every node and the outputs form a solution of $\Pi_{\mathrm{bit}}$. By the proof of \Cref{lem:lcl-bit-fault-free}, every node of such a solution on a correctly labeled $G_h$ outputs the root bit, here $b$.

Finally, we show that all leaves output $0$ in every execution of the chain of \Cref{lem:lcl-chain}. This holds in its first execution, $\calE(0,c_\infty)$, by the preceding paragraph. Suppose that all leaves output $0$ in an execution of the chain, and let $w$ be a leaf witnessing the step to the next execution. Since $w$ has the same view through round $T$ in both executions and $\mathcal A$ is deterministic, it also outputs $0$ in the next execution. By the first paragraph, all leaves in that execution must therefore output $0$. Induction along the chain implies that all leaves output $0$ in its last execution, $\calE(1,c_\infty)$. However, this execution is fault-free with root bit $1$, so every leaf must output $1$, a contradiction.

Consequently, every deterministic algorithm solving $\Pi_{\mathrm{bit}}$ under up to $f$ atomic crashes requires more than $f$ rounds in at least one of these executions. Since $f=(n_h+1)/16\geq n/32$, its worst-case round complexity is $\Omega(n)$.

More generally, let $g(n)=\Theta(n)$ be any prescribed crash budget. For sufficiently large $n$, choose $f$ as the largest power of two at most $\min\{g(n),(n+1)/16\}$, set $h=3+\log_2 f$, and pad $G_h$ with correctly labeled isolated nodes to obtain $n$ nodes. Then $n_h=16f-1\leq n$, $f\leq g(n)$, and $f=\Omega(n)$, so the same argument applies.
\end{proof}

\paragraph{Upper bound.}
It remains to prove the crash-tolerant $O(n)$ upper bound of \Cref{thm:lcl-crash-complexity}(ii). As in \Cref{thm:cn-to-lcl}(ii), we obtain it by simulating the fault-free algorithm of \Cref{lem:lcl-bit-fault-free} on consistent views.

\begin{lemma}[Linear upper bound with standard crashes]\label{lem:lcl-crash-upper-bound}
With correctness as in \Cref{def:lcl-away-from-crashes}, $\Pi_{\mathrm{bit}}$ can be solved by a deterministic $O(n)$-round \local algorithm on every input-labeled graph of maximum degree at most five, tolerating any number of standard crashes.
\end{lemma}
\begin{proof}
Let $\mathcal A_0$ be the fault-free algorithm of \Cref{lem:lcl-bit-fault-free}, and let $T'=O(\log n)$ be its round complexity. Our algorithm first runs \Cref{alg:linear-cn-from-propagation} with radius $R:=T'+3$. Each surviving node then simulates $\mathcal A_0$ on its view and outputs the label that this simulation computes for itself.

We now show that this algorithm is correct. By \Cref{thm:cn-to-lcl}(ii), it suffices that $R\geq T'+\ell+1$ for the checking radius $\ell=2$, which holds by our choice of $R$, and that $\mathcal A_0$ solves $\Pi_{\mathrm{bit}}$ on every view within $T'$ rounds. Each view is a subgraph of the input graph that keeps the input labels and local port names of its nodes (see \Cref{sec:initial-knowledge}), and the global bounds are unchanged. A label in a view may name a port whose edge is missing from the view; test~(1) then fails, so this is simply a structural defect. Hence, every view is an input-labeled graph of maximum degree at most five, on which $\mathcal A_0$ solves $\Pi_{\mathrm{bit}}$ by \Cref{lem:lcl-bit-fault-free}.

Finally, by \Cref{lem:linear-cn-from-propagation} (see also \Cref{thm:upper-bound}(i)), \Cref{alg:linear-cn-from-propagation} solves $\CN(R)$ within $n-1$ rounds under any number of standard crashes, and the simulations require only local computation. Hence, the algorithm takes $O(n)$ rounds in total.
\end{proof}

\paragraph{Putting it all together.} We now combine \Cref{lem:lcl-bit-fault-free,lem:lcl-crash-lower-bound,lem:lcl-crash-upper-bound} to prove \Cref{thm:lcl-crash-complexity}(ii).

\begin{proof}[Proof of \Cref{thm:lcl-crash-complexity}(ii)]
The problem $\Pi_{\mathrm{bit}}$ is defined on all input-labeled graphs of maximum degree at most five and has checking radius two. Its deterministic fault-free round complexity is $\Theta(\log n)$ by \Cref{lem:lcl-bit-fault-free}. The $\Omega(n)$ lower bound holds already for $\Theta(n)$ allowed atomic crashes by \Cref{lem:lcl-crash-lower-bound}, while \Cref{lem:lcl-crash-upper-bound} gives an $O(n)$-round algorithm tolerating any number of standard crashes. Since atomic crashes are a special case of standard crashes, the lower and upper bounds apply in both models, establishing deterministic crash-tolerant round complexity $\Theta(n)$.
\end{proof}

\subsubsection{Removing input labels}
\label{sec:lcl-no-input-labels}
The separation in \Cref{thm:lcl-crash-complexity}(ii) also holds without problem-specific input labels, with constant maximum degree and checking radius. We use the pendant-tree construction in the full version of~\cite[Section~3.8]{PODC:BBCORS19}; see also~\cite[Section~4 and Lemma~4.6]{CrucianiDasRaevskayaSuomela2026} for input removal in promise-free LCLs. We describe the resulting local specification and verify the crash lower bound separately.

\paragraph{Construction and fault-free complexity.}
Encode each label of $\Sigma_{\mathrm{in}}$ by its constant-size rooted tree from~\cite{PODC:BBCORS19}, attached to the original node by a single edge. The maximum degree becomes six. Use the direct decoder of that construction, before its optional radius reduction. On every graph $X$ of maximum degree six, a constant number of local pruning steps identifies a set $V_{\mathrm{label}}$. The main nodes are precisely the nodes outside $V_{\mathrm{label}}$ with exactly one neighbor in $V_{\mathrm{label}}$; denote their set by $V_{\mathrm{main}}$. Each main node decodes its input label within a constant radius, assigning a fixed label of $\Sigma_{\mathrm{in}}$ to failed or unassigned codes. Every main node has a neighbor outside $V_{\mathrm{main}}$, so the induced graph $H:=X[V_{\mathrm{main}}]$ has maximum degree at most five. At each node of $H$, number its incident edges by their order among its physical ports, and interpret the decoded port references in this order.

Define the input-free problem to use output alphabet $\Sigma_{\mathrm{out}}$, requiring the checks of $\Pi_{\mathrm{bit}}$ on $H$ at nodes of $V_{\mathrm{main}}$, and output $0$ elsewhere. Fix a constant physical checking radius $\ell'$ sufficient for the decoder and these checks. Local decoding followed by the algorithm of \Cref{lem:lcl-bit-fault-free} takes $O(\log n)$ rounds on every $X$. Thus the new LCL is promise-free: malformed encodings also have solutions, without additional error certificates.

For genuine encodings of $G_h$, $h\geq1$, choose the structural labels for the induced port order before encoding them. The decoder recovers $G_h$ and its labels exactly, so every valid output assigns the encoded root bit to all original nodes. The encoding has $\Theta(2^h)$ nodes, and the attachments cannot shorten distances between original nodes. Changing only the root attachment therefore gives the same $\Omega(\log n)$ indistinguishability argument as in \Cref{lem:lcl-bit-fault-free}. Let $s$ bound the number of attachment nodes per original node. For any sufficiently large known size bound $n$, choose $h$ maximal with $(s+1)n_h\leq n$, and pad both encodings with isolated nodes to obtain exactly $n$ nodes. Keep identifiers, ports on common edges, and all global bounds identical. Then $2^h=\Theta(n)$.

\paragraph{Crash-tolerant complexity.}
Fix a constant $L>\ell'$ and take $n$ sufficiently large that $h\geq L+1$. Crash only the original nodes at depth $h-L$; this set $F$ has size $f=2^{h-L}=\Theta(n)$. Let $S$ contain the bottom $L$ original levels and their attached trees, and put all remaining nodes outside $F$ in $C$, including the attachments of nodes in $F$. There is no $C$--$S$ edge, and every original leaf is at distance $L>\ell'$ from $F$.

As in the proof of \Cref{obs:lcl-distant-leaf}, assign each original node in $S$ the position of its ancestor in the top original level of $S$, and extend this function constantly over each attached tree. Each edge still changes this function by at most one. Hence every node of $F$ has an original leaf at distance at least $f$ from its neighbors in $S$, measured in $X[S]$. With $T=f$, the crash-time changes in the proof of \Cref{lem:lcl-chain} therefore still have leaf witnesses. For its middle step, changing the root bit changes topology only inside $C$; when all of $F$ crashes before communication, the initial states and histories throughout $S$ are identical in the two encodings.

All original leaf checks remain obligatory. Check~(a) of $\Pi_{\mathrm{bit}}$ forces a common output type along their horizontal path: $0$, $1$, or error. Each witness preserves this type across a chain step, whereas the fault-free endpoints force opposite bits. This gives $\Omega(n)$ rounds already with atomic crashes. Finally, totality on all physical subgraphs permits the general simulation of \Cref{thm:cn-to-lcl}(ii), together with \Cref{thm:upper-bound}(i), giving $O(n)$ rounds with standard crashes.

\newcommand{\false}{\ensuremath{\textsf{false}}}
\newcommand{\true}{\ensuremath{\textsf{true}}}

\newcommand{\seefailure}{\ensuremath{\textsf{see-failure}}}

\newcommand{\EGalgo}{\ensuremath{\mathcal{A}_{\textnormal{\textsc{EG}}}}}
\newcommand{\SOalgo}{\ensuremath{\mathcal{A}_{\textnormal{\textsc{SO}}}}}

\newcommand{\orientededges}{\ensuremath{\mathcal{O}}}

\section{Crash-Tolerant Algorithms for Locally Checkable Problems}\label{sec:edge-grabbing-lemmas}

In this section, we focus on two locally checkable problems, namely edge grabbing and sinkless orientation. Both are LCL problems when the maximum degree is bounded by a constant. For fixed $\Delta$, their outputs can be encoded by a local port choice (or $\bot$) for edge grabbing and one orientation bit per incident port for sinkless orientation, using port names in $\{1,\ldots,\Delta\}$. Our results hold on graphs of arbitrary maximum degree. We first present a crash-tolerant transformation for edge grabbing that preserves the round complexity of the fault-free algorithm. In \Cref{sec:sinkless-orientation}, we use this transformation to obtain a crash-tolerant transformation for sinkless orientation that increases the round complexity by one communication round, proving \Cref{thm:so-transformation}.

\subsection{Edge Grabbing} \label{sec:edge-grabbing-protocol}

The fault-free edge grabbing problem requires every node of degree at least three to \emph{grab} exactly one of its incident edges and output it, while every node of degree less than three outputs $\bot$. No edge is grabbed by both endpoints. Under crashes, we require these conditions only for surviving nodes. All degrees and incident edges below refer to the input graph $G$, including edges to nodes that crash.

On a $\Delta$-regular graph with $\Delta\geq3$ supplied with a proper edge $\Delta$-coloring, fault-free edge grabbing is precisely the \emph{sinkless coloring} problem studied in~\cite[Section~2.4]{STOC:BFHKLR16}: each node outputs the color of its grabbed edge, and no edge may have its color chosen by both endpoints. Our formulation requires no supplied edge coloring and also covers nodes of degree less than three. The same work gives the elementary conversion from sinkless orientation to sinkless coloring without communication and the reverse conversion in one round. Our contribution here is the crash-tolerant edge-grabbing transformation, which preserves the running time and enforces the requirements even at surviving nodes adjacent to crashes.

\begin{definition}[Crash-Tolerant Edge Grabbing] \label{def:edge-grabbing}
    An algorithm solves crash-tolerant edge grabbing on $G=(V,E)$ if the following conditions hold in every execution:
    \begin{enumerate}[(i)]
        \item Every surviving node $v$ with $\deg_G(v) \geq 3$ outputs an incident edge of $G$, and every surviving node $v$ with $\deg_G(v) < 3$ outputs $\bot$.
        \item No edge is output by both of its surviving endpoints.
    \end{enumerate}
\end{definition}

We now state the main result for edge grabbing.

\begin{theorem}[Edge grabbing under crash faults]\label{thm:edge-grabbing-transformation}
On graphs of arbitrary maximum degree, every deterministic $T$-round fault-free \local algorithm for edge grabbing can be transformed into a deterministic $T$-round \local algorithm solving crash-tolerant edge grabbing in the standard crash-fault model for any number of crashes.
\end{theorem}

Let $\EGalgo$ be a deterministic $T$-round fault-free \local algorithm for edge grabbing. Our transformation runs $\EGalgo$ while each node also keeps track of whether it has seen a failure during the execution of $\EGalgo$. The key idea is to propagate failure information together with the simulated messages, ensuring that nodes that never learn of a failure behave as in a fault-free execution. If the simulated algorithm halts early, we retain its state and output and send no further simulated messages; the wrapper continues exchanging flags and produces its own output only after round~$T$.

In our algorithm, each node defines  a flag $\seefailure$ that is initially $\false$.
In every round, each node sends its flag to each neighbor together with the messages that $\EGalgo$ would send to that neighbor. The flag and these messages are sent as a single message; if $\EGalgo$ has no message to send, the node sends only its flag.
If a node $v$ with $\seefailure=\false$ receives no message from a neighbor $u$, or receives $\seefailure=\true$ from $u$, it sets its own flag to $\true$ and stores the edge $\{u,v\}$. If several neighbors satisfy this condition in the same round, $v$ chooses the one with the smallest identifier. Such a node no longer runs $\EGalgo$ and, from the next round on, sends only its flag to its neighbors. A surviving node of degree at least three grabs and outputs its stored edge if its flag is $\true$, and uses $\EGalgo$'s output otherwise. We present the pseudocode of the transformation in \Cref{alg:edge-grabbing-protocol}.

\begin{algorithm}[ht]
\caption{Crash-Tolerant Edge Grabbing at node $v$}
\label{alg:edge-grabbing-protocol}
\begin{algorithmic}[1]
    \Statex \textbf{Input:} labeled star $\sigma_v$, algorithm $\EGalgo$ and its inputs, and number of rounds $T$.
    \Statex \textbf{Output:} an incident edge of $G$ or $\bot$.
    \State Initialize the simulated state of $\EGalgo$; set $e\gets\bot$ and $\seefailure\gets\false$.
    \For{$t=1,\ldots,T$}
        \State Prepare one message for each neighbor containing $\seefailure$.
        \State If $\seefailure=\false$, include the $\EGalgo$-messages for that neighbor.
        \State Send these messages and receive the messages delivered in round $t$.
        \If{$\seefailure=\false$ and (some neighbor's message is missing or has flag $\true$)}
            \State Choose such a neighbor $u$ with the smallest identifier.
            \State $\seefailure\gets\true$ and $e\gets\{u,v\}$.
        \ElsIf{$\seefailure=\false$}
            \State Update the simulated state of $\EGalgo$ using the received $\EGalgo$-messages.
        \EndIf
    \EndFor
    \State If $\deg_G(v)<3$, \Return $\bot$.
    \State If $\seefailure=\false$, \Return the output of the simulated $\EGalgo$.
    \State \Return $e$.
\end{algorithmic}
\end{algorithm}

We prove \Cref{thm:edge-grabbing-transformation} using the following three lemmas describing the guarantees of \Cref{alg:edge-grabbing-protocol}. We first show that nodes whose $\seefailure$ flags remain $\false$ produce the same outputs as in the fault-free execution of the underlying algorithm $\EGalgo$.

\begin{lemma} \label{EG:lemma:no-failure-means-happy}
Let $\calE^\star$ be the fault-free execution of $\EGalgo$ on $G$. If a node $v$ with $\deg(v) \geq 3$ completes \Cref{alg:edge-grabbing-protocol} with $\seefailure = \false$, then $v$'s output in \Cref{alg:edge-grabbing-protocol} is the same as $v$'s output in $\calE^\star$.
\end{lemma}
\begin{proof}
Let $\calE$ be an execution of \Cref{alg:edge-grabbing-protocol}. Consider a node $v$ of degree at least three that completes $\calE$ with $\seefailure=\false$. By the output rule, $v$ returns the output of its simulation of $\EGalgo$. Since $\EGalgo$ is deterministic, its output is determined by its state at the end of round $T$. We therefore compare the simulated state of $v$ in $\calE$ with its state in the fault-free execution $\calE^\star$, using the same graph, identifiers, and inputs. If these states coincide after $T$ rounds, then $v$ produces the same output in both executions.
As in \Cref{sec:lb-framework}, we use an indistinguishability argument and prove the following claim by induction on the round number $t \in \{0, \ldots, T\}$: for every node $v$, if, at the end of round $t$ of $\calE$, node $v$ has not crashed and has $\seefailure=\false$, then the $\EGalgo$-state of $v$ at the end of round $t$ is the same in $\calE$ and $\calE^\star$.

For the base case $t = 0$, the claim holds trivially since the $\EGalgo$-states are initialized as in $\calE^\star$, and no rounds are executed.

For the induction step, let $1\leq t\leq T$ and consider a node $v$ that completes round $t$ with $\seefailure=\false$ in $\calE$. Its flag was also $\false$ after $t-1$ rounds, so the induction hypothesis gives the same $\EGalgo$-state at the beginning of round $t$ in both executions.
We note that, if $v$ receives $\seefailure = \false$ from a neighbor $u$ in round $t$, then the $\EGalgo$-message received by $v$ from $u$ in round $t$ is the same in both executions $\calE$ and $\calE^\star$.  Indeed, the flag received from $u$ is the one it had at the beginning of round $t$, so $u$ completed round $t-1$ with $\seefailure=\false$. Although $u$ may crash during round $t$, it had not crashed by the end of round $t-1$. We may therefore apply the induction hypothesis to $u$ and conclude that it has the same $\EGalgo$-state at the beginning of round $t$ in both executions. Since $\EGalgo$ is deterministic, the message $v$ has received from $u$ in round $t$ contains exactly the $\EGalgo$-message sent from $u$ to $v$ in round $t$ of $\calE^\star$. When $\EGalgo$ prescribes no message from $u$ to $v$, $u$ sends only its flag, and $v$ treats this as receiving no message from $u$ in the simulation.

Since $v$ completes round $t$ with $\seefailure = \false$, it has received a message with $\seefailure = \false$ from every neighbor in round $t$. Consequently, at the beginning of round $t$, the simulated $\EGalgo$-state of $v$ in $\calE$ equals its state in $\calE^\star$. The $\EGalgo$-messages received by $v$ in that round are also identical in the two executions. Its deterministic state update gives the same $\EGalgo$-state at the end of round $t$, proving the induction step.

Finally, let $v$ satisfy the assumptions of the lemma. The claim at $t=T$ gives the same final $\EGalgo$-state in both executions. Since $\deg_G(v)\geq3$ and $\seefailure=\false$, \Cref{alg:edge-grabbing-protocol} outputs the simulated $\EGalgo$-output at $v$, which is therefore its output in $\calE^\star$.
\end{proof}

Next, we show that the outputs of \Cref{alg:edge-grabbing-protocol} satisfy property~(i) of \Cref{def:edge-grabbing}.

\begin{lemma} \label{EG:lemma:output-is-incident-edge}
    If a surviving node $v$ obtains an output $e$ in \Cref{alg:edge-grabbing-protocol}, then $e = \bot$ if  $\deg(v) < 3$, and $e$ is an incident edge of $v$ if $\deg(v) \geq 3$.
\end{lemma}
\begin{proof}
    If $\deg(v) < 3$, then $v$ outputs $e = \bot$ by the description of \Cref{alg:edge-grabbing-protocol}. In the remainder of the proof, we assume that $\deg(v) \geq 3$.

    If $v$ completes \Cref{alg:edge-grabbing-protocol} with $\seefailure = \false$, then $v$ outputs the edge it obtains from $\EGalgo$. By \Cref{EG:lemma:no-failure-means-happy}, this edge is the output of $v$ in $\calE^\star$, the fault-free execution of $\EGalgo$. As $\deg(v) \geq 3$, the correctness of $\EGalgo$ implies that $e$ is an incident edge of $v$.

    Otherwise, $v$ completes \Cref{alg:edge-grabbing-protocol} with $\seefailure = \true$. Then, in  the round in which $v$ set $\seefailure \gets \true$, it also set $e \gets \{u, v\}$ for a  neighbor $u$ that has either crashed or sent $\seefailure = \true$ to $v$. As $v$ never  modifies $e$ afterwards, $e = \{u, v\}$ is an incident edge of $v$.
\end{proof}

We now establish that adjacent nodes do not \emph{grab} the same edge, which implies property~(ii) of \Cref{def:edge-grabbing}.

\begin{lemma} \label{EG:lemma:adjacent-vertices}
    Let $u$ and $v$ be two adjacent surviving nodes that output edges $e$ and $e'$ respectively in  \Cref{alg:edge-grabbing-protocol}. Then, $e \neq e'$.
\end{lemma}
\begin{proof}
    We assume that $e = e'$. Since both $u$ and $v$ output edges,  \Cref{EG:lemma:output-is-incident-edge} guarantees that this edge is incident to both $u$ and $v$. This implies that $e = e' = \{u, v\}$.

    First, assume that both $u$ and $v$ complete \Cref{alg:edge-grabbing-protocol} with $\seefailure = \false$. By the output rule of \Cref{alg:edge-grabbing-protocol}, both $u$ and $v$ have degree at least three, since they output edges. By \Cref{EG:lemma:no-failure-means-happy}, $e$ and $e'$ are the outputs of $u$ and $v$ in the fault-free execution of $\EGalgo$, and $e = e'$ contradicts the correctness of $\EGalgo$.

    Next, suppose that exactly one of $u$ and $v$ completes \Cref{alg:edge-grabbing-protocol} with $\seefailure=\true$, and assume without loss of generality that this is $u$. Since $v$ survives and its flag remains $\false$, $u$ receives a message with flag $\false$ from $v$ in every round. Thus $u$ cannot select $v$ when setting its flag to $\true$, so its stored edge is not $\{u,v\}$. This contradicts $e=\{u,v\}$.

    Finally, suppose that both nodes complete \Cref{alg:edge-grabbing-protocol} with $\seefailure=\true$. Let $t$ be the first round in which either node sets its flag to $\true$, and assume without loss of generality that $u$ does so in round $t$. At the beginning of that round, $v$ still has flag $\false$, even if both nodes change their flags in the same round. Since $v$ survives, $u$ receives its round-$t$ message with flag $\false$. Hence $u$ cannot select $v$ in round $t$. Its stored edge therefore differs from $\{u,v\}$ and is never changed, contradicting $e=\{u,v\}$.
\end{proof}

Finally, we show that \Cref{alg:edge-grabbing-protocol} is a crash-tolerant edge grabbing algorithm, completing the proof of \Cref{thm:edge-grabbing-transformation}.

\begin{proof}[Proof of \Cref{thm:edge-grabbing-transformation}]
Let $\EGalgo$ be a deterministic $T$-round fault-free \local algorithm for edge grabbing, and consider \Cref{alg:edge-grabbing-protocol} on an input graph $G$.

The algorithm is deterministic, since $\EGalgo$ is deterministic and the smallest identifier determines the choice among eligible neighbors. Each of the $T$ iterations uses one communication round: the flag and any simulated $\EGalgo$-messages are sent together. Every surviving node completes these iterations and then produces an output using local computation.

Consider any execution in the standard crash-fault model. By \Cref{EG:lemma:output-is-incident-edge}, every surviving node $v$ outputs an edge of $G$ incident to $v$ if $\deg_G(v)\geq3$, and outputs $\bot$ otherwise. By \Cref{EG:lemma:adjacent-vertices}, no edge is output by both surviving endpoints. Thus both properties of \Cref{def:edge-grabbing} hold, proving crash-tolerant edge grabbing for any number of crashes.
\end{proof}

\subsection{Sinkless Orientation}\label{sec:sinkless-orientation}

We now focus on \Cref{thm:so-transformation}. The fault-free sinkless orientation problem requires every node to orient its incident edges so that the endpoints of each edge agree on its direction and every node of degree at least three has at least one outgoing edge. To specify these orientations, each node outputs a set $\orientededges$ of directed edges, where $(u,v)$ denotes $\{u,v\}$ directed from $u$ to $v$. Under crashes, we require these conditions only for surviving nodes. As in \Cref{sec:edge-grabbing-protocol}, all degrees and incident edges refer to the input graph $G$, including edges to nodes that crash.

\begin{definition}[Crash-Tolerant Sinkless Orientation]\label{SO:definition}
An algorithm solves crash-tolerant sinkless orientation on $G=(V,E)$ if the following conditions hold in every execution:
\begin{enumerate}[(i)]
    \item Every surviving node $v$ outputs a set $\orientededges$ consisting of exactly one orientation of each edge of $G$ incident to $v$.
    \item Every surviving node $v$ with $\deg_G(v)\geq3$ has at least one outgoing edge in its output $\orientededges$, that is, $(v,u)\in\orientededges$ for some $u\in N_G(v)$.
    \item If adjacent surviving nodes $u$ and $v$ output $\orientededges$ and $\orientededges'$, respectively, then $(u,v)\in\orientededges$ if and only if $(u,v)\in\orientededges'$.
\end{enumerate}
\end{definition}

We first show how any deterministic $T$-round fault-free edge grabbing algorithm yields a deterministic $(T+1)$-round crash-tolerant sinkless orientation algorithm. Given a deterministic $T$-round fault-free edge grabbing algorithm, \Cref{alg:edge-grabbing-protocol} makes it crash-tolerant without increasing its round complexity. We show how one additional round converts the resulting edge grabbing outputs into a crash-tolerant sinkless orientation. We then prove \Cref{thm:so-transformation} by obtaining the required fault-free edge grabbing algorithm from a fault-free sinkless orientation algorithm, without additional communication.

Let $\EGalgo$ be any deterministic $T$-round fault-free \local algorithm for edge grabbing. Each node runs \Cref{alg:edge-grabbing-protocol} with $\EGalgo$ as input for $T$ rounds, obtaining an edge grabbing output $e$; by \Cref{thm:edge-grabbing-transformation}, this procedure is crash-tolerant. In round $T+1$, each node sends $e$ to every neighbor, with $e=\bot$ if it grabbed no edge. A node $v$ first orients its grabbed edge, if any, outwards. For each remaining incident edge $\{u,v\}$, it orients the edge from $u$ to $v$ if it receives an announcement from $u$ that $u$ grabbed $\{u,v\}$, and towards the endpoint with the smaller identifier otherwise. The pseudocode is given in \Cref{alg:sinkless-orientation}, and \Cref{lemma:eg-to-so} establishes its correctness and round complexity.

\begin{algorithm}[H]
\caption{Crash-Tolerant Sinkless Orientation at node $v$}
\label{alg:sinkless-orientation}
\begin{algorithmic}[1]
    \Statex \textbf{Input:} labeled star $\sigma_v$, algorithm $\EGalgo$ and its inputs, and number of rounds $T$.
    \Statex \textbf{Output:} a set $\orientededges$ of directed edges orienting every edge incident to $v$.
    \State Run \Cref{alg:edge-grabbing-protocol} with $\EGalgo$ and $T$, and obtain output $e$.
    \State $\orientededges\gets\emptyset$.
    \State In round $T+1$, send $e$ to every neighbor and receive the delivered messages.
    \If{$e=\{u,v\}\neq\bot$}
    \State Add $(v,u)$ to $\orientededges$.
    \EndIf
    \For{every incident edge $\{u,v\}$ not yet oriented}
        \If {$v$ received $\{u,v\}$ from $u$}
        \State Add $(u,v)$ to $\orientededges$.
        \Else
        \State Orient $\{u, v\}$ towards the endpoint with the smaller identifier.
        \State Add the oriented edge to $\orientededges$.
        \EndIf
    \EndFor
    \State \Return $\orientededges$.
\end{algorithmic}
\end{algorithm}

\begin{lemma}
\label{lemma:eg-to-so}
On graphs of arbitrary maximum degree, every deterministic $T$-round \local algorithm for edge grabbing in the fault-free setting can be transformed into a deterministic $(T+1)$-round \local algorithm for crash-tolerant sinkless orientation in the standard crash-fault model for any number of crashes.
\end{lemma}
\begin{proof}
We first verify that \Cref{alg:sinkless-orientation} satisfies the properties of \Cref{SO:definition}. For property~(i), the output set $\orientededges$ is initially empty. A surviving node $v$ first orients its grabbed edge outwards, if it has one, and then orients each remaining incident edge exactly once in the loop. Only orientations of edges incident to $v$ are added. Hence its output contains exactly one orientation of every incident edge of $G$.

For property~(ii), let $v$ be a surviving node with $\deg_G(v)\geq3$. By \Cref{thm:edge-grabbing-transformation}, it grabs an incident edge. This edge is oriented outwards and its orientation is never changed, so $v$ has an outgoing edge in its output.

For property~(iii), let $u$ and $v$ be adjacent surviving nodes. By \Cref{def:edge-grabbing} and \Cref{thm:edge-grabbing-transformation}, at most one of them grabs $\{u,v\}$. If neither does, neither announces this edge, and both orient it towards the endpoint with the smaller identifier. Otherwise, assume without loss of generality that $v$ grabs $\{u,v\}$. Node $v$ orients it from $v$ to $u$ and announces it in round $T+1$. Since $u$ and $v$ survive the execution, $u$ receives this announcement. Node $u$ does not grab $\{u,v\}$, so it processes this edge in the loop and also orients it from $v$ to $u$. Thus their orientations agree in both cases.

This establishes that \Cref{alg:sinkless-orientation} satisfies the properties of \Cref{SO:definition}. It remains to show that \Cref{alg:sinkless-orientation} is a deterministic $(T+1)$-round algorithm.

By \Cref{thm:edge-grabbing-transformation}, \Cref{alg:edge-grabbing-protocol} is a deterministic $T$-round algorithm. \Cref{alg:sinkless-orientation} adds one communication round followed by deterministic local computation, and hence is a deterministic $(T+1)$-round algorithm, completing the proof.
\end{proof}

We may now conclude the section by presenting the proof of \Cref{thm:so-transformation}.

\begin{proof}[Proof of \Cref{thm:so-transformation}]
Let $\SOalgo$ be any deterministic $T$-round fault-free \local algorithm for sinkless orientation. Define an edge grabbing algorithm $\EGalgo$ as follows. On an input graph $G$, each node first executes $\SOalgo$. A node $v$ with $\deg_G(v)\geq3$ then outputs $\{u,v\}$, where $u$ has the smallest identifier among the neighbors to which $v$ has an outgoing edge. Every node of degree less than three outputs $\bot$.

In every fault-free execution, the correctness of $\SOalgo$ guarantees that each node of degree at least three has an outgoing edge, so its choice is well defined. If $v$ selects $\{u,v\}$, then both endpoints orient this edge from $v$ to $u$, and hence $u$ cannot select it as an outgoing edge. Thus no edge is selected by both endpoints, and $\EGalgo$ solves fault-free edge grabbing. Since the selection rule is deterministic and requires only local computation, $\EGalgo$ is a deterministic $T$-round algorithm.

Applying \Cref{lemma:eg-to-so} to $\EGalgo$ yields a deterministic $(T+1)$-round \local algorithm satisfying \Cref{SO:definition} in the standard crash-fault model for any number of crashes, proving the theorem.
\end{proof}

\section{Conclusions}\label{sec:conclusion}

We introduced consistent neighborhoods as a general primitive for simulating deterministic fault-free local algorithms in the presence of crash failures. For locally checkable problems, this simulation guarantees correctness at surviving nodes whose checking neighborhoods contain no crashed node, provided that the fault-free algorithm is correct on every subgraph of the input graph. We establish upper and lower bounds on the round complexity of constructing consistent neighborhoods. On general graphs, our $O(n)$-round upper bound is tight even for constant radius and a single crash. For every fixed $\Delta\geq3$, we show that exponential dependence on the radius is unavoidable in the worst case. When a crash bound $f\geq1$ is known, $O(R\cdot \Delta\cdot f)$ rounds suffice. Graphs of maximum degree two and graphs of girth greater than $2R$ admit $O(R)$-round algorithms.

Our results also show that crash failures affect locally checkable problems differently across complexity regimes. On graphs of fixed maximum degree, every LCL problem with deterministic fault-free complexity $O(\log^* n)$ retains this upper bound under crashes. In contrast, a problem with fault-free complexity $\Theta(\log n)$ can require $\Theta(n)$ rounds. Sinkless orientation demonstrates the potential of problem-specific transformations: any deterministic $T$-round fault-free algorithm yields a crash-tolerant algorithm using only one additional round, with guarantees that also apply to surviving nodes adjacent to crashes.

\section{AI Methodology}\label{sec:ai-methodology}

The authors developed the main ideas and proof sketches for all results without assistance from large language models (LLMs). We subsequently used the ChatGPT models 5.6sol, 6 Astra, and 6.1sol to help formalize and, in some cases, simplify these ideas. LLMs also contributed concrete ideas in developing the formal arguments, notably the use of propagation graphs in the consistent-neighborhood upper bounds of \Cref{sec:parameter-cn}. For writing and proof-checking throughout the paper, we used the ChatGPT models as well as Claude Fable 5.1 and Opus 5.5; the Claude models were used only for these tasks. LLMs were also used to create all figures. The authors take full responsibility for the correctness, originality, and integrity of all content, including material prepared with LLM assistance and the references.

\clearpage
\phantomsection
\addcontentsline{toc}{section}{\refname}
% Include the precompiled bibliography; no BibTeX database is needed for this version.
\newcommand{\etalchar}[1]{$^{#1}$}

\end{document}